\documentclass[%
 reprint,
amsmath,amssymb,
aps,
pra,
longbibliography
]{revtex4-2}

\usepackage[utf8]{inputenc}
\usepackage{physics}
\usepackage{mathtools}
\usepackage{dsfont}
\usepackage{amsmath}
\usepackage{upgreek}
\usepackage{xcolor}
\usepackage{tikz}
\usepackage{qcircuit}
\usepackage{soul}
\usepackage{bbold}
\usepackage{tabularray}
\usepackage{amsmath}
\usepackage{comment}

\usepackage{graphicx}
\usepackage{float}
\usepackage[export]{adjustbox}
\usepackage[breaklinks=true,colorlinks,citecolor=blue,linkcolor=blue,urlcolor=blue]{hyperref}
\usepackage{cleveref}
\usepackage{orcidlink}
\usetikzlibrary{graphs,graphs.standard}

\usepackage{booktabs}
\usepackage{defs}

\begin{document}

\title{Quantum Simulation of Strongly Correlated Fermion–Phonon Models in Circuit QED}

\author{Tim Bode\orcidlink{0000-0001-8280-3891}$^{1}$}\email[]{t.bode@fz-juelich.de}
\author{Riccardo Roma\orcidlink{0009-0003-0596-1900}$^{1, 2}$}
\author{Stefan Schmitz\orcidlink{0009-0000-3560-0856}$^{1, 3}$}
\author{Alessandro Ciani\orcidlink{0000-0002-8707-0532}$^{1}$}
\author{Dmitriy S. Shapiro\orcidlink{0000-0002-5895-2544}$^{1}$}\email[]{d.shapiro@fz-juelich.de}
\author{Dmitry Bagrets\orcidlink{0000-0002-3985-4834}$^{1, 3}$}
\author{Frank K. Wilhelm\orcidlink{0000-0003-1034-8476}$^{1, 2}$}
\affiliation{
 $^{1}$Institute for Quantum Computing Analytics (PGI-12), Forschungszentrum Jülich, 52425 Jülich, Germany\\
 $^{2}$Theoretical Physics, Saarland University, 66123 Saarbrücken, Germany\\
 $^{3}$Institute for Theoretical Physics, University of Cologne, 50937 Cologne, Germany
}

\date{\today}
 
\begin{abstract}

Gate-based digital quantum simulations offer an exciting new paradigm for studying the many-body physics of strongly correlated systems. In this context, electron-phonon models are challenging for qubit-only quantum simulators, as bosonic degrees of freedom require costly finite-dimensional encodings. Here, we elaborate on an alternative approach based on a digital-analog circuit QED architecture, where fermions are encoded in transmon qubits while bosons are represented directly by microwave resonators.
The central building block of this framework is a qubit-resonator Rabi gate that emulates strong electron-phonon coupling and can be implemented through a sequence of resonant Jaynes-Cummings gates interleaved with layers of single-qubit rotations. Using this Rabi gate as the fundamental unitary operation, we construct quantum circuits for the Hubbard-Holstein and Yukawa-Sachdev-Ye-Kitaev models, which describe, respectively, strongly correlated electrons coupled to phonons and phonon-mediated interactions among Majorana fermions.
We further demonstrate how nonclassical phonon physics and signatures of quantum chaos in these models can be probed through circuit simulations, and develop measurement and variational protocols tailored to near-term superconducting quantum hardware.

\end{abstract}

\maketitle

\section{Introduction}\label{sec:intro}

Electron-phonon (\textit{e-ph}) interactions underlie a wide range of condensed-matter phenomena, ranging from polaron formation in semiconductors~\cite{froehlich_1950, Holstein1959} and molecules~\cite{Dahnovsky_2007} to phase transitions to symmetry-broken states such as superconductivity (SC)~\cite{Marsiglio2008} and charge-density waves (CDW)~\cite{Nosarzewski_2021}. Combining a local \textit{e-ph} coupling (the Holstein model~\cite{Holstein1959}) with on-site electron–electron (\textit{e-e}) interactions of the Hubbard type~\cite{Hubbard1963} yields the Hubbard-Holstein (HH) model~\cite{Berger_1995}. Its phase diagram features a competition between antiferromagnetic, CDW, and SC states~\cite{Costa2020}. The infinite-range, random version of \textit{e-ph} interactions is described by the Yukawa–Sachdev–Ye–Kitaev (Yukawa-SYK) model~\cite{Esterlis_2019}, which predicts resilient Cooper pairing of incoherent electrons as well as quantum chaos.

From the perspective of quantum simulation~\cite{Feynman1982, Lloyd:1996,  Weimer2010, BassmanOftelie_2021,Bravyi2024,PRXQuantum.5.040320,Macridin:2018, Macridin:2018a, Fauseweh2024, castillomoreno2025}, emulating \textit{e-ph} coupling with superconducting circuits can be reduced to implementing the quantum Rabi model (QRM) as an elementary building block.
In circuit QED~\cite{RevModPhys.91.025005,FriskKockum2019,RevModPhys.93.025005,QIN20241}, this is realized by coupling a tunable transmon qubit~\cite{koch2007} to a bosonic mode, namely GHz photons in a high-$Q$ transmission-line resonator.
Utilizing physical resonators can circumvent the considerable overhead associated with encoding bosonic modes using qubits~\cite{Sawaya2020}.
Encodings of a bosonic Hilbert space truncated to $d$ levels are broadly classified into two categories: ``unary'' and ``binary''.
In unary encoding, one uses $d$ qubits to represent $d$ bosonic levels, which leads to  linear growth of the required qubit number when higher excitations are relevant.
Conversely, binary encodings map the level index to a binary string and, therefore, require $\sim \log d$ qubits~\cite{Macridin:2018, Macridin:2018a}, typically at the cost of additional multi-qubit gate overhead (e.g. $\mathsf{CNOT}$s).

In the weak-coupling regime, the QRM reduces to the Jaynes-Cummings model (JCM) by applying the rotating-wave approximation (RWA), which is valid for many experiments~\cite{Xie_2017}. Ultrastrong-coupling regimes--where the RWA breaks down--have also been realized~\cite{Niemczyk2010, Forn-Diaz2017}.
Thus, it is possible to implement a fully \textit{analog} simulator of the QRM. From a quantum-control perspective, however, a more convenient framework is the hybrid \textit{digital-analog} approach~\cite{Mezzacapo2014, Kumar2025, Leppaekangas2025} demonstrated experimentally in Refs.~\cite{Langford2017,Than:2025} for transmon and trapped-ion platforms, respectively. See also~\cite{Crane2026} for an introductory review. 
Within the scheme of Ref.~\cite{Langford2017}, the QRM time evolution is decomposed into (i) analog blocks implemented by a JCM architecture with a tunable transmon, and (ii) digital, phase-controlled single-qubit gates that effectively reproduce the counter-rotating terms. In other words, a \textit{simulated} QRM with  strong coupling is mapped onto a \textit{physical} weakly coupled JCM platform with genuine RWA-type interactions.

The \textit{analog} part of the simulation outlined above entails that the bosonic degrees of freedom are encoded via the Hilbert space of the \textit{resonator}, i.e.~one avoids a digital mapping to auxiliary qubits~\cite{Sawaya2020}. However, one could argue that this scheme is still fully \textit{digital}, as it does not rely on any \textit{analog} gate block~\cite{Kumar2025}, in the sense that the desired qubit-resonator Rabi interaction strength is not implemented \textit{physically} by the hardware. Rather, it is engineered by chaining many applications of the physical gate with varying duration.

In this work, we extend this framework substantially. We begin by providing a detailed theoretical breakdown of qubit-resonator Rabi gates and their circuit QED realizations. We then apply these gates to two representative many-body problems: 
(i) the HH model describing strongly correlated (complex) fermions interacting with phonons, and (ii) the Yukawa-SYK model describing randomly coupled (real) Majorana fermions and phonons.

For the HH model, we construct the Jordan-Wigner-mapped circuits for local \textit{e-ph} coupling and phonon-assisted hopping, analyze a finite-site system by exact diagonalization, and identify a fluctuation-dominated regime with non-Poissonian phonon statistics. We further introduce a hardware-oriented variational Hamiltonian ansatz (VHA) that follows the structure of the HH model and can prepare the relevant ground states with high fidelity. For the Yukawa-SYK model, we show how random Majorana-boson couplings can be compiled into Rabi-gate blocks and use the resulting circuits to probe signatures of quantum chaos.

The ideas developed here are closely related to those recently proposed in Ref.~\cite{Crane:2024}. However, the types of bosonic gates involved and the range of physical models under consideration differ in important respects. In particular, the authors of Ref.~\cite{Crane:2024} focus on the simulation of low-dimensional bosonic lattice gauge theories and quantum $U(1)$ link models. In addition to the coupling between transmons and data resonators, which can be realized within the circuit QED framework, such models require the implementation of bosonic beam-splitter entangling gates. To this end, Ref.~\cite{Crane:2024} proposes the use of high-quality 3D cavities coupled via so-called SNAIL junctions. By contrast, our proposal is based on a natural extension of conventional planar transmon architectures.

The paper is organized as follows. In Sec.~\ref{sec:DA_framework}, we formulate the digital-analog  approach and  introduce the Rabi-gate primitive (\ref{subsec:QRM}) and derive its first- and second-order circuit decompositions (\ref{sec:rabi_gate_1st} and \ref{sec:rabi_gate_2nd}). In Subsec.~\ref{subsec:flux_tune}, we discuss circuit QED implementations based on dispersive controlled displacements and resonant Jaynes-Cummings gates. In Sec.~\ref{sec:HH}, we construct the simulation of the \textit{e-ph} coupled  HH model (\ref{subsec:Hamiltonian_HH} and \ref{subsec:JW}), present exact-diagonalization phase diagrams (\ref{subsec:phase_diag}), and describe Hadamard-test protocols for extracting phonon statistics (\ref{subsec:H_test}). In the following Sec.~\ref{sec:VHA}, we introduce the VHA for ground-state preparation (\ref{subsec:trial_wf}) and an energy-functional estimation protocol (\ref{subsec:energy_meas}). In Sec.~\ref{sec:YSYK}, we address the Yukawa-SYK model: introduce the Hamiltonian (\ref{subsec:Hamiltonian_YSYK}) and derive the corresponding circuit for the Trotter evolution (\ref{subsec:Tr_evol_YSYK}).  In Subsec.~\ref{subsec:QuC},  we provide numerical data on  simulations of quantum-chaotic behavior, including correlation functions and their measurement protocol.   We discuss our results and conclude in Sec.~\ref{sec:discussion}.
In Appendix~\ref{appendix:optimization}, we provide details on the optimization procedure in the VHA. In Appendix~\ref{appendix:lindblad}, the emulation of dissipative processes in the VHA is discussed. Finally, the controlled Rabi gate is constructed in Appendix~\ref{appendix:ctrl_Rabi}. 

\section{Digital-analog framework}
\label{sec:DA_framework}

In this section, we introduce the Rabi gate, an elementary operation that entangles qubit and bosonic degrees of freedom. As discussed below, this gate serves as the central building block for the simulation of electron-phonon interactions in the multi-orbital Hubbard-Holstein and Yukawa-SYK models. Our approach is inspired by Ref.~\cite{Shapiro2025}, where a related gate was employed as a primitive for the quantum simulation of the spin-boson Dicke-Ising model.

We construct quantum circuits implementing the Rabi gate using first- and second-order Trotter decompositions of the underlying quantum Rabi model. We then discuss how the corresponding time evolution can be realized on transmon-resonator architectures, thereby extending the standard toolbox of single- and two-qubit gates.

\subsection{Quantum Rabi Model}
\label{subsec:QRM}

Throughout this work, $\hat X$, $\hat Y$, and $\hat Z$ denote the standard Pauli operators acting on a single qubit. We further define the ladder operators $\hat\sigma^{\pm}=(\hat X \mp i\hat Y)/2$.
The bosonic annihilation and creation operators are denoted by $\hat a$ and $\hat a^\dagger$, respectively, and satisfy the canonical commutation relation $[\hat a,\hat a^\dagger]=1$. We also set $\hbar=k_{\rm B}=1$.

In this section, we model the dynamics of a coupled qubit–resonator system using the quantum Rabi model (QRM) with Hamiltonian
\begin{align}
\label{eq:H_R}
\hat H_{\mathrm{R}} = \hat H_0 + \hat H_X,
\end{align}
where
\begin{align}\label{eq:H_X}
    \hat H_0 = \omega_0 \hat a^\dagger \hat a, \quad \hat H_X = \frac{g}{2} \hat X (\hat a^{\dagger} + \hat a).
\end{align}
Here, $\omega_0$ denotes the resonator frequency and $g$ the qubit-resonator coupling strength. We note that \cref{eq:H_R} corresponds to the QRM with the qubit frequency set to zero. This choice simplifies the resulting implementation of the Rabi gate by eliminating the need for additional single-qubit rotations.

Within the digital quantum simulation framework, time is discretized into $L$ intervals, $t_p=t_0+p\tau$, where $\tau$ denotes the Trotter step and $0\le p\le L-1$. 
The exact evolution operator can then be represented as
\begin{equation}\label{H_R_trotterized_exact}
e^{-i\hat H_{\rm R}(t_L-t_0)} = e^{-i \hat H_0 t_L}\left(\prod_{p=0}^{L-1} \hat S_{\rm R}(t_p+\tau,t_p)\right) e^{i \hat H_0 t_0} ,
\end{equation}
i.e., as a product of Trotterized evolution operators $\hat S_{\rm R}(t_p+\tau,t_p)$, which we shall refer to as Rabi gates. Their exact form is given by
\begin{equation}
\label{SR_exact}
\hat S_{\rm R}(t_p+\tau,t_p) = e^{i \hat  H_0 (t_p+\tau)} e^{-i \hat H_{\rm R}\tau} e^{-i \hat H_0 t_p}.
\end{equation}
These gates are defined in the rotating frame with respect to the free Hamiltonian $\hat H_0$.
  
\subsubsection{Rabi gate: First-order approximation and controlled-$X$ displacement}
\label{sec:rabi_gate_1st}

The first-order approximation to the exact Rabi gate,
$\hat S_{\rm R}=\hat S_{\rm R}^{(1)}+{\cal O}(\tau^2)$, 
is obtained from the Lie-Trotter decomposition
\begin{equation}\label{approx_1}
e^{-i \hat H_{\rm R}\tau}= e^{-i \hat H_{\rm 0}\tau}e^{-i \hat H_{\rm X}\tau}+{\cal O}(\tau^2). 
\end{equation}
Substituting \cref{approx_1} into \cref{SR_exact} and performing straightforward algebra yields
\begin{equation}\label{S_R_1}
\hat S_{\rm R}^{(1)}(t_p+\tau,t_p)= \exp(-i \frac{g\tau}{2}(\hat a e^{-i \omega_0 t_p} + \hat a^\dagger e^{i \omega_0 t_p})\hat X). 
\end{equation}
The resulting unitary coincides with the controlled-$X$ displacement gate~\cite{stavenger2022bosonic},
\begin{align}
\label{eq:sx}\mathsf{CD}_{\mathsf{X}}(\varphi) := \exp( \hat{X}(\varphi \hat{a}^{\dagger} - \varphi^* \hat{a})),
\end{align}
parameterized by a complex displacement amplitude $\varphi$. Consequently, the exact Rabi gate can be approximated to first order in $\tau$ as
\begin{align}
\label{eq:sx_2}\hat S_{\rm R}(t_p+\tau, t_p)\approx\mathsf{CD}_{\mathsf{X}}\left(\varphi=\frac{-i g \tau}{2}e^{i \omega_0 t_p}\right).
\end{align}
The $\mathsf{CD}_{\mathsf{X}}$ gate can be realized using a fixed-frequency transmon dispersively coupled to a microwave resonator. Its experimental implementation will be discussed below.

\subsubsection{Rabi gate: Second-order approximation.  Jaynes-Cummings gate}
\label{sec:rabi_gate_2nd}

We now turn to a more accurate construction of the Rabi gate based on a second-order Trotter decomposition,
$\hat S_{\rm R}=\hat S_{\rm R}^{(2)}+{\cal O}(\tau^3)$,
originally proposed in Refs.~\cite{Mezzacapo2014,Langford2017}. The key idea is to decompose the Rabi Hamiltonian into the sum of two non-commuting contributions,
\begin{equation}\label{H_R_JC_AJC}
\hat H_{\rm R} =\frac{1}{2}\left(\hat H_{\rm JC}+\hat H_{\rm AJC}\right).
\end{equation}
The first contribution, 
\begin{equation}\label{H_JC}
\hat H_{\rm JC} =\hat H_0 -\frac{\omega_{\rm q}}{2}\hat Z+g(\hat a \hat\sigma^+ + \hat a^\dagger \hat \sigma^-),
\end{equation}
is the Jaynes-Cummings (JC) Hamiltonian describing a qubit of frequency $\omega_{\rm q}$ coupled to a bosonic mode within the RWA, where the interaction is expressed in terms of the ladder operators $\hat\sigma^\pm$.
The second contribution in \cref{H_R_JC_AJC} reads
\begin{equation}
\label{eq:H_AJC}
\hat H_{\rm AJC} =\hat X\hat H_{\rm JC}\hat X= \hat H_0 +\frac{\omega_{\rm q}}{2}\hat Z+g(\hat a \hat\sigma^- + \hat a^\dagger \hat \sigma^+),
\end{equation}
This Hamiltonian corresponds to the anti-Jaynes-Cummings (AJC) model. It is connected to $\hat H_{\rm JC}$ by $\hat X$-transformation and describes a qubit with effective frequency 
$-\omega_{\rm q}$ coupled to the bosonic mode through counter-rotating terms.

\begin{figure*}[htb!]
	\begin{center}
		\includegraphics[width=0.9\linewidth]{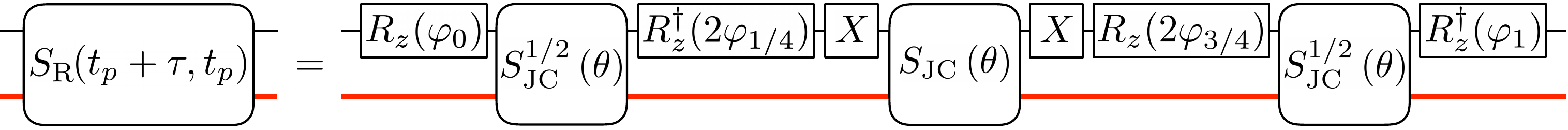}
		\caption
		{\label{fig:S_R} Circuit diagram for the Rabi gate $\hat S_{\mathrm{R}}(t_{p}+\tau, t_{p})$ introduced in \cref{S_R_2}. The phases in \textit{Z}-rotations are     $\varphi_q = \omega_0 (t_p + q \tau)$ with $q=0,  \frac{1}{4}, \frac{3}{4}, 1$ corresponding to the elapsed times in pulse sequence 
        (\ref{eq:U_R_Trotter}). Red line corresponds to the resonator mode, and the black one is the  data qubit. The phase in the JC gate defined by \cref{S_JC_def}  is $\theta=g\tau/2$.}
	\end{center}
\end{figure*}

The second-order approximation is based on the symmetrized Trotter decomposition
\begin{equation}\label{exp_2}
e^{-i \hat H_{\rm R}\tau }=
e^{-i \frac{\tau}{4}\hat H_{\rm JC}}e^{-i \frac{\tau}{2}\hat H_{\rm AJC}}e^{-i \frac{\tau}{4}\hat H_{\rm JC}} +{\cal O}(\tau^3).
\end{equation}
Introducing the evolution operator of the JC Hamiltonian,
$\hat U_{\rm JC}(t_2, t_1) := \exp(- i (t_2-t_1) \hat H_{\rm JC} )$,
the above approximation can be equivalently rewritten as
\begin{align}\label{eq:U_R_Trotter}
    \begin{split}
            \hat U_{\rm R}(t_{p} + \tau, t_{p}) &=  \hat U_{\rm JC} (t_p + \tau, t_p + \tfrac {3 \tau} 4)  \\
            &\times \hat X\, \hat U_{\rm JC} (t_p + \tfrac {3 \tau} 4 , t_p + \tfrac {\tau} 4 ) \, \hat X \\
            &\times \hat U_{\rm JC} (t_p + \tfrac {\tau} 4 , t_p) + {\cal O}(\tau^3).
    \end{split}
\end{align}
This identity follows directly from~(\ref{eq:H_AJC}), which expresses the AJC Hamiltonian in terms of the JC Hamiltonian. 
Equation~(\ref{eq:U_R_Trotter}) shows that the evolution generated by the Rabi Hamiltonian over a single Trotter interval is decomposed into three consecutive segments: evolution under the JC Hamiltonian for a duration $\tau/4$, evolution under the AJC Hamiltonian for $\tau/2$, and a final JC evolution over the remaining $\tau/4$.

Our next goal is to express this Trotterized time evolution as a sequence of quantum gates that can be implemented on superconducting quantum hardware.
To this end, we move to the interaction picture with respect to the Hamiltonian
\begin{equation}
    \hat h =\hat H_0 -\frac{1}{2} \omega_{\rm q} \hat Z,
\end{equation}
which coincides with the free part of the JC Hamiltonian.
Operators in this interaction picture will be denoted by calligraphic symbols, e.g., $\hat  {\cal S}_{\rm JC}(t_1,t_2)$ and $ \hat  {\cal S}_{\rm R}(t_1,t_2)$.
They are related to corresponding evolution operators in the stationary frame by the relation analogous
to~\cref{SR_exact}, that is
\begin{equation}
\hat  {\cal S}(t_2,t_1) = e^{ i \hat h t_2} \hat U(t_2, t_1)  e^{  - i \hat h t_1}.
\end{equation}
In particular, for the JC evolution operator one can write
\begin{align}
 \hat {\cal S}_{\rm JC}(t_2, t_1) = \mathcal{T} \exp\!\left(\! - i g\int\limits_{t_1}^{t_2}\! \left( \hat a(t) \hat\sigma^+(t) + {\rm h.c.} \right)\! dt\! \right)\! , 
\end{align}
where $\hat a (t) = \hat a e^{-i\omega_0 t}$ and $\hat\sigma^+(t) = \hat\sigma^+ e^{i\omega_{\rm q} t}$, and
$\mathcal T$ denotes the time-ordering operator.
When the qubit is far detuned from the resonator, $g\ll|\omega_0-\omega_{\rm q}|$, the operator $\hat{\cal S}_{\rm JC}(t_2,t_1)$ remains close to the identity because the rapidly oscillating terms average to zero.
By tuning the qubit into resonance, $\omega_{\rm q}(t)=\omega_0$, for a duration $\tau/2$
(see also Sec.~\ref{sec:Resonant_regime}), two phases compensate each other exactly and
the following \textit{analog} JC gate characterized by the angle $\theta = g \tau/2$ is realized,
\begin{equation}
\label{S_JC_def}
\hat S_{\rm JC} (\theta) = e^{ - i \theta ( \hat a^+ \hat  \sigma_- + \hat  a \hat  \sigma_+ )}.
\end{equation} 
Setting $\omega_{\rm q} = \omega_0$ hereafter, 
we can express the one-step Rabi evolution operator $\hat {\cal S}_{R}(t_p+\tau, t_p)$ 
via JC gates and simple one-qubit gates,
\begin{align}\label{eq:S_Rabi}
\begin{split}
  \hat {\cal S}_{\rm R}(t_p+\tau, t_{p}) &= \hat S^{1/2}_{\rm JC}(\theta) \hat X(\varphi_{3/4}) 
    \hat S_{\rm JC}(\theta) \hat X (\varphi_{1/4}) S^{1/2}_{\rm JC}(\theta) \\ 
    & + {\cal O}(\tau^3).
\end{split}
\end{align}
Here time-dependent rotated $X$-gates,  
\begin{equation}
\hat X(\varphi) = \hat R_z(\varphi) \, \hat X\, \hat R^\dagger_z(\varphi), 
\end{equation}
are defined with the help of a standard {\it digital} gate $ \hat R_z(\varphi) = e^{- i\varphi \hat Z/2}$
at particular angles
$\varphi_{1/4} = \omega_0 \tau (p+\tfrac 1 4)$ and $\varphi_{3/4} = \omega_0 \tau (p+\tfrac 3 4)$. 
The structure of the Rabi gate in \cref{eq:S_Rabi} directly reflects the second-order Trotter decomposition~\cref{exp_2}. The rotated operators $\hat X(\varphi)$ arise naturally from transforming the original $\hat  X$ gate into the interaction picture defined by $\hat h$.

At this stage it remains to relate the Rabi gate~\cref{eq:S_Rabi} to its analog~\cref{SR_exact} 
defined in the rotating frame with respect to
the free resonator Hamiltonian $\hat H_0$. 
These two representations of the Rabi gate are connected by the unitary transformation
\begin{equation}
    \hat S_{\rm R}(t_{p}+\tau , t_p)  = e^{\frac i 2 \omega_0 (t_{p}+\tau) \hat Z }
    \hat {\cal S}_{\rm R}(t_{p}+\tau , t_p) e^{-\frac i 2 \omega_0 t_{p} \hat Z },
\end{equation}
achieved by single qubit rotations along $z$-axis.
Using ~\cref{eq:S_Rabi}, we finally obtain the second-order approximation to the Rabi gate,
\begin{align}\label{S_R_2}
    \begin{split}
   \hat  S_{\rm R}^{(2)}(t_p+\tau,t_p) &= \hat R_z^\dagger(\varphi_1)\hat S_{\rm JC}^{1/2}\left(\theta\right)  \hat R_z(2\varphi_{3/4})\\  
    &\times  \hat X \hat S_{\rm JC}\left(\theta\right) \hat X \\  
    &\times \hat R^\dagger_z(2\varphi_{1/4}) \hat S_{\rm JC}^{1/2}\left(\theta\right) \hat R_z(\varphi_0),        
    \end{split}
\end{align}
where $\theta=g\tau/2$ and $\varphi_q = \omega_0 (t_p + q \tau)$.
Equation~(\ref{S_R_2}) shows that the Rabi gate can be implemented as a sequence of three \emph{analog} JC gates interleaved with the standard \emph{digital} gates $\hat X$ and $\hat R_z(\varphi)$. The corresponding rotation angles are determined entirely by the simulation time $t_p$. The resulting quantum circuit is shown in Fig.~\ref{fig:S_R}. Throughout the remainder of this work, thin black horizontal lines represent qubits, thick red horizontal lines represent resonators, and dashed lines denote qubits or resonators that remain idle during a given circuit layer.

\subsection{Transmon-resonator architectures}\label{subsec:flux_tune}

In this section, we discuss implementations of the Rabi gates in a circuit QED architecture, where a transmon is capacitively coupled to an \textit{LC} resonator. As follows from the standard circuit quantization procedure \cite{vool2017, rasmussen2021, ciani2024}, the Hamiltonian of the system reads
\begin{align}\label{eq:freqtunh}
    \begin{split}
        \hat{H}_{\rm sys} &= \omega_{\rm tr} \hat{b}^{\dagger} \hat{b} -\frac{E_{\rm C}}{2}  \hat{b}^{\dagger} \hat{b}^{\dagger} \hat{b} \hat{b} + \omega_{\rm res} \hat{a}^{\dagger} \hat{a}  \\ 
        &- \frac{\Tilde{g}}{2} {(\hat{a}^{\dagger} - \hat{a})} {(\hat{b}^{\dagger} - \hat{b})}.
    \end{split}
\end{align}
The transmon is approximated here as a bosonic oscillator with the excitation frequency $\omega_{\rm tr}$ and the quartic Kerr-type nonlinearity, corresponding to the Duffing approximation. The strength of the nonlinearity is determined by the charging energy $E_{\rm C}$ of the capacitor. The annihilation and creation operators of the transmon are $\hat{b}$ and $\hat{b}^{\dagger}$, respectively, and satisfy the commutation relation $[\hat{b}, \hat{b}^{\dagger}] = 1$.
The \textit{physical} parameters of the circuit are $\omega_{\rm tr}$ (transmon frequency), $E_{\rm C}$ (anharmonicity), $\tilde g$ (transmon-resonator coupling strength), and $\omega_{\rm res}$ (resonator frequency).

We next perform the RWA by neglecting the terms proportional to $\hat a^\dagger \hat b^\dagger$ and $\hat a \hat b$, which do not conserve the total excitation number. The RWA is valid provided that $\Tilde{g}\ll\{\omega_{\rm tr} , \omega_{\rm res}\}$, a condition that is well satisfied in typical experimental settings.
A further standard approximation consists of truncating the transmon Hilbert space to its two lowest energy levels. Under this approximation, the bosonic operators are mapped onto Pauli operators according to $\hat b^\dagger\to \hat \sigma^+$, $\hat b\to \hat \sigma^-$, and $\hat b^\dagger\hat b \to \frac{1}{2}-\frac{1}{2}\hat Z$. This two-level description is justified provided that the characteristic gate times are longer than $1/E_{\rm C}$.

In the remainder of this section, we consider two operating regimes of the transmon -- the dispersive and resonant limits -- which enable the implementation of 
the controlled-\textit{X}-displacement gate $\mathsf{CD}_{\mathsf{X}}$ in \cref{eq:sx_2} and the JC gate $\hat S_{\rm JC}$ in \cref{S_JC_def}, respectively.

\subsubsection{Dispersive regime}

\begin{figure}[t!]
    \centering
    \includegraphics[width=0.7\linewidth]{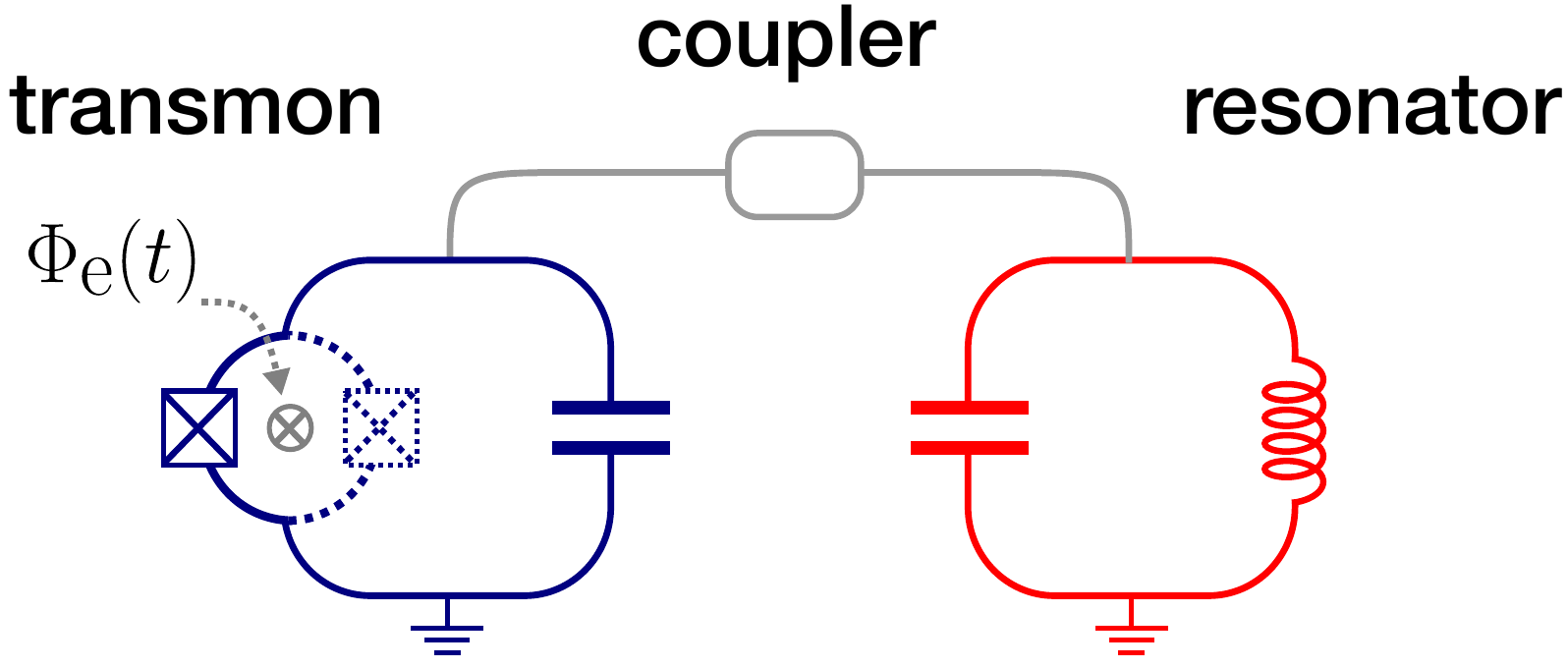}  \caption{Equivalent electric  circuit of the coupled transmon-resonator system. The transmon circuit is depicted in blue, the resonator circuit in red. The transmon and the resonator are coupled by a coupler which can be either a fixed capacitor, another bus resonator, or a tunable coupler~\cite{RevModPhys.93.025005}. Depending on the target gate set, the transmon can be either fixed-frequency with  a single Josephson junction, or a flux-tunable transmon with a SQUID loop, where the external flux  $\Phi_{\mathrm{e}}(t)$ can be used to tune the fundamental frequency of the transmon. }\label{fig:trcircuit}
\end{figure}

The $\mathsf{CD}_{\mathsf X}$ gate can be implemented using a fixed-frequency transmon operated in the dispersive regime, $\tilde g \ll |\Delta|$, where $\Delta=\omega_{\rm tr}-\omega_{\rm res}$ denotes the transmon-resonator detuning. In this limit, the system Hamiltonian reduces to~\cite{blais2004, RevModPhys.93.025005}
\begin{align} \label{H_disp}
\hat{H}_{\mathrm{disp}} = -\frac{1}{2}\omega_{\rm tr} \hat{Z} + (\omega_{\rm res} -\chi \hat{Z}) \hat{a}^{\dagger} \hat{a},
\end{align}
where
$
\chi=-\Tilde  g^{2}E_{\rm C}/\Delta^2
$
is the dispersive shift, assuming $E_{\rm C}\ll \Delta$. The corresponding circuit architecture is shown in Fig.~\ref{fig:trcircuit}.

By applying microwave drive pulses to both the resonator and the qubit, one can engineer the control Hamiltonian
\begin{align}
    \begin{split}
        \hat{H}_{\mathrm{drive}}(t) =\; &\varepsilon_{\rm d}(t) \cos(\omega_{\rm d} t +\; \phi_{\rm d}) (\hat{a} + \hat{a}^{\dagger}) \\
        + &\mathcal{E}_{\rm d}(t) \cos(\Omega_{\rm d} t + \xi_{\rm d}) \hat{X}.
    \end{split}
\end{align}
The resonator (qubit) drive is characterized by the envelope function $\varepsilon_{\rm d}(t)$ ($\mathcal{E}_{\rm d}(t)$), the drive frequency $\omega_{\rm d}$ ($\Omega_{\rm d}$), and the phase $\phi_{\rm d}$ ($\xi_{\rm d}$).
The unitary evolution generated by the Hamiltonian $\hat H_{\rm disp}+\hat H_{\rm drive}(t)$ enables the implementation of the controlled-$Z$-displacement gate~\cite{Campagne-Ibarcq2020,Eickbusch2022,Sivak2023}
\begin{align}
\label{eq:cdisp}
    \mathsf{CD}_{\mathsf{Z}}(\varphi) = \exp( \hat{Z}\left(\varphi \hat{a}^{\dagger} - \varphi^* \hat{a} \right)).
\end{align}
The desired $\mathsf{CD}_{\mathsf{X}}(\varphi)$ gate can then be obtained from $\mathsf{CD}_{\mathsf{Z}}(\varphi)$ by a basis rotation generated by the Hadamard gate
$\mathsf{H}$. 
Specifically,
$\mathsf{CD}_{\mathsf{X}}(\varphi) = \mathsf{H}\,\mathsf{CD}_{\mathsf{Z}}(\varphi)\, \mathsf{H}$.

\subsubsection{Resonant regime}
\label{sec:Resonant_regime}

\begin{figure}[t!]
    \centering
    \includegraphics[width=\linewidth]{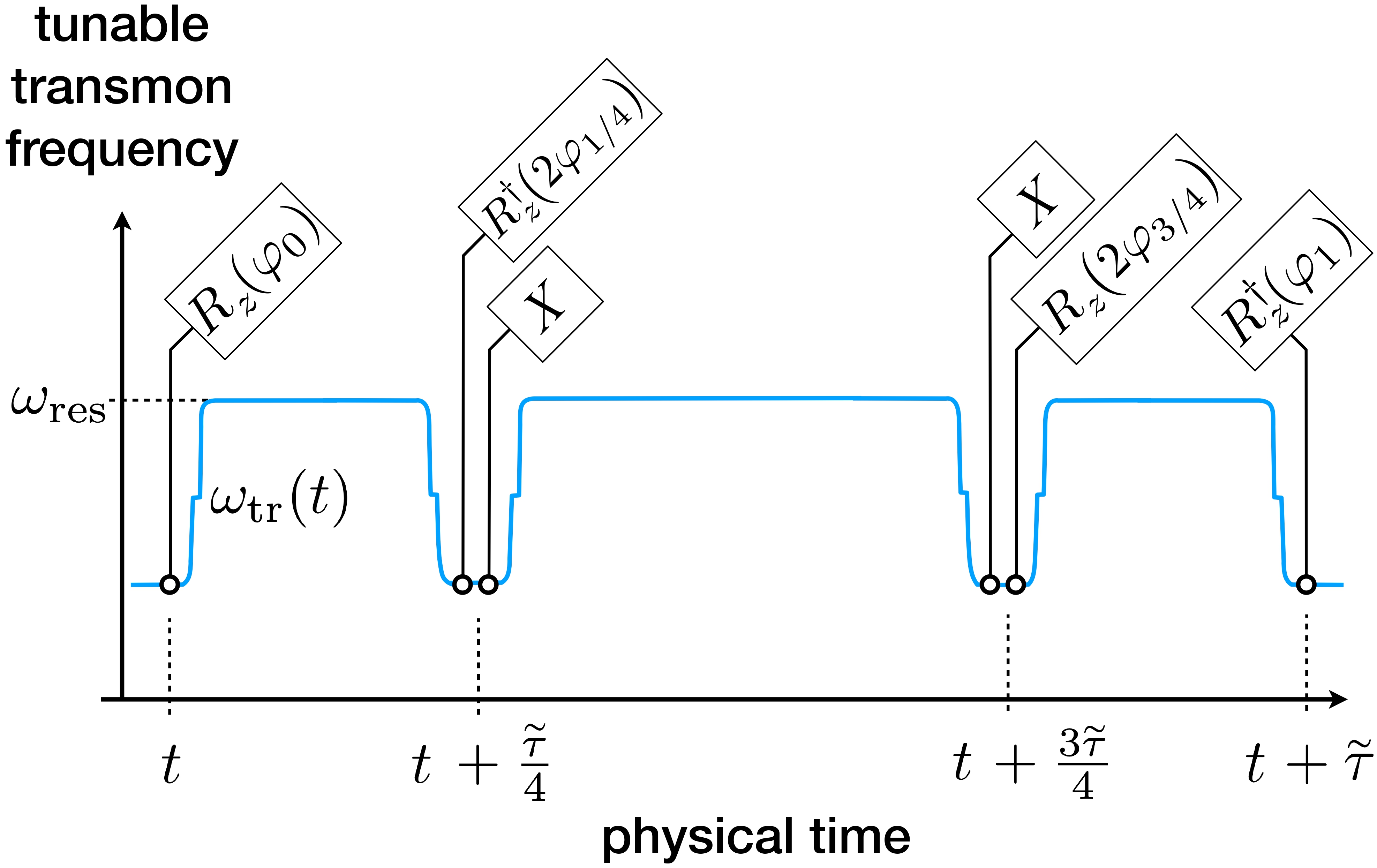}  
    \caption{
Schematic of the pulse sequence implementing the Rabi-gate protocol in \cref{S_R_2}. The tunable transmon frequency $\omega_{\rm tr}(t)$ is modulated as a function of the physical time $t$ to realize three resonant JC gates with durations $\tilde{\tau}/4$, $\tilde{\tau}/2$, and $\tilde{\tau}/4$, respectively, where $\tilde{\tau}$ denotes the total duration of the Rabi gate. The digital gates are assumed to be much faster than the JC gates. The pulse profile of $\omega_{\rm tr}(t)$ may include additional short buffer intervals at the pulse edges to compensate for dynamical phases generated by the free evolution terms in \cref{H_JC_sys}. The JC gate accumulates the phase $\theta=\tilde g \tilde \tau /2$, where $\tilde g$ and $\tilde\tau$ denote the  \textit{physical} qubit-resonator coupling strength and gate duration, respectively. This phase is identified with the simulation parameters through the relation $\theta=g\tau/2$.}\label{fig:sequence}
\end{figure}

The JC gate in \cref{S_JC_def} is implemented using the tu\-na\-ble-fre\-quency transmon architecture shown in Fig.~\ref{fig:trcircuit}, where the transmon frequency $\omega_{\rm tr}(t)$ is controlled in time by an external magnetic flux $\Phi_{\rm e}(t)$\cite{Langford2017, valadares2026}. By applying a flux pulse such that the resonance condition $\omega_{\rm tr}(t)=\omega_{\rm res}$ is satisfied, the system Hamiltonian reduces to
\begin{equation}\label{H_JC_sys}
\hat H_{\rm sys} = \omega_{\rm res}\left(\hat a ^\dagger \hat a -\frac{1}{2}\hat Z\right )+\frac{\tilde g}{2}(\hat a \hat\sigma^+ + \hat a^\dagger \hat \sigma^-).
\end{equation}
Allowing the resonantly coupled system to evolve for a duration $\Tilde\tau$ and subsequently detuning the transmon from the resonator yields the gate $\hat S_{\rm JC}(\theta=\tilde g \Tilde\tau/2)$. The corresponding pulse sequence controlling $\omega_{\rm tr}(t)$ within the Rabi-gate protocol is shown in Fig.~\ref{fig:sequence}.
By adjusting the \textit{physical} interaction time $\tilde\tau$, the phase accumulated by the hardware gate can be matched to that of the \textit{simulated} evolution, i.e.,
$\theta=\tilde g \tilde \tau /2= g \tau /2$.

\section{Hubbard-Holstein Model}
\label{sec:HH}

The Hubbard model provides a minimal setting for describing strongly
correlated lattice fermions. For $N$ orbitals, a noninteracting
Hubbard-type problem reduces to a quadratic single-particle Hamiltonian
acting on $2N$ spin-orbitals. Once local Coulomb repulsion is included, the
problem has to be formulated in the full fermionic Fock space, whose dimension
scales as $4^N$. This exponential growth sets the main computational
bottleneck for classical exact diagonalization.

This motivates the variational cluster approximation
(VCA)~\cite{Senechal2002,Potthoff2003,PotthoffAichhornDahnken2003}, where
the lattice problem is decomposed into smaller interacting clusters. Including
electron-phonon interactions, as in the Hubbard-Holstein model, further
enlarges the Hilbert space due to the additional bosonic degrees of freedom.
This model is nevertheless compatible with the VCA framework~\cite{Payeur2011}. Within VCA, the
cluster spectra and Green functions are computed exactly, while the coupling
between clusters is incorporated through the self-energy or, equivalently,
through the Luttinger-Ward functional. The accuracy of the method is
therefore limited by the largest cluster that can be solved reliably.

This limitation makes VCA a natural setting for hybrid quantum-classical
algorithms. The interacting cluster can be simulated on a quantum processor,
and the resulting Green functions can be used as input for the classical VCA
procedure~\cite{Wecker2015,Bauer2016,Cade2020,Endo2020,bishop2023quantum}.
Motivated by this perspective, we consider below a Hubbard-Holstein-type
cluster in a minimal dimer configuration. The dimer retains hopping, on-site
energies, and local Coulomb repulsion, while coupling the fermions to a single
bosonic mode. Although simplified, it captures the interplay between electronic
correlations and boson-assisted processes. It therefore provides a useful toy
model for developing quantum algorithms based on Trotterized real-time
evolution, which can be exploited for Green-function estimation.

\subsection{Hamiltonian}\label{subsec:Hamiltonian_HH}

The Hubbard-Holstein  model extends the standard Hubbard model by incorporating the interaction of electrons with vibrational degrees of freedom. Its Hamiltonian is given by
\begin{align}\label{eq:H_H}
    \hat H_{\rm HH} = \hat H_{\rm Hubbard} + \hat H_{\rm Holstein} .
\end{align}
For a one-dimensional chain, the Hubbard contribution takes the form
\begin{align}\label{eq:H_elph}
\begin{split}
    \hat H_{\rm Hubbard} = 
    V\sum_{\sigma, i=1}^{N-1}  \left(\hat{c}^{\dagger}_{i,\sigma}\hat{c}^{\phantom{\dagger}}_{i+1,\sigma} + \mathrm{h.c.}\right)  \\
+ \sum_{\sigma, i=1}^{N} \varepsilon_{i,\sigma}   \hat n^{\phantom{\dagger}}_{i,\sigma}  +
   U \sum_{i=1}^N \hat n^{\phantom{\dagger}}_{i,\uparrow} \hat n^{\phantom{\dagger}}_{i,\downarrow} .
\end{split}
\end{align}
Here, the fermionic creation and annihilation operators with spin projection $\sigma\in\{\uparrow,\downarrow\}$ satisfy the canonical anticommutation relations,
$\{c^{\phantom{\dagger}}_{i, \sigma}, c^\dagger_{j, \sigma'}\} = \delta_{ij}\delta_{\sigma\sigma'}$, $\{c^{\phantom{\dagger}}_{i, \sigma}, c^{\phantom{\dagger}}_{j, \sigma'}\} = \{c^{\dagger}_{i, \sigma}, c^{\dagger}_{j, \sigma'}\} =0  $.
The parameters $\varepsilon_{i,\sigma}$ and $V$ denote the on-site energies and nearest-neighbor hopping amplitude, respectively.
The Holstein contribution in (\ref{eq:H_H}) reads
\begin{align}\label{eq:H_HOLSTEIN}
    \begin{split}
       \hat H_{\rm Holstein}  =\; &\omega_0 \hat{a}^{\dagger} \hat{a}^{\phantom{\dagger}} + (\hat{a}^{\dagger}+\hat{a})\Bigg[g_0 \sum_{\sigma, i=1}^{N}  \hat n^{\phantom{\dagger}}_{i,\sigma}\\
       +\; &g \sum_{\sigma, i=1}^{N-1}  \left(\hat{c}^{\dagger}_{i,\sigma}\hat{c}^{\phantom{\dagger}}_{i+1,\sigma} + \mathrm{h.c.}\right)\Bigg].
    \end{split}
\end{align} 
This model is simplified in that it contains only a single phonon mode of frequency $\omega_0$. At the same time, the electron-phonon interaction is generalized beyond the conventional Holstein coupling. Besides the standard coupling between the phonon displacement field and the total electron density, with coupling constant $g_0$, the Hamiltonian includes a phonon-assisted hopping term with coupling strength $g$. As shown below, this additional interaction substantially enhances the entanglement of the many-body ground state.

\subsection{Jordan-Wigner representation. Quantum gates}\label{subsec:JW}

We employ the Jordan-Wigner transformation to map the fermionic operators
$(\hat c_{1,\uparrow},\ldots,\hat c_{N,\uparrow},\hat c_{1,\downarrow},\ldots,\hat c_{N,\downarrow})$
onto spin operators
$\hat X_{i,\sigma}$, $\hat Y_{i,\sigma}$, and $\hat Z_{i,\sigma}$,
where $1\leq i\leq N$ and $\sigma=\uparrow,\downarrow$.
For the fermions from the first half of the string corresponding to $\sigma=\uparrow$, one has
\begin{equation}
\hat c_{i,\uparrow}=\frac{1}{2}\left(\hat X_{i,\uparrow}+i \hat Y_{i,\uparrow}\right)\prod\limits_{j=1}^{i-1} \hat Z_{j,\uparrow}. \label{JW_uparrow}
\end{equation}
For the second half, corresponding to $\sigma=\downarrow$,
\begin{equation}
\hat c_{i,\downarrow}=\frac{1}{2}\left(\hat X_{i,\downarrow}+i \hat Y_{i,\downarrow}\right)\Bigg(\prod\limits_{j=1}^{i-1} \hat Z_{j,\downarrow}\Bigg) \Bigg(\prod\limits_{k=1}^{N} \hat Z_{k,\uparrow}\Bigg).
\end{equation}
This mapping casts the Hubbard-Holstein Hamiltonian into a form suitable for quantum simulation. In the remainder of this section, we consider each contribution to $\hat H_{\rm HH}$ separately and construct the corresponding unitary evolution operators, which serve as the elementary quantum gates of the simulation protocol.

\begin{figure}[t!]
	\begin{center}
		\includegraphics[width=\linewidth]{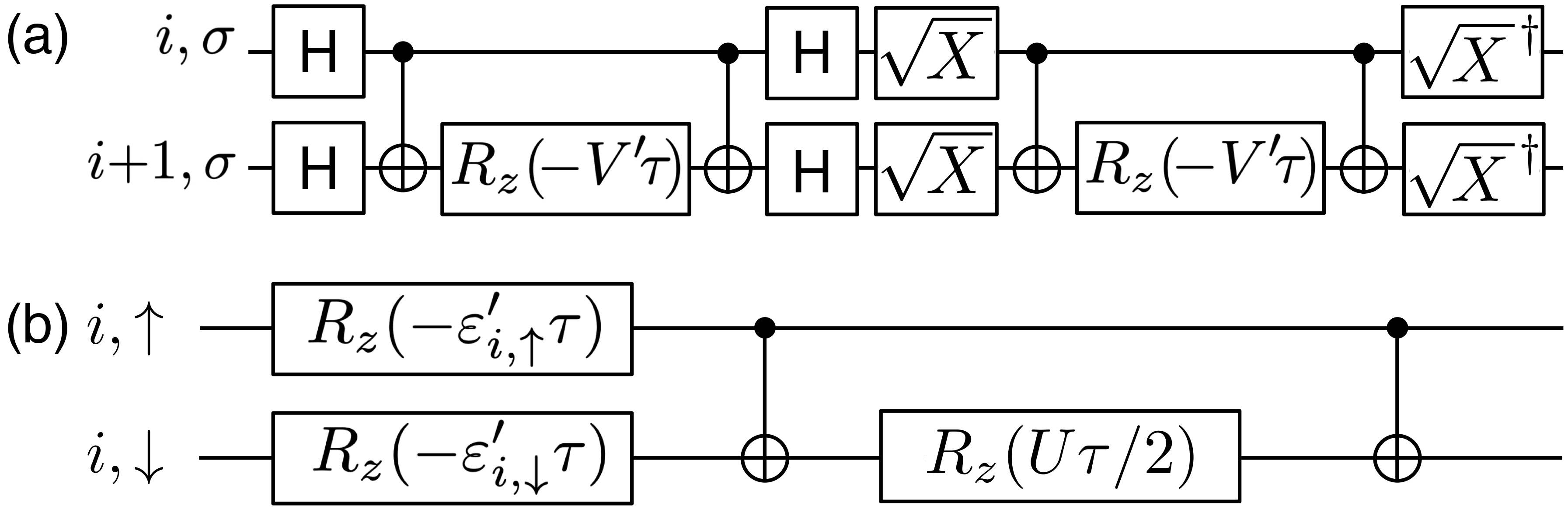}
		\caption
		{\label{fig:fermionic_hop0}  (a) Circuit diagram for implementing a single Trotter step $\sim \mathsf{XX}(\phi)\mathsf{YY}(\phi)$   that corresponds to  the hopping term in \cref{eq:HH_JWrot} with  the amplitude $V'$. Here, the phase $\phi = - V'\tau$   and   $\sqrt{  \hat X}= \exp(-i \pi \hat X/4)$. (b) Circuit diagram for implementing a single Trotter step with the on-site  and Coulomb interaction terms in \cref{eq:HH_JWrot}, i.e.\ $\mathsf{ZZ}_i(U\tau / 4) \exp\big(i \tau (\varepsilon_{i, \uparrow }'\hat Z_{i, \uparrow }+\varepsilon_{i, \downarrow }'\hat Z_{i, \downarrow} ) / 2\big)$, with the Ising-like unitary $\mathsf{ZZ}_i(\phi)$ defined in \cref{eq:U_coulomb_ZZ}.}
	\end{center}
\end{figure}

The hopping term in \cref{eq:H_elph} is mapped to
\begin{align}\label{eq:fermionic_hop0}
    \begin{split}
        \hat c_{i+1,\sigma}^\dagger \hat c_{i,\sigma}^{\phantom{\dagger}} + \mathrm{h.c.} = \frac{1}{2}\left( \hat X_{i,\sigma}\hat X_{i+1, \sigma} + \hat Y_{i, \sigma} \hat Y_{i+1, \sigma}\right).
    \end{split}
\end{align}
The corresponding Trotterized time evolution is readily decomposed into single-qubit and $\mathsf{CNOT}$ gates~\cite{Troyer_fermions_2015,bishop2023quantum}, as illustrated in Fig.~\ref{fig:fermionic_hop0}(a). This construction defines the elementary two-qubit gates
\begin{align}\label{eq:U_hop}
    \begin{split}
        \mathsf{XX}_{i,\sigma}(\phi) &:= \exp\rbs{-i \phi {\hat X_{i,\sigma}\hat X_{i+1, \sigma}   }}, \\
        \mathsf{YY}_{i,\sigma}(\phi) &:= \exp\rbs{-i \phi { \hat Y_{i, \sigma} \hat Y_{i+1, \sigma}}  }. 
    \end{split}    
\end{align}

The local fermion number operator transforms according to
$\hat n_{i,\sigma}=\frac{1}{2}\rbs{\mathbb{1}-\hat Z_{i,\sigma}}$,
while the Hubbard interaction becomes
\begin{align}\label{eq:U_coulomb}
    \begin{split}
         \hat n_{i,\uparrow} \hat n_{i,\downarrow} &= \frac{1}{4} \rbs{\mathbb{1} - \hat Z_{i, \uparrow}} \rbs{\mathbb{1} 
         - \hat Z_{i , \downarrow}}.
    \end{split}
\end{align}
The two-qubit interaction in \cref{eq:U_coulomb} gives rise to the unitary gate
\begin{align}\label{eq:U_coulomb_ZZ}
    \mathsf{ZZ}_i(\phi) := \exp\cbs{-i\phi \hat Z_{i, \uparrow} \hat Z_{i, \downarrow} },
\end{align}
which can be implemented using single-qubit and $\mathsf{CNOT}$ gates in direct analogy with the hopping terms. The only simplification is that no basis-change gates, such as Hadamard or $\sqrt{\hat X}$ gates, are required, as shown in Fig.~\ref{fig:fermionic_hop0}(b). The hybrid qubit-boson interaction is discussed in
the remainder of this subsection.

The Hamiltonian $\hat H_{\rm HH}$ takes the following form after the Jordan-Wigner transformation (up to an additive constant):
\begin{align} 
       &\hat H_{\rm HH} = \omega_0 \hat{a}^{\dagger} \hat{a}+Ng_0(\hat{a}^{\dagger}+\hat{a}) + \frac{U}{4} \sum_{i=1}^N  \hat Z_{i,\uparrow}   \hat Z_{i,\downarrow} \label{eq:HH_JW} \\
       &-\frac{1}{2}\sum_{\sigma, i=1}^{N} \left(\varepsilon_{i,\sigma}+\frac{U}{2} +g_0 (\hat{a}^{\dagger}+\hat{a})\right) Z_{i,\sigma}  \nonumber
       \\ &+\frac{1}{2}\left(V+g(\hat{a}^{\dagger}+\hat{a})\right)\hspace{-1.1mm}\sum_{\sigma, i=1}^{N-1} \hspace{-1.1mm}\left(\hat X_{i,\sigma}\hat X_{i+1,\sigma} + \hat Y_{i,\sigma}\hat Y_{i+1,\sigma}\right).  \nonumber
\end{align}
To implement this Hamiltonian in a digital-analog circuit, it is convenient to eliminate the linear bosonic term proportional to $g_0(\hat a^\dagger+\hat a)$. This is achieved by the unitary transformation
${\hat H'_{\rm HH}=\hat D^\dagger(\varphi) \hat H_{\rm HH}\hat D(\varphi) }$,
where the displacement operator is defined as
\begin{align}\label{eq:D_phi}
    \begin{split}
      \hat D(\varphi) = e^{\varphi \hat a^\dagger - \varphi^*\hat a},
    \end{split}
\end{align}
and acts according to $D^\dagger(\varphi) \hat a\hat D(\varphi)=\hat a+\varphi$.
The corresponding gate needs to be applied only once at the beginning of the circuit and once again at its end. The linear bosonic term is eliminated by choosing
\begin{align}\label{eq:phi_gauge}
    \begin{split}
      \varphi = -\frac{Ng_0}{\omega_0}.
    \end{split}
\end{align}
The transformed Hamiltonian then becomes
\begin{align} 
    &\hat H'_{\rm HH}  =  
       \omega_0 \hat{a}^{\dagger} \hat{a}-\frac{1}{2}\sum_{\sigma, i=1}^{N} \left(\varepsilon'_{i,\sigma}+g_0 (\hat{a}^{\dagger}\!+\hat{a})\,\right) \hat Z_{i,\sigma} \nonumber \\ 
       &+ \frac{1}{2}\left(V'\!+\!g(\hat{a}^\dagger+\hat{a})\right) \hspace{-1.1mm}\sum_{\sigma, i=1}^{N-1}  \hspace{-1.1mm}\left(\hat X_{i,\sigma}\hat X_{i+1,\sigma} + \hat Y_{i,\sigma}\hat Y_{i+1,\sigma}\right) \nonumber\\
       &+ \frac{U}{4} \sum_{i=1}^N  \hat Z_{i,\uparrow}\hat  Z_{i,\downarrow} ,\label{eq:HH_JWrot}
\end{align}
where the renormalized on-site energies and hopping amplitude are given by
$\varepsilon'_{i,\sigma}=\varepsilon_{i,\sigma}+ \frac 12 {U}  - 2N{g_0^2}/{\omega_0}$ and 
$V'=V-2N{g_0g}/{\omega_0}$, respectively.

\begin{figure}[t!]
	\begin{center}
		\includegraphics[width=\linewidth]
  {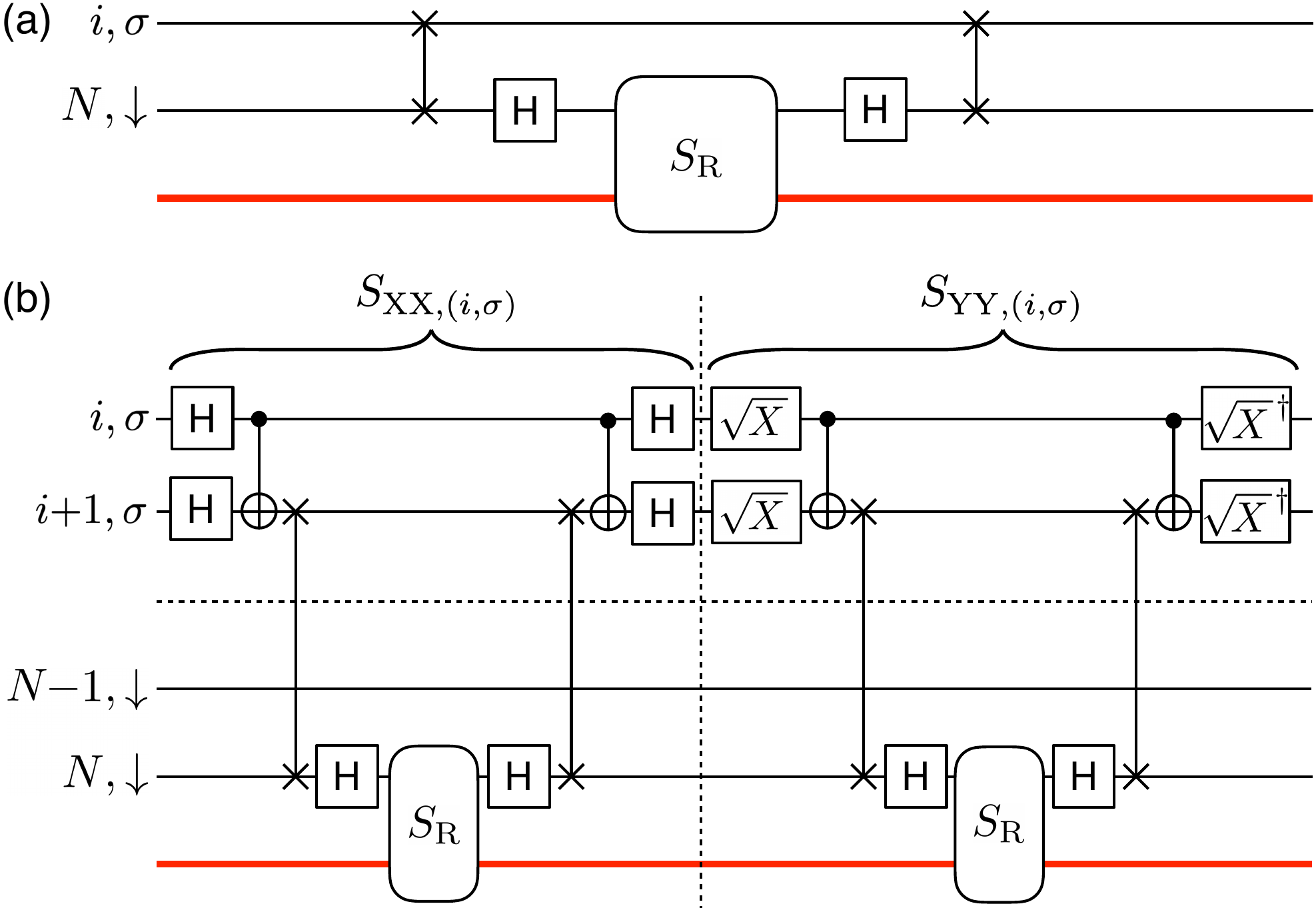}
		\caption
		{ \label{fig:a_dag_a_Z} 
(a) Circuit implementation of the gate block $\hat S_{{\rm Z},(i,\sigma)}$ corresponding to the electron-phonon interaction on lattice site $(i,\sigma)$. A sequence of $\mathsf{SWAP}$ gates transfers the quantum state of qubit $(i,\sigma)$ to the terminal qubit $(N,\downarrow)$, which is directly coupled to the resonator. The inverse $\mathsf{SWAP}$ sequence restores the original qubit ordering after the interaction. The Rabi gate $\hat S_{\rm R}(t_p+\tau,t_p)$ is implemented according to \cref{S_R_2} with phase $\theta=-g_0\tau/2$. The surrounding Hadamard gates rotate the interaction from the $\hat Z$ basis of the electron-phonon coupling to the $\hat X$ basis required by the Rabi gate. (b) Circuit diagram for  the gate $\hat S_{{\rm XY},(i,\sigma)}$ (\ref{s_XY_gate}) representing  phonon-assisted hopping term $\sim g(\hat a^\dagger +\hat a ) \big(\hat X_{i,\sigma}\hat X_{i+1,\sigma} + \hat Y_{i,\sigma}\hat Y_{i+1,\sigma}\big)$. The phase entering into $\hat S_{\rm R}$ is $\theta=-g\tau/2$.
}
\end{center}
\end{figure}

The \textit{e-ph} interaction in $\hat H'_{\rm HH}$ is implemented using layers of Rabi and Hadamard gates. To this end, we employ the identity $\hat Z =\mathsf{H} \hat X \mathsf{H}$ to rewrite the elementary \textit{e-ph} interaction term proportional to $g_0(\hat a^\dagger+\hat a)\hat Z$ in terms of the modified Rabi gate
\begin{equation}\label{S_Z}
\hat S_{\rm Z} = \mathsf{H}\hat S_{\rm R} \mathsf{H} .
\end{equation}
The remainder of the construction closely follows the digital-analog simulation of the quantum Rabi model extended to a $2N$-qubit architecture, in direct analogy with the Dicke-Ising model~\cite{Shapiro2025}. We assume a linear qubit array in which only the terminal qubit, corresponding to the $(N,\downarrow)$ orbital, is directly coupled to the resonator. 
Consequently, implementing the gate $ \hat  S_{{\rm Z},(i,\sigma)}$ on an arbitrary qubit $(i,\sigma)$ requires a sequence of $\mathsf{SWAP}$ gates that transfers its quantum state to the terminal qubit, followed by the application of $\hat  S_{\rm Z}$ from \cref{S_Z} with phase $\theta=-g_0\tau/2$, and finally by the inverse $\mathsf{SWAP}$ sequence, which restores the original qubit ordering:
\begin{equation}
  \label{s_Z_gate}
   \hat S_{{\rm Z},(i,\sigma)}  = \mathsf{SWAP}^{(N,\downarrow|i,\sigma)}\
   \mathsf{H}\,\hat S_{\rm R} \,\mathsf{H} \, \mathsf{ SWAP}^{(N,\downarrow|i,\sigma)}.
\end{equation}
The corresponding circuit implementation is shown in Fig.~\ref{fig:a_dag_a_Z}(a).

The two contributions to the phonon-assisted hopping term, proportional to $g(\hat a^\dagger+\hat a)\hat X_{i,\sigma}\hat X_{i+1,\sigma}$ and $g(\hat a^\dagger+\hat a)\hat Y_{i,\sigma}\hat Y_{i+1,\sigma}$, are implemented by the hybrid gates $\hat S_{{\rm XX},(i,\sigma)}$ and $\hat S_{{\rm YY},(i,\sigma)}$, respectively. Both hybrid gates are constructed from the elementary Rabi gate with phase $\theta=-g\tau/2$. Their sequential application defines the composite gate
\begin{equation}
\hat S_{{\rm XY}, (i,\sigma)}= \hat S_{{\rm YY}, (i,\sigma)}  \hat S_{{\rm XX}, (i,\sigma)}, \label{s_XY_gate}
\end{equation}
which implements the complete phonon-assisted hopping interaction. The corresponding quantum circuit is shown in Fig.~\ref{fig:a_dag_a_Z}(b).

Finally, the Trotter step over the interval $t\in [t_p,t_{p}{+}\tau]$ generated by the bosonic contribution $\hat H_{\rm Holstein}'=\hat D^\dagger(\varphi)H_{\rm Holstein}D(\varphi)$ to the Hubbard-Holstein Hamiltonian $\hat H'_{\rm HH}$ is given by
\begin{align}
& e^{-i \hat H_{\rm Holstein}' \tau } 
\approx e^{-i \hat H_0 (t_p+\tau)} \Bigg(\prod\limits_\sigma\hat S_{{\rm Z},(N,\sigma)}\Bigg)\nonumber\\
& \times\Bigg(\prod\limits_{\sigma}\prod\limits_{i=1}^{N-1} \hat S_{{\rm Z},(i,\sigma)}   \hat S_{{\rm XY},(i,\sigma)} \Bigg)\,\!e^{i \hat H_0 t_p}.
\end{align}
We conclude that a single Trotter step of the Hubbard-Holstein evolution can be decomposed into a universal set of digital single- and two-qubit gates supplemented by the elementary analog Rabi gate $\hat S_{\rm R}(\theta)$. This decomposition forms the basis of our digital-analog simulation protocol for electron-phonon models.

\subsection{Phase diagrams}
\label{subsec:phase_diag}

\begin{figure}[b!]
\includegraphics[width=0.95\linewidth]
{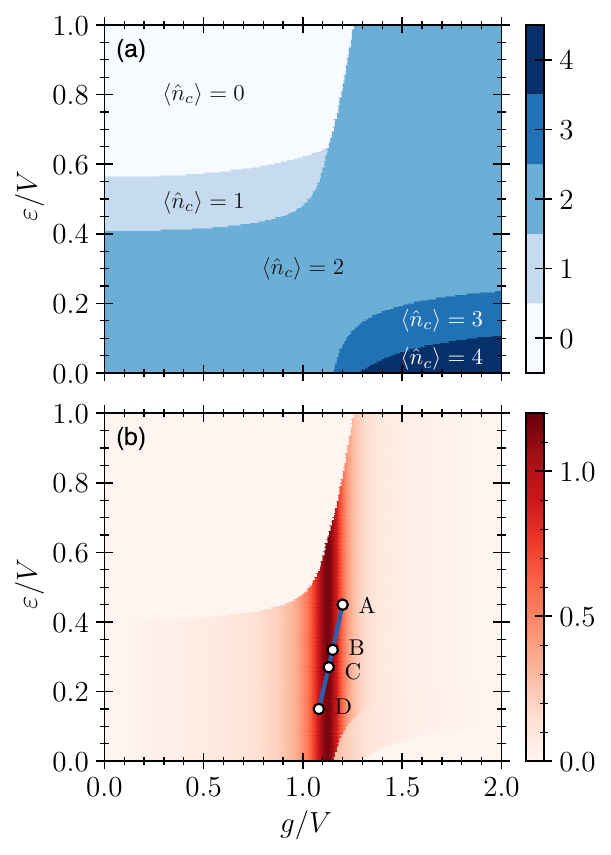}
\vspace{-0.4cm}\caption{(a) Ground-state phase diagram showing the fermion filling for the Hamiltonian $\hat H_{\rm HH}$ in \cref{eq:HH_JW} for $N=2$, $\omega_0/V=5$, $g_0=g$, $\varepsilon_{1, \uparrow \downarrow }=\varepsilon\pm 10^{-3}V$, $\varepsilon_{2, \uparrow \downarrow }=\varepsilon_{1, \uparrow \downarrow }+1.2V$, and $U/V=0.25$.
(b) Magnitude of the irreducible correlator $\mathcal{C}_{\mathrm{irr}}$ from \cref{eq:corr}, showing nonzero fermion--boson entanglement in the fluctuation-dominated region near $g\sim V$. Points A, B, C, and D support non-Poissonian phonon distributions; see the histograms in Fig.~\ref{fig:histograms0}. The parameters for these points are ($g_{\rm A}/V, \varepsilon_{\rm A}/V$) = (1.20, 0.45), ($g_{\rm B}/V, \varepsilon_{\rm B}/V$) = (1.15, 0.33), ($g_{\rm C}/V, \varepsilon_{\rm C}/V$) = (1.13, 0.28), and ($g_{\rm D}/V, \varepsilon_{\rm D}/V$) = (1.08, 0.15).}
 \label{fig_phase_diagram}
\end{figure}

In Fig.~\ref{fig_phase_diagram}~(a), we present the phase diagram of the model $\hat H_{\rm HH}$ in \cref{eq:HH_JW}, showing the fermion occupation number $\langle \hat n_c\rangle$, where ${\hat n_c=\sum_{i,\sigma} \hat c^\dagger_{i, \sigma}\hat c_{i, \sigma} }$. 
Domains with even fermion fillings ${\langle \hat n_c\rangle=0,2,4}$ have zero magnetization ${\langle \hat m\rangle=0}$, where ${ \hat m =\sum_{i}(  \hat c^\dagger_{i, \uparrow}\hat c_{i, \uparrow}  -   \hat c^\dagger_{i, \downarrow}\hat c_{i, \downarrow})}$. The sectors with odd filling, $\langle \hat n_c\rangle=1,3$, have nonzero magnetization, which is $\langle \hat m\rangle=-1$ for the parameter set given in the caption of Fig.~\ref{fig_phase_diagram}.

The $\langle \hat n_c\rangle=2$ region in the phase diagram [Fig.~\ref{fig_phase_diagram}~(a)], which has the form of an avoided crossing, is particularly important. It supports a fluctuation-dominated region with nonzero boson--fermion entanglement, visible as a bright vertical stripe in the data for the irreducible correlator
\begin{align}\label{eq:corr}
    \begin{split}
\mathcal{C}_{\mathrm{irr}} = \left\langle\!\left\langle  \left(\hat a +\!\hat a^{ \dagger }\right)  \left(\hat X_{1, \uparrow }\hat X_{2, \uparrow } + \hat Y_{1, \uparrow }\hat Y_{2, \uparrow }\right)\right\rangle\!\right\rangle
\end{split}
\end{align}
in Fig.~\ref{fig_phase_diagram}~(b), where we define $\langle\!\langle \hat A \hat B\rangle\!\rangle := \langle \hat A \hat B \rangle - \langle \hat A \rangle \langle \hat B \rangle$.
In analogy with the Dicke model, this bright fluctuation-dominated region separates a normal phase with vanishing phonon occupation ($g\lesssim V$) from a superradiant-like phase~\cite{popov1988functional,emary2003chaos, eastham2001bose, PhysRevA.94.061802, kirton2018introduction,PhysRevA.102.023703} with condensed phonons ($g\gtrsim V$). 
The constriction width in the $\langle \hat n_c\rangle=2$ region is determined by the difference between $\varepsilon_{1,\sigma}$ and $\varepsilon_{2,\sigma}$. A small Zeeman splitting between $\varepsilon_{i,\uparrow}$ and $\varepsilon_{i,\downarrow}$ is added to lift the degeneracy and avoid frustration between $\langle m\rangle=\pm 1$ in the regions with odd fermion fillings. 

\begin{figure}[b!]
	\begin{center}
\includegraphics[width=\linewidth]
{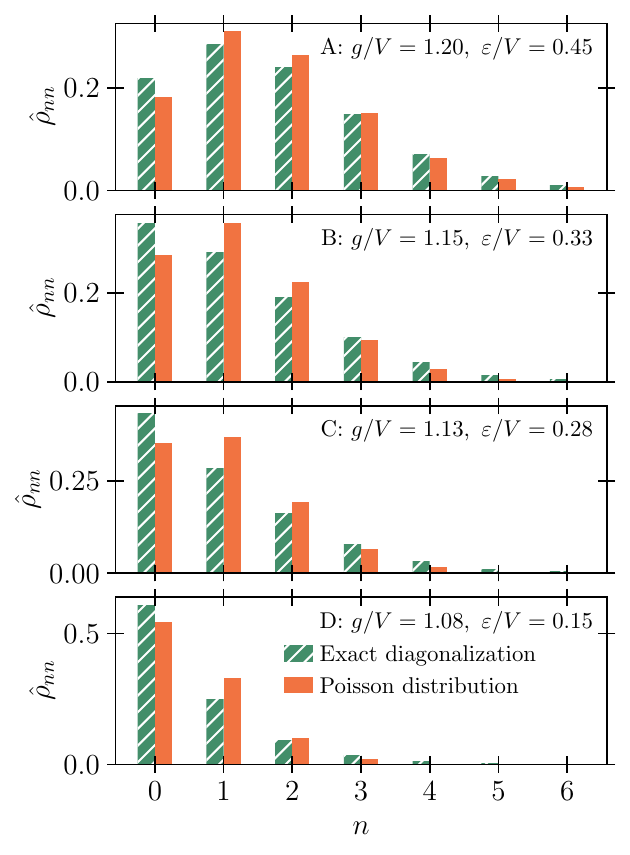}\vspace{-0.5cm}
\caption{\label{fig:histograms0} Histograms of the non-Poissonian phonon distributions, shown by hatched bars, corresponding to points A--D in the fluctuation-dominated region of Fig.~\ref{fig_phase_diagram}~(b). The distributions are calculated for the ground state of the rotated Hamiltonian $\hat H_{\rm HH}'$. Bold colored bars show Poissonian distributions with the same average phonon numbers in the rotated frame: $\langle\hat n_{\rm ph}\rangle_{\rm A}=0.61$, $\langle\hat n_{\rm ph}\rangle_{\rm B}=1.04$, $\langle\hat n_{\rm ph}\rangle_{\rm C}=1.26$, and $\langle\hat n_{\rm ph}\rangle_{\rm D}=1.71$. 
}
	\end{center}
\end{figure}

Solving the Schr\"odinger equation for the rotated Hamiltonian, $\hat H'_{\rm HH}|\psi_{\rm GS}\rangle=E_0|\psi_{\rm GS}\rangle$, we construct the full density matrix 
\begin{equation}
\hat \rho_{\rm GS}=|\psi_{\rm GS}\rangle \langle\psi_{\rm GS}|.
\end{equation}
Tracing out the fermionic degrees of freedom, we obtain the reduced density matrix
\begin{equation}\label{eq:rho_ph}
\hat\rho={\rm tr}_{i,\sigma}\left(\rho_{\rm GS}\right),
\end{equation} 
in the basis suitable for the quantum-circuit implementation.
Fig.~\ref{fig:histograms0} shows histograms of the phonon number distributions $\rho_{n,n}$ from \cref{eq:rho_ph} for the points A, B, C, and D in Fig.~\ref{fig_phase_diagram}~(b), which lie in the fluctuation-dominated region with large entanglement. The data in Fig.~\ref{fig:histograms0} show non-Poissonian distributions, indicating a nonclassical resonator state in this region.

\subsection{Hadamard test for the phonon distribution}
\label{subsec:H_test}
\subsubsection{Discrete Fourier transform}
To obtain the distribution $\rho_{n,n}$, we use a Hadamard-test protocol adapted to a dispersively coupled auxiliary qubit and resonator.
The central idea of the protocol is to measure $z_\Phi={\rm Re}\left[{\rm tr}\left(\hat \rho e^{i\Phi \hat a ^\dagger \hat a}\right)\right]$, which is the real part of the expectation value of the phonon phase-rotation operator $e^{i \Phi \hat a^\dagger \hat a}$. Rewriting the trace in $z_\Phi$ and using the normalization condition $\rho_{0,0}= 1-\sum_{n\geq 1} \rho_{n,n}$, one obtains
\begin{equation}\label{z_Phi}
z_\Phi= 1-\sum_{n\geq 1}^\infty \rho_{n,n}(1-\cos(n\Phi)).
\end{equation}
In a real circuit, we are interested in extracting a finite number of probabilities $\rho_{n,n}$ with $0{\leq} n{\leq} n_{\rm max}$, corresponding to a phonon Hilbert-space cutoff of dimension $n_{\rm max}{+}1$. 
By measuring the values $z_{\Phi_k}$ for $n_{\rm max}$ distinct phases $\Phi_k$ with $0{<}\Phi_k{<}2\pi$, we obtain a set of $n_{\rm max}$ linear equations from \cref{z_Phi},
\begin{equation} \label{rho_nn}
z_{\Phi_k}  = 1- \sum\limits_{n\geq 1}^{n_{\rm max}} \mathcal{M}_{k,n}\rho_{n,n} , \quad 1\leq k\leq n_{\rm max}.
\end{equation}
Here, $\mathcal{M}$ is a square, nondegenerate, $n_{\rm max}$-dimensional matrix with elements
\begin{equation}
\mathcal{M}_{k,n}=1-\cos(n\Phi_k ), \quad  k,n\in[1,n_{\rm max}].
\end{equation}
Inverting $\mathcal{M}$ amounts to performing the discrete Fourier transform and yields the array of $n_{\rm max}$ probabilities,
\begin{equation} \label{rho_nn_FT}
\rho_{n,n} = \sum\limits_{k\geq 1}^{n_{\rm max}} \left[\mathcal{M}^{-1}\right]_{n,k} (1-z_{\Phi_k} ), \quad 1\leq n\leq n_{\rm max};
\end{equation}
the zeroth element is given by $\rho_{0,0}=1-\sum_{n\geq 1}^{n_{\rm max}}\rho_{n,n}$.

\subsubsection{Quantum circuit}
We now describe a circuit for measuring the value of $z_\Phi$ entering the discrete Fourier transform in \cref{rho_nn_FT}. This protocol combines the ideas of Wigner tomography~\cite{lutterbach1997, vlastakis2013,Langford2017} and Ramsey interferometry. 

First, we consider the interacting part of the dispersive Hamiltonian in \cref{H_disp}, which can be written as  
$\hat H_{\rm disp}= 2\chi \hat a^\dagger\hat a |1\rangle\langle 1|$.
Evolution under this Hamiltonian for a tunable time $t_0$ yields the qubit-controlled phase-rotation operator $\mathsf{CR}_\Phi= e^{-i \hat H_{\rm disp} t_0}$, where the accumulated phase is $\Phi=-2\chi t_0$. This gate reads
\begin{equation}\label{C_Phi}
 \mathsf{CR}_\Phi=|0\rangle\langle 0|+ \mathsf{R}_\Phi|1\rangle\langle 1|, 
\end{equation}
where     
\begin{equation}
\mathsf{R}_{\Phi}=e^{i\Phi \hat a^\dagger \hat a}
\end{equation}
is the phase-rotation gate.

\begin{figure}[t!]
    \centering
    \includegraphics[width=0.75\linewidth]{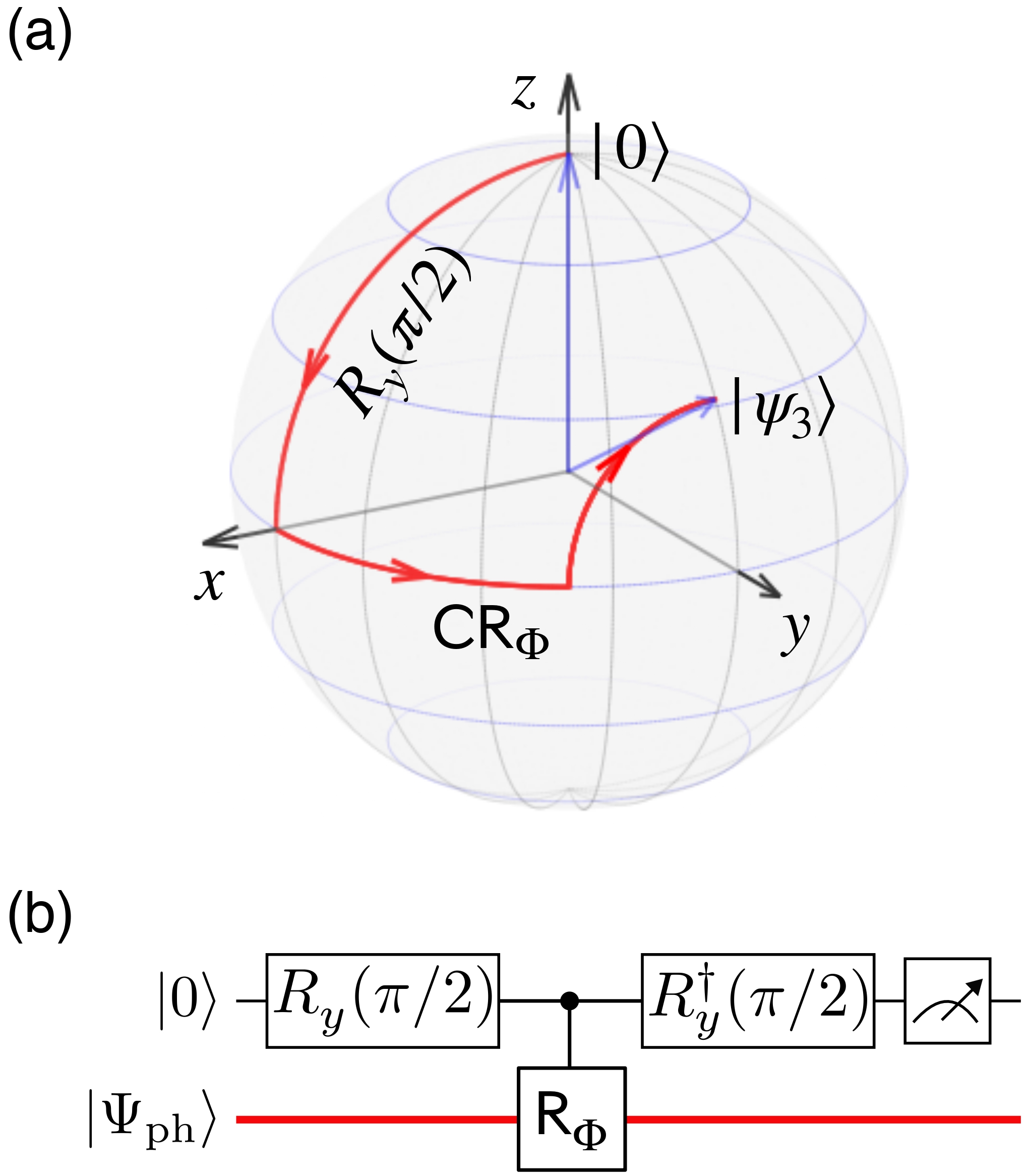}
    \caption{Hadamard-test protocol for measuring $z_\Phi$: Bloch-sphere representation (a) and the corresponding quantum circuit (b).}
    \label{fig:resonator_measurement}
\end{figure}

The full protocol proceeds as follows. Applying the $R_y(\pi/2)$ gate transfers the qubit from the top of the Bloch sphere, with initial state $|\psi_0\rangle=|0\rangle\otimes |\Psi_{\rm ph}\rangle$, to the equator, as shown in Fig.~\ref{fig:resonator_measurement}~(a). Here, the phonon state is denoted by $|\Psi_{\rm ph}\rangle$. The resulting state is $|\psi_1\rangle=\frac{1}{\sqrt2}(|0\rangle+ |1\rangle)\otimes |\Psi_{\rm ph}\rangle$, i.e., parallel to the $x$ axis. The state then rotates along the equator during the evolution with $\hat H_{\rm disp}$ over the time $t_0$. Equivalently, applying the $\hat C_\Phi$ gate to $|\psi_1\rangle$ yields
\begin{equation}
|\psi_2\rangle=\frac{1}{\sqrt 2}\big( |0\rangle\otimes |\Psi_{\rm ph}\rangle+  |1\rangle\otimes e^{i \Phi \hat a^\dagger \hat a} |\Psi_{\rm ph}\rangle\big ).
\end{equation}
At this step, the phase of the qubit component depends on the phonon state $|\Psi_{\rm ph}\rangle$. The final rotation is given by $R_y(-\pi/2)$, which brings the qubit state out of the equator to the $yz$ plane, as in Ramsey-interferometry protocols. The final state reads
\begin{equation}
|\psi_3\rangle=\frac{1}{2}\big( (|0\rangle - |1\rangle )\otimes |\Psi_{\rm ph}\rangle + (|0\rangle + |1\rangle )\otimes e^{i \Phi \hat a^\dagger \hat a} |\Psi_{\rm ph}\rangle \big ).
\end{equation}
Measuring the $Z$ projection, one obtains
\begin{equation}\label{Hadamard_test}
\langle\psi_3|\hat Z|\psi_3\rangle= \langle\Psi_{\rm ph}|\cos(\Phi\hat a^\dagger \hat a )|\Psi_{\rm ph}\rangle.
\end{equation}
The scalar product in \cref{Hadamard_test} can be written as
$\langle\Psi_{\rm ph}|\cos(\Phi\hat a^\dagger \hat a )|\Psi_{\rm ph}\rangle=\sum_{n\geq0} \rho_{n,n} \cos(n\Phi)$,
which coincides with the definition of $z_\Phi$ in \cref{z_Phi}. Thus, the protocol implements a Hadamard-test measurement of the phonon phase-rotation operator through the auxiliary qubit.
The corresponding quantum circuit is shown in Fig.~\ref{fig:resonator_measurement}~(b).
 
\section{Variational Hamiltonian ansatz}
\label{sec:VHA}

\begin{figure*}[htb!]
\centering\includegraphics[width=\textwidth]{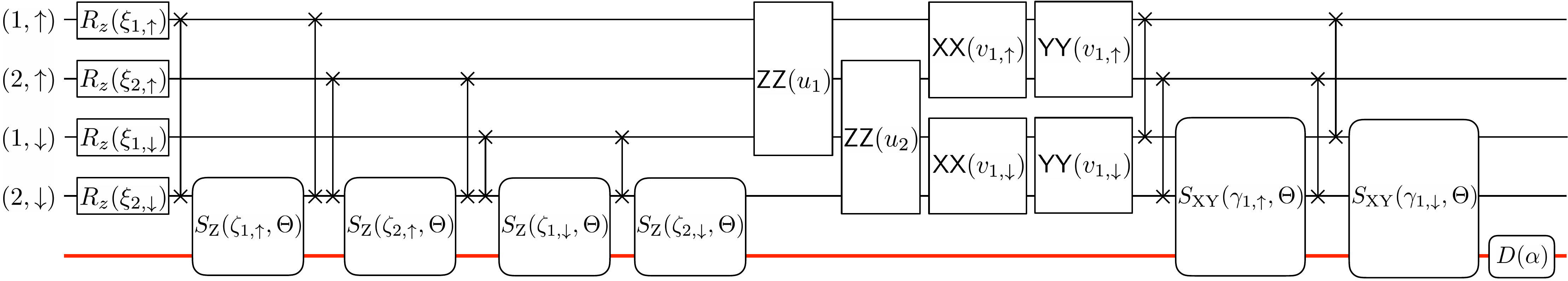}
    \caption{Gate sequence for the VHA $\mathsf{A}(\mathbf{x})$ defined in \cref{A_def}.}
    \label{fig:ansatz}
\end{figure*}

In this section, we introduce the VHA algorithm used to approximate the ground state using the quantum circuits introduced above. As a representative application, we apply the algorithm to the HH model. 

The VHA is a hybrid quantum-classical method: a parameterized quantum circuit prepares a family of trial states $\Psi(\mathbf{x})$, while a classical optimizer updates the set of circuit parameters $\mathbf x$ to minimize the measured energy,
\begin{equation}
E(\mathbf{x})=\langle\Psi(\mathbf{x})|\hat H_{\rm HH}|\Psi(\mathbf{x})\rangle.
\end{equation}
In this work, the parameters \(\mathbf{x}\) are updated using gradient descent based on \(\nabla_{\mathbf{x}} E(\mathbf{x})\).

\subsection{Trial wave function}
\label{subsec:trial_wf}

In our setting, the trial states are realized as
\begin{equation}
|\Psi(\mathbf{x})\rangle=\mathsf{A}(\mathbf{x})|\Psi_0\rangle,
\end{equation}
where $|\Psi_0\rangle$ is the eigenstate of the noninteracting HH Hamiltonian in \cref{eq:HH_JW} for $U=g=g_0=0$, and $\mathsf{A}(\mathbf{x})$ is the variational Hamiltonian ansatz (VHA) operator,
\begin{align} 
\mathsf{A}(\mathbf{x}) &=\hat D(\alpha) \left(\prod_\sigma\hat R_z(\xi_{N,\sigma})\hat S_{{\rm Z},(N,\sigma)}^{\rm (VHA)}(\zeta_{N,\sigma},\Theta) \right)  \label{A_def}  \\ &\times\left(\prod_{i=1}^N \mathsf {ZZ}(u_i)  \right)\prod_{\sigma}\prod_{i=1}^{N-1}\left(\hat S_{{\rm Z},(i,\sigma)}^{\rm (VHA)}(\zeta_{i,\sigma},\Theta) \right.\nonumber \\
&\left.\times   \ \hat S_{{\rm XY},(i,\sigma)}^{\rm (VHA)}(\gamma_{i,\sigma}, \Theta)   \mathsf{XX}(v_{i, \sigma})\mathsf{YY}(v_{i, \sigma})
\hat R_z(\xi_{i,\sigma})  \right). \nonumber 
\end{align}
The gates in \cref{A_def} correspond to the different components of $\hat H_{\rm HH}$: (i) the phonon displacement gate $\hat D(\alpha)$ corresponds to the term linear in the electric field, $\sim (\hat a^\dagger + \hat a) $; (ii) the $\mathsf{ZZ}(u_i)$ gates correspond to the Coulomb interaction; (iii) $\hat S_{\rm Z}^{\rm (VHA)}(\zeta_{i,\sigma},\Theta)$ corresponds to the \textit{e-ph} interaction; (iv) $\hat S_{\rm XY}^{\rm (VHA)}(\gamma_{i,\sigma},\Theta)$ corresponds to phonon-assisted hopping; and (v) the $\mathsf{XX}(v_{i,\sigma})$, $\mathsf{YY}(v_{i,\sigma})$, and $\hat R_z(\xi_{i,\sigma})$ gates correspond to the fermion hopping and on-site-potential terms. 

We note that in $\mathsf{A}(\textbf{x})$, the spin-boson entangling gates, $\hat S_{{\rm Z},(i,\sigma)}^{\rm (VHA)}$ and $\hat S_{{\rm XY}, (i,\sigma)}^{\rm (VHA)}$, are defined similarly to those in Fig.~\ref{fig:a_dag_a_Z}, in panel (a) and (b) respectively, with the difference that the Rabi gate is modified to
\begin{align} \label{S_R_VHA}
    \begin{split}
        \hat S_{\mathrm{R}}^{\rm (VHA)}(\nu, \Theta)
        =  \hat R^\dagger_z(\Theta)\,\hat S_{\mathrm{JC}}(\nu/4) \hat R_z(2 \Theta) \\
        \times \hat X \hat S_{\mathrm{JC}}(\nu/2)\hat X  R^\dagger_z(2 \Theta)\,\hat S_{\mathrm{JC}}(\nu/4) \hat R_z(\Theta),
    \end{split} 
\end{align}
where, depending on the term, $\nu=\zeta_{i,\sigma}$ for $\hat S_{{\rm Z},(i,\sigma)}^{\rm (VHA)}$ or $\nu=\gamma_{i,\sigma}$ for $\hat S_{{\rm XY},(i,\sigma)}^{\rm (VHA)}$.
The operator structure of $S_{\mathrm{R}}^{\rm (VHA)}$ is identical to that of the original $\hat S_{\rm R}$ in \cref{S_R_2}, except that the phase shifts in the $\hat R_z$ gates, originating from the time delays $ \tau/4$, $3\tau/4$, and $\tau$, are omitted.
This is because the variational state $|\Psi(\mathbf{x})\rangle$
is a \textit{postulated} wave function, rather than the result of a Trotterized time evolution. Hence, there is no VHA control parameter analogous to the evolution time $\tau$ in the Trotterization scheme. The above-mentioned phase shifts in the VHA version of the Rabi gate \cref{S_R_VHA} can therefore be safely omitted. The set of parameters $\mathbf{x}$ includes $\alpha$, $\Theta$, $u_i$, $v_{i,\sigma}$, $\gamma_{i,\sigma}$, $\zeta_{i,\sigma}$ and $\xi_{i,\sigma}$.

Up to single-qubit rotations, the total gate count in $\mathsf{A}(\mathbf{x})$
for $N$ sites is as follows. It involves $18N-16$ $\mathsf{CNOT}$s, of which
$2N$ come from the $\mathsf{ZZ}$ terms, $8(N-1)$ from the $\mathsf{XX{\cdot}YY}$
terms, and $8(N-1)$ from $\hat S_{\rm XY}$. It further requires $8N-8$
$\mathsf{SWAP}$s, split as $2(2N-1)$ in $\hat S_{\rm Z}$ and $2(2(N-1)-1)$ in
$\hat S_{\rm XY}$, together with $6N-4$ Rabi gates $\hat S_{\rm R}$. The qubit
ordering for this count is chosen as $\{1_{\uparrow}, 1_{\downarrow}, 2_{\uparrow}, 2_{\downarrow}, \dots\}$, which guarantees optimal scaling with the Jordan-Wigner string length. We assume all-to-all connectivity among the qubits and a single  qubit  coupled to the resonator, requiring SWAP gates to implement the interaction between the other qubits and the resonator.

\begin{figure}[b!]
\centering\includegraphics[width=0.5\linewidth]{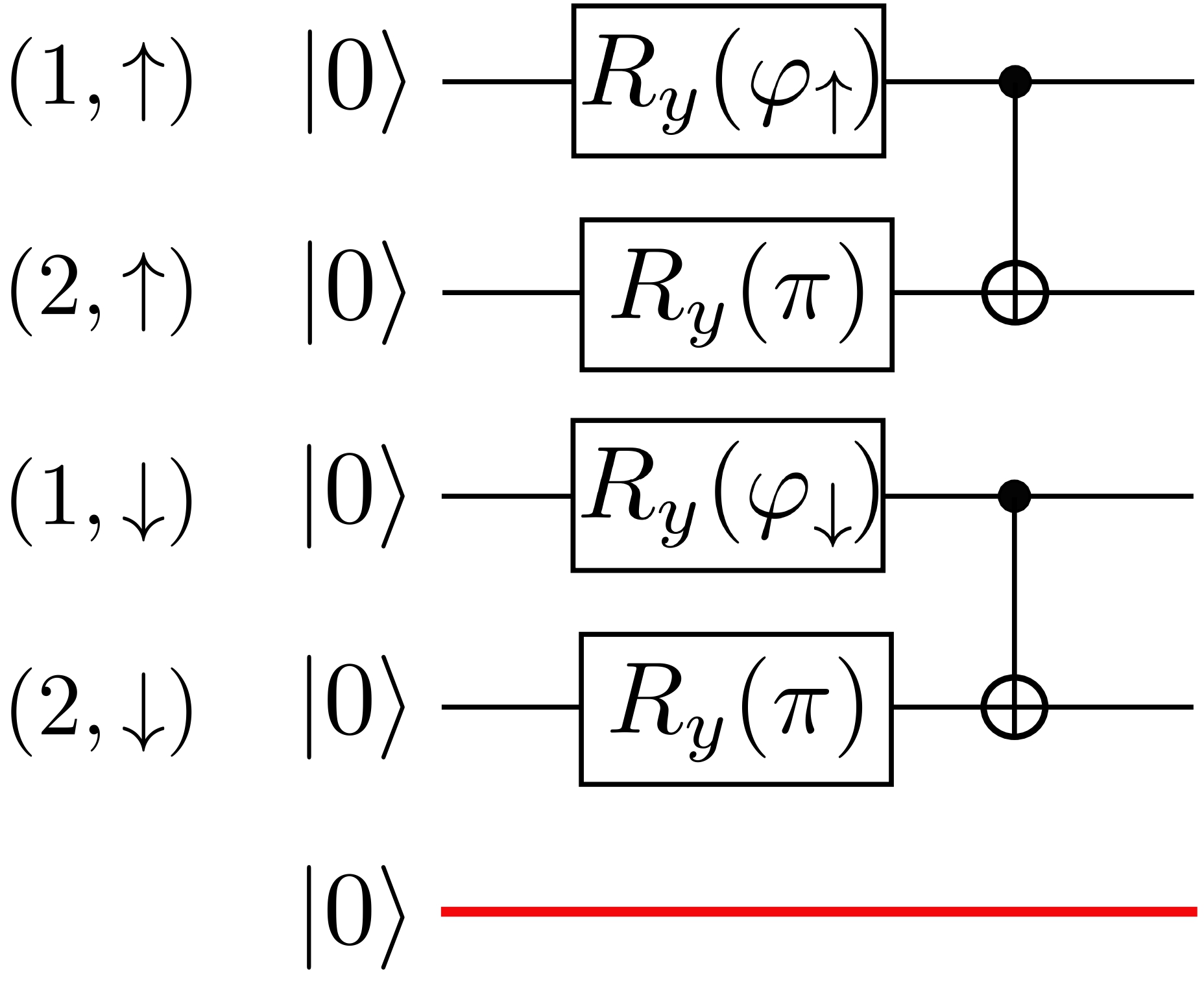}
    \caption{Quantum circuit preparing the trial wave function $|\Psi_0\rangle$ in \cref{psi_0_wf} for $N=2$.}
    \label{psi_0}
\end{figure}

The circuit representing the ansatz $\mathsf{A}(\mathbf{x})$ for $N=2$ is shown in Fig.~\ref{fig:ansatz}. 
For this case, the initial state is
\begin{align}
    |\Psi_0\rangle = |\psi_0\rangle_\uparrow \otimes |\psi_0\rangle_\downarrow \otimes |0_{\rm ph}\rangle, \label{psi_0_wf}
\end{align}
where \(|0_{\rm ph}\rangle\) denotes the empty resonator state, and \(|\psi_0\rangle_{\sigma}\) with $\sigma= \uparrow,\downarrow$ are the ground states of the decoupled spin systems in the subspace with $\langle \hat n_c\rangle=2$. This subspace is spanned by the states $|0_11_2\rangle_\sigma$ and $|1_10_2\rangle_\sigma$ for a given spin projection $\sigma$:
\begin{equation}
|\psi_0\rangle_{\sigma}= \cos(\varphi_\sigma/2) |0_11_2\rangle_\sigma + \sin(\varphi_\sigma/2)|1_10_2\rangle_\sigma,
\end{equation}
with 
\begin{equation}
    \varphi_{\sigma} = -2\arctan\left(\frac{2V}{\Delta\varepsilon_{\sigma} +\sqrt{\Delta\varepsilon_{\sigma}^2+4V^2} }\right)
\end{equation}
and $\Delta\varepsilon_\sigma = \varepsilon_{1,\sigma} - \varepsilon_{2,\sigma}$.
The circuit preparing $|\Psi_0\rangle$ is shown in Fig.~\ref{psi_0}. 

Simulation data for the VHA algorithm are presented in
Appendix~\ref{appendix:optimization}, where the approximate ground states
$|\Psi(\mathbf{x})\rangle$ are calculated at the points A--D indicated
in the phase diagram in Fig.~\ref{fig_phase_diagram}~(b). In
Appendix~\ref{appendix:lindblad}, we describe a modification of the VHA
algorithm that incorporates realistic loss rates by including Lindblad dynamics in the simulations. 

\subsection{Energy measurement}
\label{subsec:energy_meas}
The estimation of the energy terms on a quantum chip can be grouped into two contributions,
\begin{equation}
E(\mathbf{x})=\langle \hat H_{\rm Hubbard} \rangle + \langle \hat H_{\rm Holstein} \rangle .
\end{equation}
Here, the notation \(\langle \dots \rangle\) denotes the quantum-mechanical expectation value over the prepared trial state \(\Psi(\mathbf{x})\).
The terms entering \(\langle \hat H_{\rm Hubbard}\rangle\) are obtained straightforwardly by measuring the \(Z_{i,\sigma}\)-, \(X_{i,\sigma}\)-, and \(Y_{i,\sigma}\)-projections in single shots, and then averaging their first and second moments over many shots. In circuit QED, these measurements are implemented using dispersive readout. The data qubits are weakly coupled to individual dedicated readout resonators. By observing dips in the transmission spectra of the readout resonators, which appear at two different frequencies, one can determine the state of each data qubit. In this way, one obtains the energies of the local terms, which are proportional to \(\langle \hat Z_{i,\sigma}\rangle\), the hopping energy, proportional to
$
\langle \hat X_{i,\sigma}\hat X_{i+1,\sigma}\rangle + \langle \hat Y_{i,\sigma}\hat Y_{i+1,\sigma}\rangle$,
and the Coulomb interaction energy,
$
\sim \langle \hat Z_{i,\uparrow}\hat Z_{i,\downarrow}\rangle $.

Measuring \(\langle \hat H_{\rm Holstein} \rangle\) reduces to averaging four operators: (i) the free-phonon term, \(\langle \hat a^\dagger \hat a \rangle\); (ii) the \textit{e-ph} interaction, \(\langle \hat Z_{i,\sigma}(\hat a^\dagger {+} \hat a)\rangle\); and the phonon-assisted hopping terms involving (iii) $XX$ and (iv) $YY$,
$\langle 
X_{i,\sigma}\hat X_{i+1,\sigma}(\hat a^\dagger {+} \hat a)\rangle$ and $\langle \hat Y_{i,\sigma}\hat Y_{i+1,\sigma} (\hat a^\dagger {+} \hat a) \rangle$, respectively.
The expectation values of these operators are not directly accessible through unitary gates. However, their averages can be accessed indirectly using the phonon displacement gate $\hat D(i \eta)$ and the phase-rotation gate
\begin{equation}
\hat{\mathsf{R}}_{\eta}=e^{i\eta \hat a^\dagger \hat a}.
\end{equation}
Here, the parameter $\eta\in \mathbb{R}$ plays the role of a counting field in statistical mechanics and determines the moment-generating function. 
Formally, to estimate the averages in (i)--(iv), one takes the first derivative with respect to $\eta$ at $\eta=0$, thereby extracting the first statistical moment. For the phonon number, case (i), this gives
\begin{equation}
\langle \hat a^\dagger \hat a\rangle=\lim_{\eta\to 0}
\partial_\eta \, {\rm Im}\langle \hat {\mathsf{R}}_\eta \rangle.
\label{aa}
\end{equation}
For the other terms (ii)-(iv), we have:
\begin{align}
\langle \hat Z_{i,\sigma}(\hat a^\dagger + \hat a)\rangle &= \lim_{\eta\to 0}
\partial_\eta \,{{\rm Im}\langle Z_{i,\sigma}
\hat D(i\eta)
\rangle}, \label{Zaa} \\
\langle X_{i,\sigma} X_{i+1,\sigma}(\hat a^\dagger + \hat a)\rangle &=\lim_{\eta\to 0} \partial_\eta\,{\rm Im}\langle X_{i,\sigma} X_{i+1,\sigma}\hat D(i\eta) \rangle,\label{XXaa}
\end{align}
and similarly for a product of $Y$-Pauli operators:
\begin{align}\label{YYaa}
\langle Y_{i,\sigma} Y_{i+1,\sigma}(\hat a^\dagger + \hat a)\rangle=\lim_{\eta\to 0} \partial_\eta\, {\rm Im} \langle Y_{i,\sigma}Y_{i+1,\sigma}\hat D(i\eta)\rangle.
\end{align}

\begin{figure}
    \centering
    \includegraphics[width=1\linewidth]
{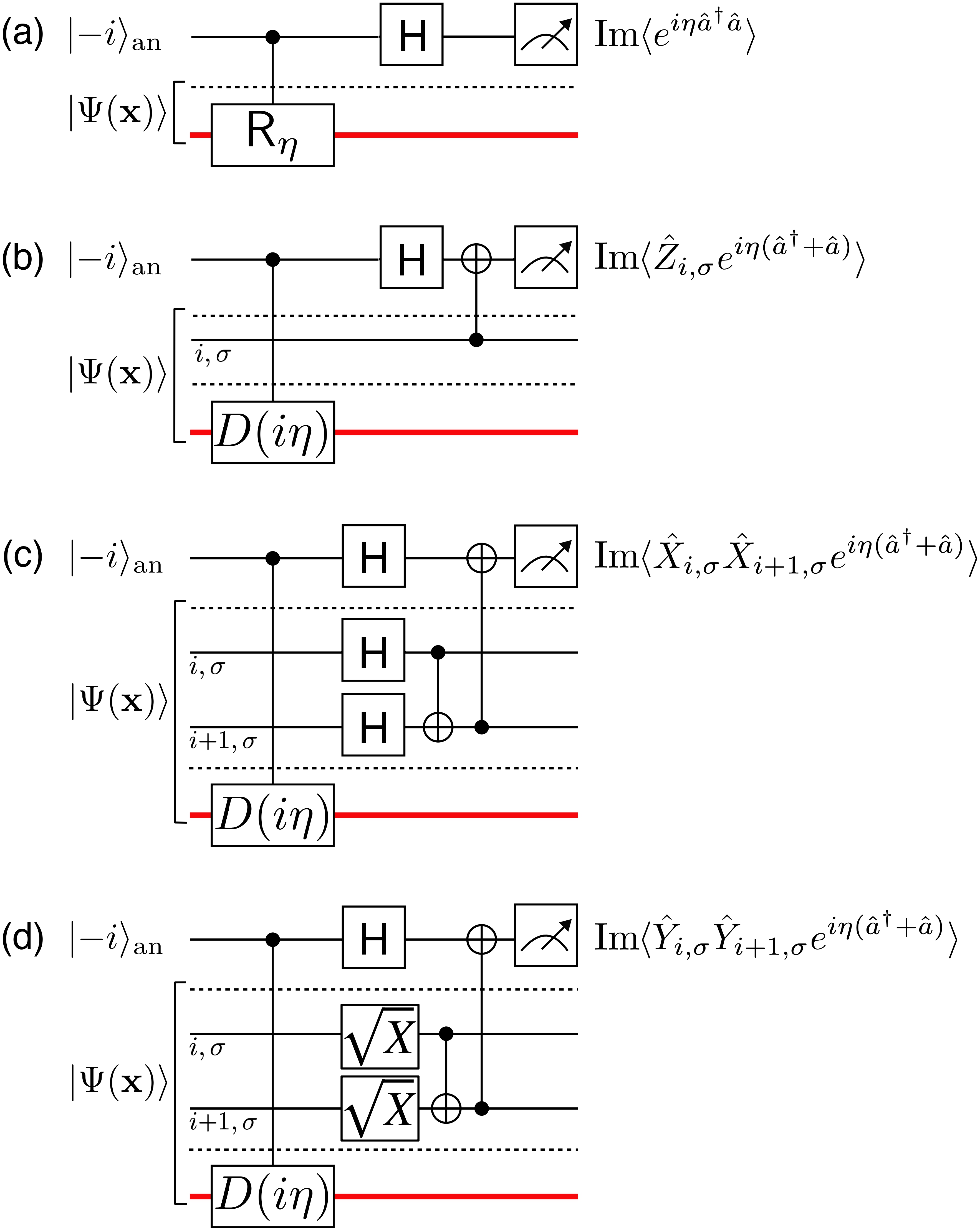}
    \caption{Quantum circuits for measuring the unitary operators in \cref{aa,Zaa,XXaa,YYaa}. The outputs yield the generating functions for all terms in $\hat H_{\rm Holstein}$: (a) the free-phonon term, (b) the \textit{e-ph} interaction, and (c,d) the phonon-assisted hopping terms. The counting variable $\eta$ is a tunable parameter that must be optimized to maximize accuracy while minimizing the required number of samples.}
    \label{fig:zqmeas_circuit}
\end{figure}

The imaginary parts of the operators appearing in \cref{aa,Zaa,XXaa,YYaa} are measured using the Hadamard-test protocol. First, consider the estimation of $\langle \hat a^\dagger \hat a \rangle$ in \cref{aa}.
The core idea is to use the phase rotation \(\hat{\mathsf{R}}_\eta\) controlled by an ancilla qubit, denoted by the subscript ``$\rm an$'':
\begin{equation}
\hat{\mathcal U}_\eta = |0\rangle\langle0|_{\rm an} \otimes\hat{\mathbb{1}}+ |1\rangle\langle1|_{\rm an} \otimes \hat{\mathsf{R}}_\eta.
\end{equation}
The input state in this protocol is $|-i\rangle_{\rm an} \otimes|\Psi(\mathbf{x})\rangle$, i.e.,  
a direct product of the prepared trial state $|\Psi(\mathbf{x})\rangle$ and the ancilla in the $-1$ eigenstate of the Pauli $Y$ operator,
\begin{equation}
{\color{black}|{-}i\rangle_{\rm an}}=\frac{1}{\sqrt2}(|0\rangle_{\rm an} - i |1\rangle_{\rm an}).
\end{equation}
Applying \(\hat{\mathcal U}_\eta\) and the Hadamard gate to the input state yields
\begin{equation}
| \Phi_\eta \rangle = \mathsf{H}_{\rm an}\ \hat{\mathcal U}_\eta \, |{-}i\rangle_{\rm an}\otimes|\Psi(\mathbf{x})\rangle .
\end{equation}
Upon measuring the \(Z\)-projection of the ancilla, one estimates the imaginary part of the rotation operator,
\begin{equation}
\langle \Phi_\eta|\hat Z_{\rm an}|\Phi_\eta\rangle = {\rm Im}\langle \hat{\mathsf{R}}_\eta\rangle . \label{H_test_0}
\end{equation}
If the ancilla were instead initialized in the $+1$ eigenstate of the $X$ operator,
$|+\rangle_{\rm an}=\frac{1}{\sqrt{2}}(|0\rangle_{\rm an}+|1\rangle_{\rm an})$,
the real part of the same average would be obtained at the output.
The corresponding quantum circuit is shown in Fig.~\ref{fig:zqmeas_circuit}(a).
Measuring \cref{H_test_0} for a set of small values of $\eta$, one obtains a discrete derivative that approximates $\langle\hat a^\dagger \hat a \rangle$.

The above protocol can be extended to the averages in \cref{Zaa,XXaa,YYaa}. The corresponding quantum circuits are shown in Fig.~\ref{fig:zqmeas_circuit}(b)--(d). Here, the corresponding Hadamard tests use the ancilla-controlled displacement gate $\hat D(i\eta)$ instead of $\hat {\mathsf{R}}_\eta$. Additional layers of single- and two-qubit rotations are added to implement the Pauli operators. 
We note that the scheme based on the controlled-displacement gate is inspired by protocols for the preparation and stabilization of Gottesman-Kitaev-Preskill states in circuit QED~\cite{Campagne-Ibarcq2020, Flühmann2019}. 

In practice, $\eta$ should be chosen small enough for the
$\eta \to 0$ approximation in \cref{aa,Zaa,XXaa,YYaa} to be accurate, but
not so small that the induced changes in the measured averages become
indistinguishable from statistical fluctuations. The optimal value of
$\eta$ therefore balances the errors from using a finite $\eta$
against sampling errors due to a weak measurement signal.

\begin{figure}[t!]
    \includegraphics[width=0.8\linewidth]{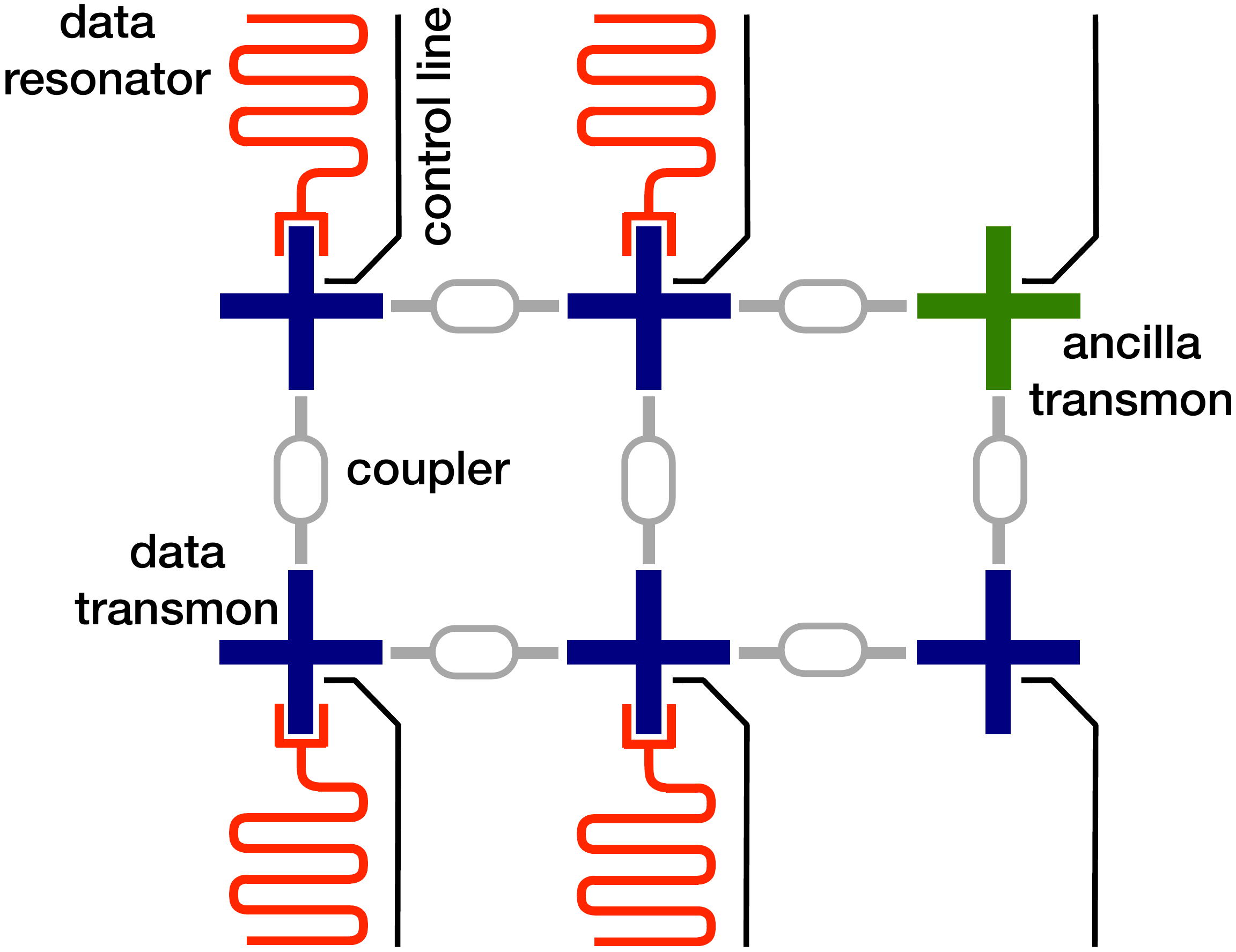}\vspace{-0.1cm}
  \caption{Transmon-resonator architecture tailored for simulations of the Yukawa-SYK model with $2N=10$ Majorana fermions and $M=4$ phonons. The couplers provide two-qubit gates. The ancilla transmon can be used to measure the energy of the system as described in Sec.~\ref{subsec:energy_meas}.}
\label{architecture_YSYK}
\end{figure}

\begin{figure*}[htb!]
	\begin{center}
		\includegraphics[width=1\textwidth]
        {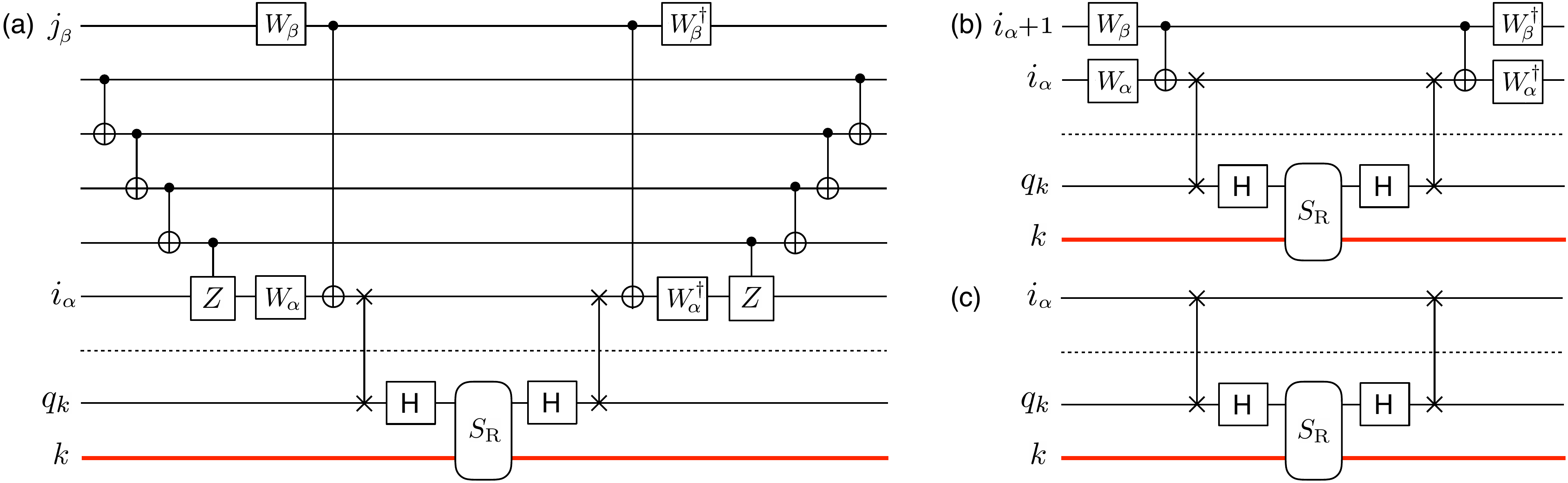}
		\caption
		{\label{fig:S_R_YSYK} Quantum circuit representing the Yukawa-SYK gates $\hat S^{(\alpha \beta k) }_{\rm YSYK}$ in \cref{H_YSYK_trotterized} for all possible combinations of Majorana fermion indices $1\leq \alpha < \beta\leq 2N$. The qubit with index $q_k$ is physically coupled to resonator $k$; the condition $M\leq N$ is assumed. The qubit indices, corresponding to the Jordan-Wigner spins, are related to $\{\alpha,\beta\}$ via the floor function as $i_\alpha=\left\lfloor\frac{\alpha+1}{2}\right\rfloor$ and $j_\beta=\left\lfloor\frac{\beta+1}{2}\right\rfloor$, with $1\leq i_\alpha\leq j_\beta\leq N$. (a) Case $i_\alpha+2\leq j_\beta$. The sequence of $\mathsf{CNOT}$s and the final $\mathsf{CZ}$ represent the Jordan-Wigner string; the gate $\hat W_\alpha=\mathsf{H}$ if $\alpha\in {\rm odd}$ and $\hat W_\alpha=\hat X^{1/2}$ if $\alpha\in {\rm even}$. The constituent blocks of the Rabi gates, $\hat S_{\rm JC}(\theta)$ and $\hat R_z(\varphi_q)$, have phases $\theta =  g_{\alpha \beta k}\tau/2$ and $\varphi_q=\omega_k(t_p +q \tau)$ with $q~=~0,  \frac{1}{4}, \frac{3}{4}, 1$. (b) Case $i_\alpha+1=j_\beta$, with a single two-qubit gate $\mathsf{CZ}$ in the Jordan-Wigner string. (c) Special case $i_\alpha=j_\beta$ ($\alpha+1=\beta$ with $\alpha\in {\rm odd}$), corresponding to \cref{chi_chi_Z}, which reduces to the $\hat S_{\rm Z}$ gate.}
	\end{center}
\end{figure*}

\section{Yukawa-Sachdev-Ye-Kitaev Model}\label{sec:YSYK}

The Yukawa-SYK model was independently formulated between 2018 and 2019~\cite{marcus_2019,Esterlis_2019,wang_2020}. 
Unlike the original SYK$_4$ models~\cite{maldacena_2016,gu_2020}, which are purely fermionic Hamiltonians with all-to-all quartic interactions, the Yukawa-SYK Hamiltonian replaces the bare four-fermion vertex with a dynamically mediated interaction via exchange of bosonic degrees of freedom. 
The motivation comes from strange-metal phenomenology: the complex SYK model~\cite{gu_2020} exhibits a non-Fermi-liquid state akin to the strange-metal phase of cuprate superconductors~\cite{phillips_2022}, and the Yukawa-SYK model extends this by incorporating boson-mediated pairing in direct analogy with Migdal–Eliashberg theory~\cite{marsiglio_2020}. 
A systematic series of works by Esterlis, Schmalian, and collaborators~\cite{hauck_2020,esterlis_2021,guo_2022,inkof_2022,valentinis_2023,patel_2023,li_2024,valentinis_2026} has since explored the thermodynamic, transport, and holographic properties of the complex Yukawa-SYK system and its extension to two spatial dimensions.

Despite this progress, the quantum chaotic properties of the Yukawa-SYK model remain incompletely understood. 
An analytic approach~\cite{Davis:2023} addresses the short-time chaotic regime, while a recent numerical study~\cite{Hauke:2026} has begun to explore statistical aspects of many-body chaos. 
Here, we address the question of late-time chaotic behavior within a gate-based quantum simulation framework.

\subsection{Hamiltonian}
\label{subsec:Hamiltonian_YSYK}

In this part of the work, we address the Yukawa-SYK model~\cite{marcus_2019}, which realizes an alternative form of \textit{e-ph} interaction: $M$ bosonic modes are coupled to bilinear combinations of $2N$ Majorana fermions. The Yukawa-SYK Hamiltonian reads
\begin{equation}
\label{eq:H_YSYK}
    \hat {H}_{\mathrm{YSYK}} =  \hat {\mathcal{H}}_0 +  \hat {H}_{\rm int},
\end{equation}
where
\begin{equation}
\label{eq:H0_YSYK}
\hat {\mathcal{H}}_0= \sum_{k=1}^{M}  \omega_{k}\hat a^{\dagger}_{k} \hat a^{\phantom{\dagger}}_{k}
\end{equation}
is the free-boson Hamiltonian. The modes have frequencies $\omega_k$ and are described by the annihilation and creation operators $\hat a^{\phantom{\dagger}}_{k}$ and $\hat a^{\dagger}_{k}$, respectively, obeying the standard commutation relations $[\hat a^{\phantom{\dagger}}_{k}, \hat a^{\dagger}_{k'}]=\delta_{kk'}$ and $[\hat a^{\phantom{\dagger}}_{k}, \hat a^{\phantom{\dagger}}_{k'}] = [\hat a^{\dagger}_{k}, \hat a^{\dagger}_{k'}] = 0$. The interacting part of the Hamiltonian $\hat H_{\rm YSYK}$ is
\begin{equation}
\label{H_int_YSYK}
\hat H_{\rm int}=\sum_{k=1}^{M} \sum_{1 \leq \alpha < \beta \leq 2N} \hspace{-1.5mm}  g_{\alpha\beta k}\ i \hat\chi_\alpha \hat\chi_\beta \rbs{\hat a^{\dagger}_{k} + \hat a^{\phantom{\dagger}}_{k}},
\end{equation}
where ${\hat \chi_\alpha}$ are real Majorana-fermion operators, ${\hat \chi_\alpha = \hat \chi_\alpha^\dagger }$, satisfying the anticommutation relations $\{\hat\chi_{\alpha},\hat\chi_{\beta}\}=2\delta_{\alpha\beta}$. 
The couplings are i.i.d. Gaussian random variables with the constraint $g_{\alpha\beta k}=-g_{\beta\alpha k}$.
Each coupling has zero mean, $\mathbb{E}[g_{\alpha\beta k}]=0$, where $\mathbb{E}[\cdots]$ denotes the average over random configurations. 
The variance of the couplings is ${\mathbb{E}\left[\,g_{\alpha\beta k}^2\,\right] = g^2/8 N^2}$
where $g$ is the characteristic interaction energy scale.
A sketch of a possible circuit-QED architecture for the Yukawa-SYK model is shown in Fig.~\ref{architecture_YSYK}.

\subsection{Trotter evolution}
\label{subsec:Tr_evol_YSYK}
We are interested in constructing a quantum circuit for a single Trotter step associated with each interaction term in $\hat H_{\rm YSYK}$.
To this end, we first represent the $2N$ Majorana fermions $\hat \chi_\alpha$ with $1\leq \alpha \leq 2N$ through $N$ pairs of {\it spinless} complex fermions, $\hat c_j^{\phantom{\dagger}}$ and $\hat c_j^\dagger$, with $1\leq j \leq N$:
\begin{align}
    \begin{split}
        \hat \chi_{2j-1} &= \phantom{-i(}\hat c_j^{\phantom{\dagger}}+\hat c_j^\dagger, \\
        \hat \chi_{2j\phantom{-1}} &= -i(\hat c_j^{\phantom{\dagger}}-\hat c_j^\dagger).
    \end{split}
\end{align}
The second step is to apply the Jordan-Wigner transformation, which is given by \cref{JW_uparrow} without the spin index:
\begin{equation}
\hat c_j = \frac{\hat X_j+i\hat Y_j}{2}\hat Z^{\rm (JW)}_{1,j-1}.
\end{equation}
Here, the operator $\hat Z^{\rm (JW)}_{i,j}$ is the Jordan-Wigner string, with $i\leq j$,
\begin{equation}
\label{JW_string_YSYK}
\hat Z^{\rm (JW)}_{i,j} = 
\prod\limits_{k=i}^{j}\hat Z_{k} .
\end{equation}

Using the above representation, the Majorana-fermion bilinears in $\hat H_{\rm int}$ are transformed into spin operators. For $i<j$, we obtain four types of combinations:
\begin{align}\label{chi_alpha_chi_beta}
\begin{split}
    i \hat \chi_{2i-1} 
    \hat \chi_{2j-1} &=\phantom{-}\hat Y_i 
    \hat Z^{\rm (JW)}_{i+1,j-1}
    \hat X_j ,   \\
    i \hat \chi_{2i} 
    \hat \chi_{2j-1} &=-\hat X_i  
    \hat Z^{\rm (JW)}_{i+1,j-1}
    \hat X_j,  \\
    i \hat \chi_{2i-1} 
    \hat \chi_{2j} &=\phantom{-}\hat Y_i 
    \hat Z^{\rm (JW)}_{i+1,j-1}\hat Y_j,   \\ 
    i \hat \chi_{2i} 
    \hat \chi_{2j} &=-\hat X_i  
    \hat Z^{\rm (JW)}_{i+1,j-1}
    \hat Y_j.
\end{split}
\end{align}
These relations are supplemented by one additional combination for $i=j$:
\begin{align}
i\hat \chi_{2i-1}\hat \chi_{2i}=-Z_i . \label{chi_chi_Z}
\end{align}
In analogy with the logic of the QRM simulations in Sec.~\ref{subsec:QRM}, we define the total evolution operator as
\begin{align}\label{H_YSYK_trotterized}
&e^{-i\hat H_{\rm YSYK}(t_L-t_0)} = e^{-i \hat {\mathcal{H}}_0 t_L} \prod_{p=0}^{L-1} \hat S_{\rm YSYK}(t_p+\tau,t_p) e^{i \hat {\mathcal{H}}_0 t_0}
\nonumber \\ 
& \hat S_{\rm YSYK}(t_p+\tau,t_p) = \prod\limits_{k=1}^{M}  
\prod_{\beta=2}^{2N} \prod_{\alpha=1}^{\beta-1}
\hat S_{\rm YSYK}^{(  \alpha\beta k)}(t_p+\tau,t_p).
\end{align}
Here, the Trotterized evolution operators
$ \hat S_{\rm YSYK}^{(  \alpha\beta k)}$ 
encode individual interaction terms from $\hat H_{\rm int}$. Their products form
$\hat S_{\rm YSYK}(t_p+\tau,t_p)$, which is the full evolution operator for a single Trotter step $p$. 
In analogy with the HH model, the unitaries
$\hat S_{\rm YSYK}^{(\alpha\beta k)}$ are decomposed into analog Rabi
gates interleaved with layers of digital gates. 

Assuming a circuit-QED geometry as in Fig.~\ref{architecture_YSYK}, with $M\leq N$ and each resonator $k$ connected to a qubit $q_k$, the gates $ \hat S_{\rm YSYK}^{(  \alpha\beta k)}$ are represented as shown in Fig.~\ref{fig:S_R_YSYK}. The Jordan-Wigner strings, which appear in the exponentials of the evolution operators, are simulated by layers of $\mathsf{CNOT}$s followed by a final $\mathsf{CZ}$, as suggested in Ref.~\cite{Troyer_fermions_2015}.

\subsection{Quantum Chaos}\label{subsec:QuC}

A hallmark of many-body quantum-chaotic systems is the emergence of random-matrix-theory (RMT) level-spacing statistics, as it was originally conjectured in the seminal works of Berry and Tabor~\cite{Berry:1977} and Bohigas, Giannoni, and Schmit~\cite{Bohigas:1984}. Depending on the symmetries of the underlying microscopic Hamiltonian, the spectral statistics is expected to follow one of the three canonical random-matrix ensembles. 
Such structure of energy levels, as well as the Gaussian statistics
of related eigenvectors, determines the universal long-time behaviour of many-body correlation functions -- the so-called
ergodic regime.

\begin{figure}[t!]
    \includegraphics[width=0.65\linewidth]{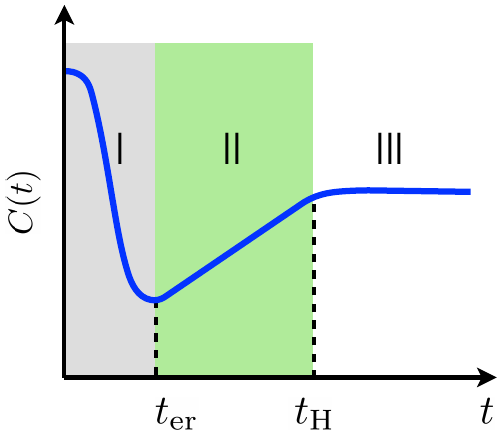}\vspace{-0.25cm}
  \caption{Typical behaviour of a generic two-point correlator $C(t)$ of a random chaotic many-body Hamiltonian 
averaged either over disorder realizations or fast time fluctuations.
The initial time decay (region I) originates from the relaxation of non-universal massive modes,
describing the energy diffusion in the many-body Fock space, see Ref.~\cite{Altland_2021} for details.
The linear growth (region II) followed by the plateau (region III) is the universal ergodic behavior which is attributed to the RMT-like spectral statistics of a many-body chaotic quantum system. By definition, the onset of the linear ramp starts at the ergodic time $t_{\rm er}$. This scale is system dependent. 
On the other hand, the plateau begins at the universal Heisenberg time $t_{\rm H}\simeq 2\pi/\bar\Delta$, with $\bar\Delta$ being the average many-body mean level spacing.
}
\label{corr_sketch}
\end{figure}

Two basic quantities, which can be used as clear indicators of ergodicity, are the following. First, it is the many-body two-point correlation function for a generic few-body operator $\hat O$, defined as
\begin{equation}\label{eq:correlator}
    C(t)=\mathbb{E}\left[{\rm tr}\left(\hat{\rho}_0 \hat{O}(t)\hat{O}^\dagger(0)\right)\right],
\end{equation}
with $\hat{O}(t)=e^{i \hat{H}t}\hat{O}e^{-i\hat{H}t}$ being the time evolution of $\hat{O}$ in the Heisenberg picture. Second, one considers the spectral form factor
\begin{equation}
\label{eq:sff}
    K(t) = \mathbb{E}\left[\left|{\rm tr}\left(\hat\rho_0 e^{- i \hat H t}\right)\right|^2\right].
\end{equation}
Both definitions for $C(t)$ and $K(t)$ assume averaging over the random configurations; $\hat \rho_0$ is the Gibbs density matrix, $\hat \rho_0=e^{-\beta \hat H}/Z_\beta$, where $\beta$ is the inverse temperature and $Z_\beta={\rm tr} \left(e^{-\beta \hat H}\right)$ is the partition sum.

For a quantum chaotic model with late-time ergodicity, the correlator $C(t)$, as well as the spectral form factor $K(t)$, show a dip-ramp-plateau structure~\cite{Altland_2021}, as indicated schematically in Fig.~\ref{corr_sketch}. 
Here, the dip marks the onset of ergodicity and the universal regime governed by eigenvalue repulsion in random matrix theory.

\begin{figure}[t!]
	\begin{center}
		\includegraphics[width=0.95\linewidth]{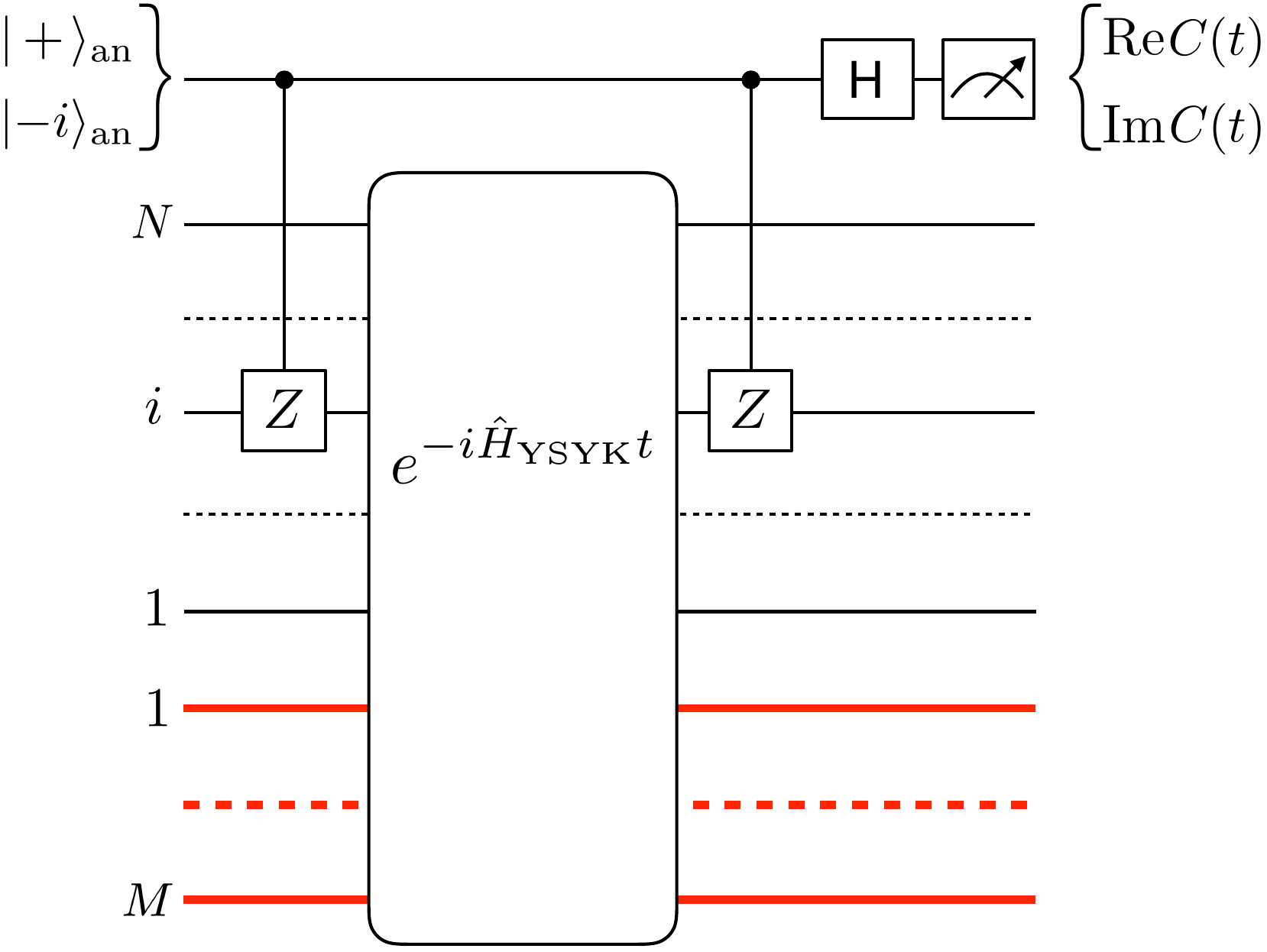}
  \end{center}
    \caption{Hadamard test protocol measuring  the real and imaginary parts of the correlator $C(t)=\langle\psi| \hat O(t)\hat O^\dagger(0)|\psi\rangle $ with  
    $\hat O=i\hat \chi_{2i}\hat\chi_{2i-1}=\hat Z_i$ and the input state  $|\psi\rangle$. Depending on the   ancilla input state $|-i\rangle_{\rm an}$  or $|  + \rangle_{\rm an}$, the imaginary or real  part  is measured, respectively. }
	\label{hadamard_corr}
\end{figure}

\begin{figure*}[htb!]
	\begin{center}
		\includegraphics[width=\textwidth]
        {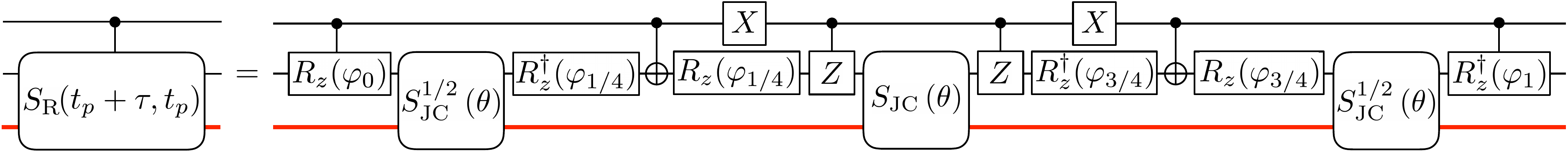}
		\caption
		{\label{fig:C_S_R} {Quantum circuit representing the qubit-controlled Rabi gate;
see Appendix~\ref{appendix:ctrl_Rabi} for the proof.   Importantly, this circuit does not require controlled JC gates. 
        Only standard controlled $Z$-rotations and ${\rm CNOT}$ gates are used.}}
	\end{center}
\end{figure*}

Our goal is to design measurement protocols for $C(t)$ and $K(t)$ using quantum circuits that implement the Yukawa-SYK model (Fig.~\ref{fig:S_R_YSYK}). 
These circuits via \cref{H_YSYK_trotterized} encode the Trotterized time-evolution operator associated with the Yukawa-SYK Hamiltonian.
Because the Hamiltonian is time-independent, the circuit in the stationary time frame is constructed from repeated applications of an identical Trotter layer. The resulting Trotterized dynamics can therefore be regarded as a particular instance of a random Floquet circuit, a class of systems that has attracted considerable attention 
in recent years~\cite{Chalker:2018, Friedman:2019, Chalker2024}. We will appeal to this relation shortly. 
The regime most amenable to quantum simulation corresponds to $g/\omega_0 \ll 1$. In this limit, the many-body spectrum decomposes into a series of bands with a typical width 
$\mathcal{W}=\sqrt{M} g^2/\omega_0$, while adjacent bands are separated by energy gaps of order $\omega_0$~\cite{SchmitzS:2026}. One can understand the formation of such bands as a 
broadening due to interaction of multiply degenerate bare phonon levels 
$E\left(\{ n_k\}\right) = \omega_0 \sum_{k=1}^M n_k$, where $n_k \geq 0$ are positive integers. 
For this reason we assume $\mathcal{W} \ll T \lesssim \omega_0$ and simplify the 
Gibbs density matrix by
\begin{equation}
\label{eq:desity_matrix_YSYK}
    \hat\rho_0 = \frac{\mathds{1}_q}{2^N} \otimes \prod_{k=1}^M |0^{(k)}_{\rm ph}\rangle \langle 0_{\rm ph}^{(k)}|.
\end{equation}
It means that all resonators are initialized in their ground states, while averaging over Majorana modes is reduced to the trace over related fermion Fock states
or, equivalently, over all possible bit strings.

To be specific, we focus on the fermionic bilinears $\hat O = i\hat \chi_{2i}\hat \chi_{2i-1} \equiv \hat Z_i$, see \cref{chi_chi_Z}. A measurement protocol for the two-point correlation function $C(t)$ corresponding to a given initial state $|\psi\rangle$ can then be constructed as a straightforward extension of the Hadamard-test circuit shown in Fig.~\ref{hadamard_corr}. Similar circuits have been widely employed in the literature devoted to the quantum simulation of Hubbard-type fermionic models; see Ref.~\cite{bishop2023quantum} and references therein. For the density matrix defined in \cref{eq:desity_matrix_YSYK}, one must average the measurement outcomes obtained from the Hadamard-test circuit over all initial bit strings of the form $|\psi\rangle =  |{\bf s}\rangle \otimes  \prod\limits_{k=1}^M|{0}^{(k)}_{\rm ph}\rangle$, where ${\bf s}\in \{0,1\}^N$, sampled with equal probability $1/2^N$. It is straightforward to verify that the presence of the additional bosonic Hilbert space does not affect the validity of the measurement protocol.

We now turn to the measurement of the spectral form factor $K(t)$. In this case, the standard Hadamard test requires the use of controlled unitary evolution operators at every Trotter step. This can be achieved by promoting the Rabi gate $S_{\rm R}(t_p+\tau,t_p)$ to its controlled version, as illustrated in Fig.~\ref{fig:C_S_R}. The Hadamard test then provides access to both the real and imaginary parts of the matrix element 
$\langle\psi|\exp(-iH_{\rm YSYK}(t_L-t_0))|\psi\rangle$.
Since all resonators are assumed to be initialized in their ground states, one may equivalently employ the evolution operator
\begin{equation}
\hat S_{\rm YSYK}(t_L, t_0)=\prod_{p=0}^{L-1}\hat S_{\rm YSYK}(t_p+\tau,t_p)
\end{equation}
defined in the rotating frame, where all elementary Yukawa-SYK quantum gates shown in Fig.~\ref{fig:S_R_YSYK} are implemented. Indeed, the rotating frame is defined with respect to the free bosonic Hamiltonian 
${\cal H}_0$, see \cref{H_R_trotterized_exact} and~\cref{eq:H0_YSYK}. 
Since evolution with ${\cal H}_0$ acts trivially on the chosen family of initial states,
$e^{-iH_0t}|\psi\rangle = |\psi\rangle$,
the corresponding matrix elements remain unchanged.

 \begin{figure}[b!]
	\includegraphics[width=0.95\linewidth]{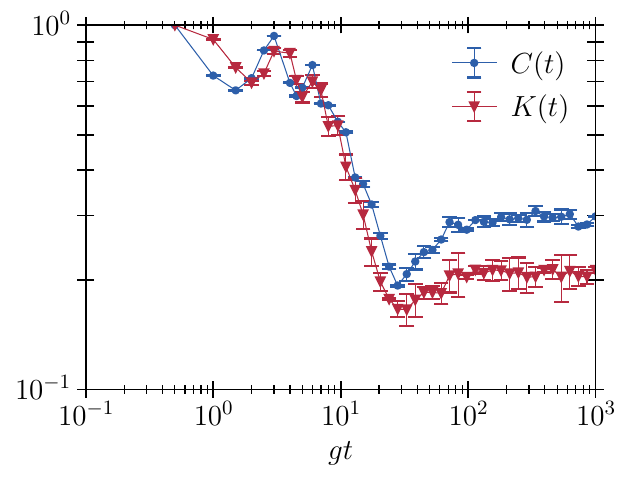}\vspace{-0.25cm}
    \caption{Circuit-based simulation data for quantum-chaotic behavior in the Yukawa-SYK model. Plots of the correlation functions: $C(t)$ for the operator $\hat O=i\hat \chi_1\hat\chi_2=-\hat Z_1$ (blue points)
    and $K(t)$ (red points). 
    Parameters of the model are: $2N=4$   Majorana fermions and $M=2$ bosons,  $\omega_{1,2}=2g$, averaging over 1000 configurations with error bars indicating  standard deviation. The Trotter step time $\tau=1/(2g)$, and the initial density matrix is specified in~\cref{eq:desity_matrix_YSYK}.}
	\label{fig:C}
\end{figure}

Numerical results obtained using the measurement protocols for $C(t)$ and $K(t)$ described above are shown in Fig.~\ref{fig:C} for the minimal instance of the model. Despite the small system size $(M=N=2)$, both quantities clearly exhibit the cha\-ra\-cteristic dip – ramp – plateau structure at long times, 
$t \gtrsim 2\pi g^{-1}$, in agreement with the schematic behavior illustrated in Fig.~\ref{corr_sketch}.
It is important to emphasize that these results probe many-body chaos in a Floquet realization of the Yukawa-SYK model. The corresponding Floquet Hamiltonian, $\hat H_{\rm F}$, differs from the target Hamiltonian~\cref{eq:H_YSYK} due to the finite Trotter step $\tau$. Specifically, the $\tau$-dependent $\hat H_{\rm F}(\tau)$ is defined through
\begin{equation}
    e^{ - i \tau \hat  H_{\rm F}(\tau)} :=  e^{ - i \hat  {\cal H}_0 (t_p+\tau)} \hat S_{\rm YSYK}(t_p+\tau,t_p)
     e^{ i \hat  {\cal H}_0 t_p}.
\end{equation}
It follows from the construction of the elementary Yukawa-SYK gates $\hat S^{(\alpha\beta k)}$ that the Floquet Hamiltonian defined in this way depends only on the Trotter step $\tau$ and is independent of the reference time $t_p$. Furthermore, because we employ a second-order Trotter decomposition, the Floquet Hamiltonian satisfies
\begin{equation}
    \hat H_{\rm F}(\tau) = \hat H_{\rm YSYK} + {\cal O}(\tau^2).
\end{equation}

The simulations presented in Fig.~\ref{fig:C} were performed using a Trotter step $\tau = 1/(2g)$, chosen to lie slightly below the so-called Trotterization threshold; see Ref.~\cite{Kargi2025quantumchaos} for a recent review. For the parameter regime considered here, we estimate the threshold to be $\tau_c \simeq 1/g$.
According to the qualitative picture developed in Ref.~\cite{Kargi2025quantumchaos}, Floquet eigenstates remain localized in the eigenbasis of the target Hamiltonian $\hat {\mathcal{H}}$ when the Trotter step satisfies $\tau \ll \tau_c$. In this regime, the Floquet spectrum is continuously connected to the spectrum of $\hat{\mathcal{H}}$, and the Trotterized dynamics faithfully approximates the exact time evolution over a broad temporal window. As the Trotter step approaches the threshold $\tau_c$, the Floquet eigenstates become strongly hybridized and the spectrum of $H_F$ reorganizes. As a result, the Floquet dynamics becomes chaotic even when the target Hamiltonian $\hat{\mathcal{H}}$ is integrable.

The continuous-time Yukawa-SYK model provides a particularly illustrative example. 
For $N=M=2$, the high degree of symmetry prevents the onset of fully developed quantum chaos: 
in this case, $\hat H_{\rm YSYK}$ belongs to the Gaussian symplectic ensemble. Signatures of nonintegrability emerge only for larger system sizes, $N\geq 3$. By contrast, its Floquet counterpart exhibits clear signatures of quantum chaos already at $N=2$ when the Trotter step approaches $\tau_c$, as evidenced by the data shown in Fig.~\ref{fig:C}. The pronounced oscillations observed in both $C(t)$ and $K(t)$ at short times, 
$t \lesssim 2\pi g^{-1}$, reflect the close correspondence between the Floquet dynamics and the exact integrable dynamics of the underlying Yukawa-SYK model in this regime.

To conclude this section, we note an interesting proposal for the analog simulation of the Yukawa-SYK model using trapped ultracold atoms in an optical cavity-QED setup~\cite{Hauke:2026}. This approach is complementary to the digi\-tal-analog framework discussed here.

\section{Discussion \& Conclusions}
\label{sec:discussion}

We have developed a  framework for simulating electron-phonon and Majorana fermion-phonon models in circuit QED architectures. The central idea is to emulate phonons through the photon  degrees of freedom in    microwave resonators. It enables one to avoid  the  overhead associated with  qubit encoding of bosons and  to  access superradiant-like ground states with a nonzero photon occupation number. The central unitary primitive of the circuits is the qubit-resonator Rabi gate, which is 
   decomposed into   three analog Jaynes-Cummings gates supplemented by single-qubit digital rotations. In this    digital-analog approach,         simulations of    arbitrary strong   fermion-phonon interactions become available  on  circuits with moderate or weak physical couplings.  These gates are available  via dispersive and resonant circuit QED settings with tunable qubits.

In the first part of the work, we formulated the digital-analog quantum circuit, which emulates Trotter evolution  in the  Hubbard-Holstein model  describing    generic electron-phonon interactions. In the second part, we introduced   protocols for the  variational quantum eigensolver for this model. The proposed variational Hamiltonian ansatz mirrors the structure of the Hubbard-Holstein Hamiltonian -- it contains displacement, Hubbard, hopping, electron-phonon, and phonon-assisted hopping layers. We show the efficiency of the ansatz in preparing the ground state near the Dicke-like quantum critical point, which is characterized by a highly entangled many-body ground state.  Including conservative estimates of resonator and qubit relaxation rates in the Lindblad description, we show a reduction of    the fidelity, but the variational states still retain the main qualitative features of the exact ground states. These results suggest that near-term circuit QED devices can prepare    nonclassical  states, specifically at the quantum critical points,  before full fault tolerance is available. 
  
 A key  ingredient of the variational eigensolver is the measurement of mixed qubit-resonator observables. We therefore proposed Hadamard-test protocols based on  controlled phase rotations and controlled displacement gates of the resonator. These measurements provide access to the bosonic contributions entering the variational energy functional,  and to other characteristics like full counting statistics. Since the same dispersive coupling techniques are  routinely used for readout and control in circuit QED, the proposed measurement schemes are compatible with existing hardware capabilities.

In the third part, we addressed another variant of fermion-boson interaction -- the Yukawa-Sachdev-Ye-Kitaev model with random couplings between Majorana fermions and phonons. The quantum circuit representing the Trotterized evolution in this model corresponds to random Floquet dynamics. Simulations of the correlation function and form-factor reveal universal random-matrix signatures of many-body quantum chaos even for a small-size system. The Hadamard-test circuits for probing real-time correlation functions are presented  as well.

Overall, our results identify hybrid qubit-resonator processors as a natural platform for digital-analog simulations of electron-phonon physics. The approach is resource efficient because it uses the resonator as a genuine bosonic degree of freedom rather than replacing it by many auxiliary qubits. At the same time, the protocol remains programmable: different fermionic Hamiltonians, local or random electron-phonon couplings, and variational or dynamical experiments can be implemented by changing the surrounding qubit gates.

\begin{acknowledgments}
This work is supported by DPG under Germany’s Excellence Strategy – Cluster of Excellence Matter and Light for Quantum Computing (ML4Q) EXC 2004/2 – 390534769. We further acknowledge 
a financial support from the German Federal Ministry of Education and Research (BMBF) in the funding program “Quantum technologies–from basic research to market,” Contract No. 13N16149 (project QSolid) and from the EU Commission grant number  101113946 (Project OpenSuperQPlus).
A. C. acknowledges funding from the BMBF program
‘German Quantum Computer based on Superconducting Qubits’ MUNIQCSC (Nr. 13N16188).
D.~S.~S. acknowledges a financial support  from the BMFTR grant number 13N15688 (Project DAQC).
\end{acknowledgments}

\section*{Code and data availability}

The code and data associated with this article will be made publicly available upon publication.

\twocolumngrid
\appendix

\section{VHA algorithm} \label{appendix:optimization}
\begin{figure}[htb!]
	\begin{center}
\includegraphics[width=\linewidth]
{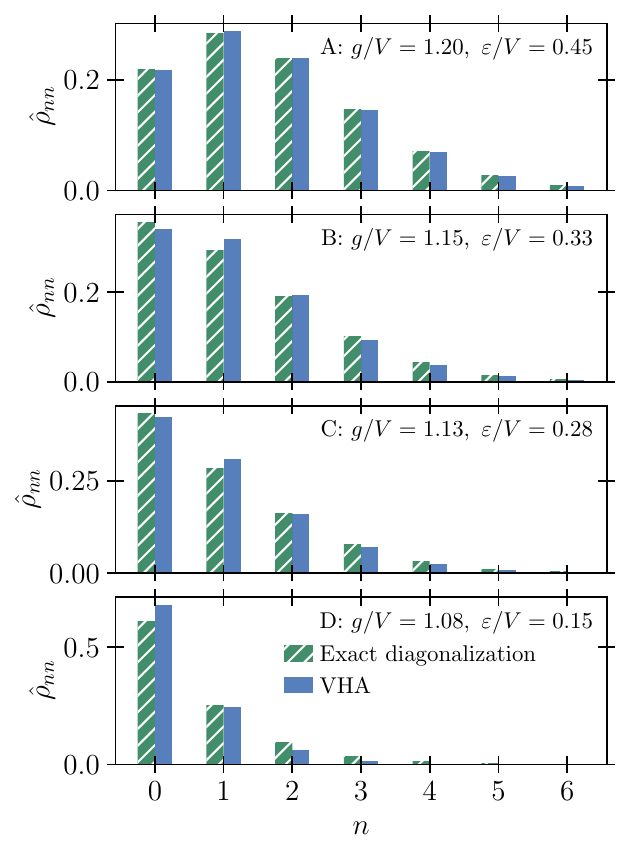}\vspace{-0.5cm}
\caption{\label{fig:histograms_VHA} Histograms of the phonon distributions $\rho_{n,n}$ corresponding to points A--D in the fluctuation-dominated region of Fig.~\ref{fig_phase_diagram}~(b). The distributions are calculated for the rotated Hamiltonian $\hat H_{\rm HH}'$. Hatched bars show exact-diagonalization results; bold colored bars show data from the VHA protocol without dissipation. 
}
	\end{center}
\end{figure}

The VHA algorithm for the \textit{e-ph}-coupled HH Hamiltonian is implemented as a two-step optimization procedure. First, we use the VHA to prepare the ground state of the spin subsystem governed by $\hat H_{\rm Hubbard}$, for which the algorithm achieves a fidelity above $99.9\%$. In this step, the parameters $\xi_{i,\sigma}^{(0)}$, $u_i^{(0)}$, and $v_{i,\sigma}^{(0)}$ are obtained. In the second step, we include the full $\hat H_{\mathrm{HH}}$ with bosonic terms, using the parameters $\xi_{i,\sigma}^{(0)}$, $u_i^{(0)}$, and $v_{i,\sigma}^{(0)}$ as a warm start for optimizing the full set $\mathbf{x}$ with the VHA circuit in Fig.~\ref{fig:ansatz}.
We apply this algorithm to calculate the approximate ground states $|\Psi(\mathbf{x})\rangle$ at points A--D in the phase diagram of Fig.~\ref{fig_phase_diagram}~(b). In Fig.~\ref{fig:histograms_VHA}, we show histograms of the phonon distributions in these states $|\Psi(\mathbf{x})\rangle$ and compare them with exact-diagonalization results. The values of the variational parameters are listed in Table~\ref{tab:VHA_params}.

\begin{table*}[htb!]
    \centering
    \begin{tabular}{|c|c|c|c|c|c|c|c|c|c|c|c|c|c|c|c|c|}
    \hline
           & $\xi_{1,\uparrow}$ & $\xi_{1,\downarrow}$ & $\xi_{2,\uparrow}$ &  $\xi_{2,\downarrow}$ & $\zeta_{1,\uparrow}$ & $\zeta_{2,\uparrow}$ & $\zeta_{1,\downarrow}$ & $\zeta_{2,\downarrow}$ & $v_{1,\uparrow}$ & $v_{1,\downarrow}$ & $u_1$ & $u_2$ & $\Theta$ & $\alpha$ & $\gamma_{1,\uparrow}$ & $\gamma_{1,\downarrow}$ \\
         \hline
         A & 0.053 & 1.412 & 0.077 & 1.435 & 0.235 & 0.178 & 0.202 & 0.145 & 0.182 & 0.182 & -0.248 & 0.089 & 1.506 & 1.122 & 0.116 & 0.116  \\
         \hline
         B & 1.235 & 0.229 & 0.246 & 1.266 & 0.121 & 0.201 & 0.235 & 0.099 & -0.247 & 0.240 & -0.187 & 0.148 & 1.080 & 1.302 & 0.092 & 0.067 \\
         \hline
         C & 1.162 & 0.302 & 0.316 & 1.195 & 0.103 & 0.201 & 0.223 & 0.085 & -0.255 & 0.252 & -0.185 & 0.150 & 1.050 & 1.185 & 0.085 & 0.076 \\
         \hline
         D & 0.948 & 0.517 & 0.549 & 0.963 & 0.132 & 0.190 & 0.200 & 0.083 & -0.222 & 0.220 & -0.189 & 0.147 & 1.106 & 0.723 & 0.042 & 0.082 \\
         \hline
    \end{tabular}
    \caption{Optimal parameters $\mathbf{x}$ of the VHA circuit in Fig.~\ref{fig:ansatz}, obtained by minimizing the ground-state energy of the Hamiltonian $\hat H_{\rm HH}$. The data correspond to points A--D in the phase diagram of Fig.~\ref{fig_phase_diagram}~(b).}
    \label{tab:VHA_params}
\end{table*}
In the next Appendix, we report simulations of the VHA algorithm in the presence of dissipation in the qubits and resonators.

\section{Lindblad noise in the VHA}\label{appendix:lindblad}
 
We assume that the dominant losses arise from relaxation during the two-qubit gates, $\mathsf{SWAP}$s and $\mathsf{CNOT}$s, and during the Rabi gates, whose operation times are much longer than those of single-qubit gates. Apart from these relaxation channels, all gates are assumed to be ideal and do not introduce additional errors. 

Losses are emulated by adding purely dissipative evolution after each $\mathsf{SWAP}$, $\mathsf{CNOT}$, and $  S_{\rm R}$ gate in the VHA circuit of Fig.~\ref{fig:ansatz}. The added dissipative dynamics is governed by the Lindblad master equation
\begin{align}\label{lindblad}
    \begin{split}
        \frac{d\hat \rho}{dt} &= \frac{1}{T_1^{(\rm res)} }\left( \hat a\hat \rho \hat a^\dagger - \frac{1}{2} \{ \hat a^\dagger \hat a, \hat \rho \} \right) \\ 
        &+ \frac{1}{T_1^{\rm (qubit)}} \sum_\sigma\sum_{j=1}^N\left( \hat \sigma_{j,\sigma}^- \hat \rho \hat \sigma_{j,\sigma}^+ - \frac{1}{2} \{ \hat \sigma_{j,\sigma}^+ \hat \sigma_{j,\sigma}^-, \hat \rho \} \right),        
    \end{split}
\end{align}
where $T_1^{\rm (res)}$ and $T_1^{\rm (qubit)}$ are the relaxation times of the resonator and qubits, respectively.

The time interval over which the many-body density matrix $\hat\rho$ evolves according to \cref{lindblad} is set by the duration of the preceding gate. These intervals are chosen as follows: (i) the Rabi-gate time $\tau_{\rm Rabi}$, (ii) the $\mathsf{CNOT}$ gate time $\tau_{\mathsf{CNOT}}$, and (iii) $3\tau_{\mathsf{CNOT}}$ for $\mathsf{SWAP}$ gates decomposed into three $\mathsf{CNOT}$s. Experimentally relevant time parameters used in the simulations are collected in Table~\ref{tab:loss_parameters}. The gates are assumed to be relatively slow.

In Table~\ref{tab:VHA_results}, we report the following characteristics of the wave functions obtained with the VHA: energy, fidelity with respect to the exact ground-state density matrix $\hat\rho_{\rm ex}$,
\begin{equation}
F_{\rm VHA} = \left({\rm tr}(\sqrt{\sqrt{\hat \rho}\hat \rho_{\rm ex}\sqrt{\hat \rho}})\right)^2,
\end{equation}
and phonon-field magnitude. There are three sets of data: (i) exact-diagonalization results, (ii) the VHA algorithm without dissipation, and (iii) the VHA algorithm with dissipation. In the absence of loss processes, the fidelity of the ground-state preparation is estimated to be above $95\%$. This accuracy drops by roughly $15\%$ when qubit and resonator losses are included in the model.

\begin{table}[htb!]
    \centering
    \renewcommand{\arraystretch}{1.5}
    \begin{tabular}{|c|c|c|}
        \hline
        & {Qubit} & {Resonator} \\
        \hline
        Relaxation time & $T_1^{\rm (qubit)} = 50 \mu$s & $T_1^{(\rm res)} = 200  \mu$s \\ \hline
        Gate time & $\tau_{\mathsf{CNOT}}=\frac{1}{3}\tau_{\mathsf{SWAP}} = 100$ns & $\tau_{\rm Rabi} = 200$ns \\
        \hline
    \end{tabular}
    \caption{Relaxation times and gate-operation times for the qubits and resonator.}
    \label{tab:loss_parameters}
\end{table}

\begin{table}[htb!]
    \centering
    \begin{tabular}{|c|c|c|c|c|c|c|c|c|}
    \hline
    \rotatebox[origin=c]{90} & $ \frac{1}{V}E_{\rm ex} $& $ \frac{1}{V}E_{\rm VHA}$ & $\frac{1}{V}E_{\rm VHA}^{\rm (diss)}$ & $F_{\rm VHA}$ & $F_{\rm VHA}^{\rm (diss)}$ & $\mathcal{E}_{\rm ex}$ & $\mathcal{E}_{\rm VHA} $ & $\mathcal{E}_{\rm VHA}^{\rm (diss)}$ \\
         \hline
         A & -0.759 & -0.754 & -0.317 & 0.999 & 0.849 & 0.803 & 0.799 & 0.741\\
         \hline
         B & -0.749 & -0.649 & -0.328 & 0.980 & 0.831 & 0.612 & 0.619 & 0.591\\
         \hline
         C & -0.771 & -0.667 & -0.389 & 0.979 & 0.830 & 0.509 & 0.511 & 0.493\\
         \hline
         D & -0.897 & -0.823 & -0.617 & 0.972 & 0.827 & 0.287 & 0.193 & 0.196\\
         \hline
    \end{tabular}
    \caption{VHA results for points A--D in the phase diagram of Fig.~\ref{fig_phase_diagram}~(b). The data are calculated for the  Hamiltonian $\hat H_{\rm HH}$. The ground-state energies are $E_{\rm ex}$ from exact diagonalization and $E_{\rm VHA}$ and $E_{\rm VHA}^{\rm (diss)}$ from the VHA without and with dissipation channels, respectively. The other entries include the corresponding fidelities $F_{\rm VHA}$ and
$F_{\rm VHA}^{\rm (diss)}$.
The last three columns show the corresponding magnitudes of the phonon-field
displacement, $\mathcal{E} = |\langle \hat a \rangle|$.}
\label{tab:VHA_results}
\end{table}

\section{Controlled Rabi gate}\label{appendix:ctrl_Rabi}

We verify below that the quantum circuit in Fig.~\ref{fig:C_S_R} realizes the controlled Rabi gate. To this end, it is sufficient to consider the two possible initial states of the ancilla qubit separately. 
If the ancilla is in the state $|0\rangle_{\rm an}$, the unitary operator acting on the computational degrees of freedom becomes the identity, as required:
\begin{align}
    \begin{split}
        \hat S_{\rm R}^{[0]}(t_p + \tau,t_p) &=  \hat S_{\rm JC}(\theta/2) \hat Z \hat S_{\rm JC}(\theta) \hat Z S_{\rm JC}(\theta/2) \\
        &= \hat S_{\rm JC}(\theta/2) \hat S_{\rm JC}(-\theta) \hat S_{\rm JC}(\theta/2) =  \mathds{1}.         
    \end{split}
\end{align}
Here, we use the fact that when the $ X$ gate is applied to the ancilla, it is flipped to the state $|1\rangle_{\rm an}$, and therefore the controlled-$Z$ gates become effective. 
Similarly, if the ancilla is initially in the state $|1\rangle_{\rm an}$, the corresponding unitary Rabi operator reads
\begin{align}
    \begin{split}
         \hat S_{\rm R}^{[1]}(t_p + \tau,t_p) &=  \hat R^\dagger_z(\varphi_{1}) \hat S_{\rm JC}(\theta/2) \\
         &\times \hat R_z(\varphi_{3/4})\hat X \hat R^\dagger_z(\varphi_{3/4})  \\
         &\times \hat S_{\rm JC}(\theta) \hat R_z(\varphi_{1/4}) \hat X \hat R^\dagger_z(\varphi_{1/4})  \\
         &\times \hat S_{\rm JC}(\theta/2) \hat R_z(\varphi_{0}) = \hat S_{\rm R}^{(2)}(t_p + \tau,t_p).        
    \end{split}
\end{align}
This expression is equal to the second-order approximation to the Rabi gate in \cref{S_R_2}, using the identity $\hat R_z(\varphi_{3/4})\hat X  \hat R^\dagger_z(\varphi_{3/4}) \equiv \hat R_z(2\varphi_{3/4})\hat X$, and analogously for the second layer of rotations with the phase $\varphi_{1/4}$.


\begin{thebibliography}{88}%
\makeatletter
\providecommand \@ifxundefined [1]{%
 \@ifx{#1\undefined}
}%
\providecommand \@ifnum [1]{%
 \ifnum #1\expandafter \@firstoftwo
 \else \expandafter \@secondoftwo
 \fi
}%
\providecommand \@ifx [1]{%
 \ifx #1\expandafter \@firstoftwo
 \else \expandafter \@secondoftwo
 \fi
}%
\providecommand \natexlab [1]{#1}%
\providecommand \enquote  [1]{``#1''}%
\providecommand \bibnamefont  [1]{#1}%
\providecommand \bibfnamefont [1]{#1}%
\providecommand \citenamefont [1]{#1}%
\providecommand \href@noop [0]{\@secondoftwo}%
\providecommand \href [0]{\begingroup \@sanitize@url \@href}%
\providecommand \@href[1]{\@@startlink{#1}\@@href}%
\providecommand \@@href[1]{\endgroup#1\@@endlink}%
\providecommand \@sanitize@url [0]{\catcode `\\12\catcode `\$12\catcode
  `\&12\catcode `\#12\catcode `\^12\catcode `\_12\catcode `\%12\relax}%
\providecommand \@@startlink[1]{}%
\providecommand \@@endlink[0]{}%
\providecommand \url  [0]{\begingroup\@sanitize@url \@url }%
\providecommand \@url [1]{\endgroup\@href {#1}{\urlprefix }}%
\providecommand \urlprefix  [0]{URL }%
\providecommand \Eprint [0]{\href }%
\providecommand \doibase [0]{https://doi.org/}%
\providecommand \selectlanguage [0]{\@gobble}%
\providecommand \bibinfo  [0]{\@secondoftwo}%
\providecommand \bibfield  [0]{\@secondoftwo}%
\providecommand \translation [1]{[#1]}%
\providecommand \BibitemOpen [0]{}%
\providecommand \bibitemStop [0]{}%
\providecommand \bibitemNoStop [0]{.\EOS\space}%
\providecommand \EOS [0]{\spacefactor3000\relax}%
\providecommand \BibitemShut  [1]{\csname bibitem#1\endcsname}%
\let\auto@bib@innerbib\@empty
%</preamble>
\bibitem [{\citenamefont {H.~Fröhlich}\ and\ \citenamefont
  {Zienau}(1950)}]{froehlich_1950}%
  \BibitemOpen
  \bibfield  {author} {\bibinfo {author} {\bibfnamefont {H.~P.}\ \bibnamefont
  {H.~Fröhlich}}\ and\ \bibinfo {author} {\bibfnamefont {S.}~\bibnamefont
  {Zienau}},\ }\bibfield  {title} {\bibinfo {title} {{XX. Properties of slow
  electrons in polar materials}},\ }\href
  {https://doi.org/10.1080/14786445008521794} {\bibfield  {journal} {\bibinfo
  {journal} {The London, Edinburgh, and Dublin Philosophical Magazine and
  Journal of Science}\ }\textbf {\bibinfo {volume} {41}},\ \bibinfo {pages}
  {221} (\bibinfo {year} {1950})}\BibitemShut {NoStop}%
\bibitem [{\citenamefont {Holstein}(1959)}]{Holstein1959}%
  \BibitemOpen
  \bibfield  {author} {\bibinfo {author} {\bibfnamefont {T.}~\bibnamefont
  {Holstein}},\ }\bibfield  {title} {\bibinfo {title} {{Studies of polaron
  motion: Part II. The “small” polaron}},\ }\href
  {https://doi.org/https://doi.org/10.1016/0003-4916(59)90003-X} {\bibfield
  {journal} {\bibinfo  {journal} {Annals of Physics}\ }\textbf {\bibinfo
  {volume} {8}},\ \bibinfo {pages} {343} (\bibinfo {year} {1959})}\BibitemShut
  {NoStop}%
\bibitem [{\citenamefont {Dahnovsky}(2007)}]{Dahnovsky_2007}%
  \BibitemOpen
  \bibfield  {author} {\bibinfo {author} {\bibfnamefont {Y.}~\bibnamefont
  {Dahnovsky}},\ }\bibfield  {title} {\bibinfo {title} {{{Ab initio electron
  propagators in molecules with strong electron-phonon interaction. I. Phonon
  averages}}},\ }\href {https://doi.org/10.1063/1.2741528} {\bibfield
  {journal} {\bibinfo  {journal} {The Journal of Chemical Physics}\ }\textbf
  {\bibinfo {volume} {126}},\ \bibinfo {pages} {234111} (\bibinfo {year}
  {2007})}\BibitemShut {NoStop}%
\bibitem [{\citenamefont {Marsiglio}\ and\ \citenamefont
  {Carbotte}(2008)}]{Marsiglio2008}%
  \BibitemOpen
  \bibfield  {author} {\bibinfo {author} {\bibfnamefont {F.}~\bibnamefont
  {Marsiglio}}\ and\ \bibinfo {author} {\bibfnamefont {J.~P.}\ \bibnamefont
  {Carbotte}},\ }\bibinfo {title} {{{Electron-Phonon Superconductivity}}},\ in\
  \href {https://doi.org/10.1007/978-3-540-73253-2_3} {\emph {\bibinfo
  {booktitle} {{{Superconductivity: Conventional and Unconventional
  Superconductors}}}}},\ \bibinfo {editor} {edited by\ \bibinfo {editor}
  {\bibfnamefont {K.~H.}\ \bibnamefont {Bennemann}}\ and\ \bibinfo {editor}
  {\bibfnamefont {J.~B.}\ \bibnamefont {Ketterson}}}\ (\bibinfo  {publisher}
  {Springer Berlin Heidelberg},\ \bibinfo {address} {Berlin, Heidelberg},\
  \bibinfo {year} {2008})\ pp.\ \bibinfo {pages} {73--162}\BibitemShut
  {NoStop}%
\bibitem [{\citenamefont {Nosarzewski}\ \emph {et~al.}(2021)\citenamefont
  {Nosarzewski}, \citenamefont {Huang}, \citenamefont {Dee}, \citenamefont
  {Esterlis}, \citenamefont {Moritz}, \citenamefont {Kivelson}, \citenamefont
  {Johnston},\ and\ \citenamefont {Devereaux}}]{Nosarzewski_2021}%
  \BibitemOpen
  \bibfield  {author} {\bibinfo {author} {\bibfnamefont {B.}~\bibnamefont
  {Nosarzewski}}, \bibinfo {author} {\bibfnamefont {E.~W.}\ \bibnamefont
  {Huang}}, \bibinfo {author} {\bibfnamefont {P.~M.}\ \bibnamefont {Dee}},
  \bibinfo {author} {\bibfnamefont {I.}~\bibnamefont {Esterlis}}, \bibinfo
  {author} {\bibfnamefont {B.}~\bibnamefont {Moritz}}, \bibinfo {author}
  {\bibfnamefont {S.~A.}\ \bibnamefont {Kivelson}}, \bibinfo {author}
  {\bibfnamefont {S.}~\bibnamefont {Johnston}},\ and\ \bibinfo {author}
  {\bibfnamefont {T.~P.}\ \bibnamefont {Devereaux}},\ }\bibfield  {title}
  {\bibinfo {title} {{Superconductivity, charge density waves, and bipolarons
  in the Holstein model}},\ }\href
  {https://doi.org/10.1103/PhysRevB.103.235156} {\bibfield  {journal} {\bibinfo
   {journal} {Phys. Rev. B}\ }\textbf {\bibinfo {volume} {103}},\ \bibinfo
  {pages} {235156} (\bibinfo {year} {2021})}\BibitemShut {NoStop}%
\bibitem [{\citenamefont {Hubbard}(1963)}]{Hubbard1963}%
  \BibitemOpen
  \bibfield  {author} {\bibinfo {author} {\bibfnamefont {J.}~\bibnamefont
  {Hubbard}},\ }\bibfield  {title} {\bibinfo {title} {Electron correlations in
  narrow energy bands},\ }\href {https://doi.org/10.1098/rspa.1963.0204}
  {\bibfield  {journal} {\bibinfo  {journal} {Proceedings of the Royal Society
  of London. A. Mathematical and Physical Sciences}\ }\textbf {\bibinfo
  {volume} {276}},\ \bibinfo {pages} {238} (\bibinfo {year}
  {1963})}\BibitemShut {NoStop}%
\bibitem [{\citenamefont {Berger}\ \emph {et~al.}(1995)\citenamefont {Berger},
  \citenamefont {Val\'a\ifmmode~\check{s}\else \v{s}\fi{}ek},\ and\
  \citenamefont {von~der Linden}}]{Berger_1995}%
  \BibitemOpen
  \bibfield  {author} {\bibinfo {author} {\bibfnamefont {E.}~\bibnamefont
  {Berger}}, \bibinfo {author} {\bibfnamefont {P.}~\bibnamefont
  {Val\'a\ifmmode~\check{s}\else \v{s}\fi{}ek}},\ and\ \bibinfo {author}
  {\bibfnamefont {W.}~\bibnamefont {von~der Linden}},\ }\bibfield  {title}
  {\bibinfo {title} {{Two-dimensional Hubbard-Holstein model}},\ }\href
  {https://doi.org/10.1103/PhysRevB.52.4806} {\bibfield  {journal} {\bibinfo
  {journal} {Phys. Rev. B}\ }\textbf {\bibinfo {volume} {52}},\ \bibinfo
  {pages} {4806} (\bibinfo {year} {1995})}\BibitemShut {NoStop}%
\bibitem [{\citenamefont {Costa}\ \emph {et~al.}(2020)\citenamefont {Costa},
  \citenamefont {Seki}, \citenamefont {Yunoki},\ and\ \citenamefont
  {Sorella}}]{Costa2020}%
  \BibitemOpen
  \bibfield  {author} {\bibinfo {author} {\bibfnamefont {N.~C.}\ \bibnamefont
  {Costa}}, \bibinfo {author} {\bibfnamefont {K.}~\bibnamefont {Seki}},
  \bibinfo {author} {\bibfnamefont {S.}~\bibnamefont {Yunoki}},\ and\ \bibinfo
  {author} {\bibfnamefont {S.}~\bibnamefont {Sorella}},\ }\bibfield  {title}
  {\bibinfo {title} {{Phase diagram of the two-dimensional Hubbard-Holstein
  model}},\ }\href {https://doi.org/10.1038/s42005-020-0342-2} {\bibfield
  {journal} {\bibinfo  {journal} {Commun. Phys.}\ }\textbf {\bibinfo {volume}
  {3}},\ \bibinfo {pages} {80} (\bibinfo {year} {2020})}\BibitemShut {NoStop}%
\bibitem [{\citenamefont {Esterlis}\ and\ \citenamefont
  {Schmalian}(2019)}]{Esterlis_2019}%
  \BibitemOpen
  \bibfield  {author} {\bibinfo {author} {\bibfnamefont {I.}~\bibnamefont
  {Esterlis}}\ and\ \bibinfo {author} {\bibfnamefont {J.}~\bibnamefont
  {Schmalian}},\ }\bibfield  {title} {\bibinfo {title} {{Cooper pairing of
  incoherent electrons: An electron-phonon version of the Sachdev-Ye-Kitaev
  model}},\ }\href {https://doi.org/10.1103/PhysRevB.100.115132} {\bibfield
  {journal} {\bibinfo  {journal} {Phys. Rev. B}\ }\textbf {\bibinfo {volume}
  {100}},\ \bibinfo {pages} {115132} (\bibinfo {year} {2019})}\BibitemShut
  {NoStop}%
\bibitem [{\citenamefont {Feynman}(1982)}]{Feynman1982}%
  \BibitemOpen
  \bibfield  {author} {\bibinfo {author} {\bibfnamefont {R.~P.}\ \bibnamefont
  {Feynman}},\ }\bibfield  {title} {\bibinfo {title} {Simulating physics with
  computers},\ }\href {https://doi.org/10.1007/BF02650179} {\bibfield
  {journal} {\bibinfo  {journal} {International Journal of Theoretical
  Physics}\ }\textbf {\bibinfo {volume} {21}},\ \bibinfo {pages} {467}
  (\bibinfo {year} {1982})}\BibitemShut {NoStop}%
\bibitem [{\citenamefont {Lloyd}(1996)}]{Lloyd:1996}%
  \BibitemOpen
  \bibfield  {author} {\bibinfo {author} {\bibfnamefont {S.}~\bibnamefont
  {Lloyd}},\ }\bibfield  {title} {\bibinfo {title} {Universal quantum
  simulators},\ }\href {https://doi.org/10.1126/science.273.5278.1073}
  {\bibfield  {journal} {\bibinfo  {journal} {Science}\ }\textbf {\bibinfo
  {volume} {273}},\ \bibinfo {pages} {1073} (\bibinfo {year}
  {1996})}\BibitemShut {NoStop}%
\bibitem [{\citenamefont {Weimer}\ \emph {et~al.}(2010)\citenamefont {Weimer},
  \citenamefont {M{\"u}ller}, \citenamefont {Lesanovsky}, \citenamefont
  {Zoller},\ and\ \citenamefont {B{\"u}chler}}]{Weimer2010}%
  \BibitemOpen
  \bibfield  {author} {\bibinfo {author} {\bibfnamefont {H.}~\bibnamefont
  {Weimer}}, \bibinfo {author} {\bibfnamefont {M.}~\bibnamefont {M{\"u}ller}},
  \bibinfo {author} {\bibfnamefont {I.}~\bibnamefont {Lesanovsky}}, \bibinfo
  {author} {\bibfnamefont {P.}~\bibnamefont {Zoller}},\ and\ \bibinfo {author}
  {\bibfnamefont {H.~P.}\ \bibnamefont {B{\"u}chler}},\ }\bibfield  {title}
  {\bibinfo {title} {A rydberg quantum simulator},\ }\href
  {https://doi.org/10.1038/nphys1614} {\bibfield  {journal} {\bibinfo
  {journal} {Nat. Phys.}\ }\textbf {\bibinfo {volume} {6}},\ \bibinfo {pages}
  {382} (\bibinfo {year} {2010})}\BibitemShut {NoStop}%
\bibitem [{\citenamefont {Bassman~Oftelie}\ \emph {et~al.}(2021)\citenamefont
  {Bassman~Oftelie}, \citenamefont {Urbanek}, \citenamefont {Metcalf},
  \citenamefont {Carter}, \citenamefont {Kemper},\ and\ \citenamefont
  {de~Jong}}]{BassmanOftelie_2021}%
  \BibitemOpen
  \bibfield  {author} {\bibinfo {author} {\bibfnamefont {L.}~\bibnamefont
  {Bassman~Oftelie}}, \bibinfo {author} {\bibfnamefont {M.}~\bibnamefont
  {Urbanek}}, \bibinfo {author} {\bibfnamefont {M.}~\bibnamefont {Metcalf}},
  \bibinfo {author} {\bibfnamefont {J.}~\bibnamefont {Carter}}, \bibinfo
  {author} {\bibfnamefont {A.~F.}\ \bibnamefont {Kemper}},\ and\ \bibinfo
  {author} {\bibfnamefont {W.~A.}\ \bibnamefont {de~Jong}},\ }\bibfield
  {title} {\bibinfo {title} {Simulating quantum materials with digital quantum
  computers},\ }\href {https://doi.org/10.1088/2058-9565/ac1ca6} {\bibfield
  {journal} {\bibinfo  {journal} {Quantum Sci. Technol.}\ }\textbf {\bibinfo
  {volume} {6}},\ \bibinfo {pages} {043002} (\bibinfo {year}
  {2021})}\BibitemShut {NoStop}%
\bibitem [{\citenamefont {Bravyi}\ \emph {et~al.}(2024)\citenamefont {Bravyi},
  \citenamefont {Cross}, \citenamefont {Gambetta}, \citenamefont {Maslov},
  \citenamefont {Rall},\ and\ \citenamefont {Yoder}}]{Bravyi2024}%
  \BibitemOpen
  \bibfield  {author} {\bibinfo {author} {\bibfnamefont {S.}~\bibnamefont
  {Bravyi}}, \bibinfo {author} {\bibfnamefont {A.~W.}\ \bibnamefont {Cross}},
  \bibinfo {author} {\bibfnamefont {J.~M.}\ \bibnamefont {Gambetta}}, \bibinfo
  {author} {\bibfnamefont {D.}~\bibnamefont {Maslov}}, \bibinfo {author}
  {\bibfnamefont {P.}~\bibnamefont {Rall}},\ and\ \bibinfo {author}
  {\bibfnamefont {T.~J.}\ \bibnamefont {Yoder}},\ }\bibfield  {title} {\bibinfo
  {title} {High-threshold and low-overhead fault-tolerant quantum memory},\
  }\href {https://doi.org/10.1038/s41586-024-07107-7} {\bibfield  {journal}
  {\bibinfo  {journal} {Nature}\ }\textbf {\bibinfo {volume} {627}},\ \bibinfo
  {pages} {778} (\bibinfo {year} {2024})}\BibitemShut {NoStop}%
\bibitem [{\citenamefont {Miessen}\ \emph {et~al.}(2024)\citenamefont
  {Miessen}, \citenamefont {Egger}, \citenamefont {Tavernelli},\ and\
  \citenamefont {Mazzola}}]{PRXQuantum.5.040320}%
  \BibitemOpen
  \bibfield  {author} {\bibinfo {author} {\bibfnamefont {A.}~\bibnamefont
  {Miessen}}, \bibinfo {author} {\bibfnamefont {D.~J.}\ \bibnamefont {Egger}},
  \bibinfo {author} {\bibfnamefont {I.}~\bibnamefont {Tavernelli}},\ and\
  \bibinfo {author} {\bibfnamefont {G.}~\bibnamefont {Mazzola}},\ }\bibfield
  {title} {\bibinfo {title} {Benchmarking digital quantum simulations above
  hundreds of qubits using quantum critical dynamics},\ }\href
  {https://doi.org/10.1103/PRXQuantum.5.040320} {\bibfield  {journal} {\bibinfo
   {journal} {PRX Quantum}\ }\textbf {\bibinfo {volume} {5}},\ \bibinfo {pages}
  {040320} (\bibinfo {year} {2024})}\BibitemShut {NoStop}%
\bibitem [{\citenamefont {Macridin}\ \emph
  {et~al.}(2018{\natexlab{a}})\citenamefont {Macridin}, \citenamefont
  {Spentzouris}, \citenamefont {Amundson},\ and\ \citenamefont
  {Harnik}}]{Macridin:2018}%
  \BibitemOpen
  \bibfield  {author} {\bibinfo {author} {\bibfnamefont {A.}~\bibnamefont
  {Macridin}}, \bibinfo {author} {\bibfnamefont {P.}~\bibnamefont
  {Spentzouris}}, \bibinfo {author} {\bibfnamefont {J.}~\bibnamefont
  {Amundson}},\ and\ \bibinfo {author} {\bibfnamefont {R.}~\bibnamefont
  {Harnik}},\ }\bibfield  {title} {\bibinfo {title} {Electron-phonon systems on
  a universal quantum computer},\ }\href
  {https://doi.org/10.1103/PhysRevLett.121.110504} {\bibfield  {journal}
  {\bibinfo  {journal} {Phys. Rev. Lett.}\ }\textbf {\bibinfo {volume} {121}},\
  \bibinfo {pages} {110504} (\bibinfo {year} {2018}{\natexlab{a}})}\BibitemShut
  {NoStop}%
\bibitem [{\citenamefont {Macridin}\ \emph
  {et~al.}(2018{\natexlab{b}})\citenamefont {Macridin}, \citenamefont
  {Spentzouris}, \citenamefont {Amundson},\ and\ \citenamefont
  {Harnik}}]{Macridin:2018a}%
  \BibitemOpen
  \bibfield  {author} {\bibinfo {author} {\bibfnamefont {A.}~\bibnamefont
  {Macridin}}, \bibinfo {author} {\bibfnamefont {P.}~\bibnamefont
  {Spentzouris}}, \bibinfo {author} {\bibfnamefont {J.}~\bibnamefont
  {Amundson}},\ and\ \bibinfo {author} {\bibfnamefont {R.}~\bibnamefont
  {Harnik}},\ }\bibfield  {title} {\bibinfo {title} {Digital quantum
  computation of fermion-boson interacting systems},\ }\href
  {https://doi.org/10.1103/PhysRevA.98.042312} {\bibfield  {journal} {\bibinfo
  {journal} {Phys. Rev. A}\ }\textbf {\bibinfo {volume} {98}},\ \bibinfo
  {pages} {042312} (\bibinfo {year} {2018}{\natexlab{b}})}\BibitemShut
  {NoStop}%
\bibitem [{\citenamefont {Fauseweh}(2024)}]{Fauseweh2024}%
  \BibitemOpen
  \bibfield  {author} {\bibinfo {author} {\bibfnamefont {B.}~\bibnamefont
  {Fauseweh}},\ }\bibfield  {title} {\bibinfo {title} {Quantum many-body
  simulations on digital quantum computers: State-of-the-art and future
  challenges},\ }\href {https://doi.org/10.1038/s41467-024-46402-9} {\bibfield
  {journal} {\bibinfo  {journal} {Nat. Commun.}\ }\textbf {\bibinfo {volume}
  {15}},\ \bibinfo {pages} {2123} (\bibinfo {year} {2024})}\BibitemShut
  {NoStop}%
\bibitem [{\citenamefont {Castillo-Moreno}\ \emph {et~al.}(2025)\citenamefont
  {Castillo-Moreno}, \citenamefont {Sépulcre}, \citenamefont {Hillmann},
  \citenamefont {Amin}, \citenamefont {Kervinen},\ and\ \citenamefont
  {Gasparinetti}}]{castillomoreno2025}%
  \BibitemOpen
  \bibfield  {author} {\bibinfo {author} {\bibfnamefont {C.}~\bibnamefont
  {Castillo-Moreno}}, \bibinfo {author} {\bibfnamefont {T.}~\bibnamefont
  {Sépulcre}}, \bibinfo {author} {\bibfnamefont {T.}~\bibnamefont {Hillmann}},
  \bibinfo {author} {\bibfnamefont {K.~R.}\ \bibnamefont {Amin}}, \bibinfo
  {author} {\bibfnamefont {M.}~\bibnamefont {Kervinen}},\ and\ \bibinfo
  {author} {\bibfnamefont {S.}~\bibnamefont {Gasparinetti}},\ }\href@noop {}
  {\bibinfo {title} {{Experimental observation of multimode quantum phase
  transitions in a superconducting Bose-Hubbard simulator}}} (\bibinfo {year}
  {2025}),\ \Eprint {https://arxiv.org/abs/2508.20116} {arXiv:2508.20116
  [cond-mat.quant-gas]} \BibitemShut {NoStop}%
\bibitem [{\citenamefont {Forn-D\'{\i}az}\ \emph {et~al.}(2019)\citenamefont
  {Forn-D\'{\i}az}, \citenamefont {Lamata}, \citenamefont {Rico}, \citenamefont
  {Kono},\ and\ \citenamefont {Solano}}]{RevModPhys.91.025005}%
  \BibitemOpen
  \bibfield  {author} {\bibinfo {author} {\bibfnamefont {P.}~\bibnamefont
  {Forn-D\'{\i}az}}, \bibinfo {author} {\bibfnamefont {L.}~\bibnamefont
  {Lamata}}, \bibinfo {author} {\bibfnamefont {E.}~\bibnamefont {Rico}},
  \bibinfo {author} {\bibfnamefont {J.}~\bibnamefont {Kono}},\ and\ \bibinfo
  {author} {\bibfnamefont {E.}~\bibnamefont {Solano}},\ }\bibfield  {title}
  {\bibinfo {title} {{Ultrastrong coupling regimes of light-matter
  interaction}},\ }\href {https://doi.org/10.1103/RevModPhys.91.025005}
  {\bibfield  {journal} {\bibinfo  {journal} {Rev. Mod. Phys.}\ }\textbf
  {\bibinfo {volume} {91}},\ \bibinfo {pages} {025005} (\bibinfo {year}
  {2019})}\BibitemShut {NoStop}%
\bibitem [{\citenamefont {Frisk~Kockum}\ \emph {et~al.}(2019)\citenamefont
  {Frisk~Kockum}, \citenamefont {Miranowicz}, \citenamefont {De~Liberato},
  \citenamefont {Savasta},\ and\ \citenamefont {Nori}}]{FriskKockum2019}%
  \BibitemOpen
  \bibfield  {author} {\bibinfo {author} {\bibfnamefont {A.}~\bibnamefont
  {Frisk~Kockum}}, \bibinfo {author} {\bibfnamefont {A.}~\bibnamefont
  {Miranowicz}}, \bibinfo {author} {\bibfnamefont {S.}~\bibnamefont
  {De~Liberato}}, \bibinfo {author} {\bibfnamefont {S.}~\bibnamefont
  {Savasta}},\ and\ \bibinfo {author} {\bibfnamefont {F.}~\bibnamefont
  {Nori}},\ }\bibfield  {title} {\bibinfo {title} {{Ultrastrong coupling
  between light and matter}},\ }\href
  {https://doi.org/10.1038/s42254-018-0006-2} {\bibfield  {journal} {\bibinfo
  {journal} {Nat. Rev. Phys.}\ }\textbf {\bibinfo {volume} {1}},\ \bibinfo
  {pages} {19} (\bibinfo {year} {2019})}\BibitemShut {NoStop}%
\bibitem [{\citenamefont {Blais}\ \emph {et~al.}(2021)\citenamefont {Blais},
  \citenamefont {Grimsmo}, \citenamefont {Girvin},\ and\ \citenamefont
  {Wallraff}}]{RevModPhys.93.025005}%
  \BibitemOpen
  \bibfield  {author} {\bibinfo {author} {\bibfnamefont {A.}~\bibnamefont
  {Blais}}, \bibinfo {author} {\bibfnamefont {A.~L.}\ \bibnamefont {Grimsmo}},
  \bibinfo {author} {\bibfnamefont {S.~M.}\ \bibnamefont {Girvin}},\ and\
  \bibinfo {author} {\bibfnamefont {A.}~\bibnamefont {Wallraff}},\ }\bibfield
  {title} {\bibinfo {title} {{Circuit quantum electrodynamics}},\ }\href
  {https://doi.org/10.1103/RevModPhys.93.025005} {\bibfield  {journal}
  {\bibinfo  {journal} {Rev. Mod. Phys.}\ }\textbf {\bibinfo {volume} {93}},\
  \bibinfo {pages} {025005} (\bibinfo {year} {2021})}\BibitemShut {NoStop}%
\bibitem [{\citenamefont {Qin}\ \emph {et~al.}(2024)\citenamefont {Qin},
  \citenamefont {Kockum}, \citenamefont {Muñoz}, \citenamefont {Miranowicz},\
  and\ \citenamefont {Nori}}]{QIN20241}%
  \BibitemOpen
  \bibfield  {author} {\bibinfo {author} {\bibfnamefont {W.}~\bibnamefont
  {Qin}}, \bibinfo {author} {\bibfnamefont {A.~F.}\ \bibnamefont {Kockum}},
  \bibinfo {author} {\bibfnamefont {C.~S.}\ \bibnamefont {Muñoz}}, \bibinfo
  {author} {\bibfnamefont {A.}~\bibnamefont {Miranowicz}},\ and\ \bibinfo
  {author} {\bibfnamefont {F.}~\bibnamefont {Nori}},\ }\bibfield  {title}
  {\bibinfo {title} {Quantum amplification and simulation of strong and
  ultrastrong coupling of light and matter},\ }\href
  {https://doi.org/https://doi.org/10.1016/j.physrep.2024.05.003} {\bibfield
  {journal} {\bibinfo  {journal} {Phys. Rep.}\ }\textbf {\bibinfo {volume}
  {1078}},\ \bibinfo {pages} {1} (\bibinfo {year} {2024})}\BibitemShut
  {NoStop}%
\bibitem [{\citenamefont {Koch}\ \emph {et~al.}(2007)\citenamefont {Koch},
  \citenamefont {Yu}, \citenamefont {Gambetta}, \citenamefont {Houck},
  \citenamefont {Schuster}, \citenamefont {Majer}, \citenamefont {Blais},
  \citenamefont {Devoret}, \citenamefont {Girvin},\ and\ \citenamefont
  {Schoelkopf}}]{koch2007}%
  \BibitemOpen
  \bibfield  {author} {\bibinfo {author} {\bibfnamefont {J.}~\bibnamefont
  {Koch}}, \bibinfo {author} {\bibfnamefont {T.~M.}\ \bibnamefont {Yu}},
  \bibinfo {author} {\bibfnamefont {J.}~\bibnamefont {Gambetta}}, \bibinfo
  {author} {\bibfnamefont {A.~A.}\ \bibnamefont {Houck}}, \bibinfo {author}
  {\bibfnamefont {D.~I.}\ \bibnamefont {Schuster}}, \bibinfo {author}
  {\bibfnamefont {J.}~\bibnamefont {Majer}}, \bibinfo {author} {\bibfnamefont
  {A.}~\bibnamefont {Blais}}, \bibinfo {author} {\bibfnamefont {M.~H.}\
  \bibnamefont {Devoret}}, \bibinfo {author} {\bibfnamefont {S.~M.}\
  \bibnamefont {Girvin}},\ and\ \bibinfo {author} {\bibfnamefont {R.~J.}\
  \bibnamefont {Schoelkopf}},\ }\bibfield  {title} {\bibinfo {title}
  {{Charge-insensitive qubit design derived from the Cooper pair box}},\ }\href
  {https://doi.org/10.1103/PhysRevA.76.042319} {\bibfield  {journal} {\bibinfo
  {journal} {Phys. Rev. A}\ }\textbf {\bibinfo {volume} {76}},\ \bibinfo
  {pages} {042319} (\bibinfo {year} {2007})}\BibitemShut {NoStop}%
\bibitem [{\citenamefont {Sawaya}\ \emph {et~al.}(2020)\citenamefont {Sawaya},
  \citenamefont {Menke}, \citenamefont {Kyaw}, \citenamefont {Johri},
  \citenamefont {Aspuru-Guzik},\ and\ \citenamefont {Guerreschi}}]{Sawaya2020}%
  \BibitemOpen
  \bibfield  {author} {\bibinfo {author} {\bibfnamefont {N.~P.~D.}\
  \bibnamefont {Sawaya}}, \bibinfo {author} {\bibfnamefont {T.}~\bibnamefont
  {Menke}}, \bibinfo {author} {\bibfnamefont {T.~H.}\ \bibnamefont {Kyaw}},
  \bibinfo {author} {\bibfnamefont {S.}~\bibnamefont {Johri}}, \bibinfo
  {author} {\bibfnamefont {A.}~\bibnamefont {Aspuru-Guzik}},\ and\ \bibinfo
  {author} {\bibfnamefont {G.~G.}\ \bibnamefont {Guerreschi}},\ }\bibfield
  {title} {\bibinfo {title} {{Resource-efficient digital quantum simulation of
  $d$-level systems for photonic, vibrational, and spin-$s$ Hamiltonians}},\
  }\href {https://doi.org/10.1038/s41534-020-0278-0} {\bibfield  {journal}
  {\bibinfo  {journal} {npj Quantum Inf.}\ }\textbf {\bibinfo {volume} {6}},\
  \bibinfo {pages} {49} (\bibinfo {year} {2020})}\BibitemShut {NoStop}%
\bibitem [{\citenamefont {Xie}\ \emph {et~al.}(2017)\citenamefont {Xie},
  \citenamefont {Zhong}, \citenamefont {Batchelor},\ and\ \citenamefont
  {Lee}}]{Xie_2017}%
  \BibitemOpen
  \bibfield  {author} {\bibinfo {author} {\bibfnamefont {Q.}~\bibnamefont
  {Xie}}, \bibinfo {author} {\bibfnamefont {H.}~\bibnamefont {Zhong}}, \bibinfo
  {author} {\bibfnamefont {M.~T.}\ \bibnamefont {Batchelor}},\ and\ \bibinfo
  {author} {\bibfnamefont {C.}~\bibnamefont {Lee}},\ }\bibfield  {title}
  {\bibinfo {title} {{The quantum Rabi model: solution and dynamics}},\ }\href
  {https://doi.org/10.1088/1751-8121/aa5a65} {\bibfield  {journal} {\bibinfo
  {journal} {J. Phys. A: Math. Theor.}\ }\textbf {\bibinfo {volume} {50}},\
  \bibinfo {pages} {113001} (\bibinfo {year} {2017})}\BibitemShut {NoStop}%
\bibitem [{\citenamefont {Niemczyk}\ \emph {et~al.}(2010)\citenamefont
  {Niemczyk}, \citenamefont {Deppe}, \citenamefont {Huebl}, \citenamefont
  {Menzel}, \citenamefont {Hocke}, \citenamefont {Schwarz}, \citenamefont
  {Garcia-Ripoll}, \citenamefont {Zueco}, \citenamefont {H{\"u}mmer},
  \citenamefont {Solano}, \citenamefont {Marx},\ and\ \citenamefont
  {Gross}}]{Niemczyk2010}%
  \BibitemOpen
  \bibfield  {author} {\bibinfo {author} {\bibfnamefont {T.}~\bibnamefont
  {Niemczyk}}, \bibinfo {author} {\bibfnamefont {F.}~\bibnamefont {Deppe}},
  \bibinfo {author} {\bibfnamefont {H.}~\bibnamefont {Huebl}}, \bibinfo
  {author} {\bibfnamefont {E.~P.}\ \bibnamefont {Menzel}}, \bibinfo {author}
  {\bibfnamefont {F.}~\bibnamefont {Hocke}}, \bibinfo {author} {\bibfnamefont
  {M.~J.}\ \bibnamefont {Schwarz}}, \bibinfo {author} {\bibfnamefont {J.~J.}\
  \bibnamefont {Garcia-Ripoll}}, \bibinfo {author} {\bibfnamefont
  {D.}~\bibnamefont {Zueco}}, \bibinfo {author} {\bibfnamefont
  {T.}~\bibnamefont {H{\"u}mmer}}, \bibinfo {author} {\bibfnamefont
  {E.}~\bibnamefont {Solano}}, \bibinfo {author} {\bibfnamefont
  {A.}~\bibnamefont {Marx}},\ and\ \bibinfo {author} {\bibfnamefont
  {R.}~\bibnamefont {Gross}},\ }\bibfield  {title} {\bibinfo {title} {{Circuit
  quantum electrodynamics in the ultrastrong-coupling regime}},\ }\href
  {https://doi.org/10.1038/nphys1730} {\bibfield  {journal} {\bibinfo
  {journal} {Nat. Phys.}\ }\textbf {\bibinfo {volume} {6}},\ \bibinfo {pages}
  {772} (\bibinfo {year} {2010})}\BibitemShut {NoStop}%
\bibitem [{\citenamefont {Forn-D{\'i}az}\ \emph {et~al.}(2017)\citenamefont
  {Forn-D{\'i}az}, \citenamefont {Garc{\'i}a-Ripoll}, \citenamefont
  {Peropadre}, \citenamefont {Orgiazzi}, \citenamefont {Yurtalan},
  \citenamefont {Belyansky}, \citenamefont {Wilson},\ and\ \citenamefont
  {Lupascu}}]{Forn-Diaz2017}%
  \BibitemOpen
  \bibfield  {author} {\bibinfo {author} {\bibfnamefont {P.}~\bibnamefont
  {Forn-D{\'i}az}}, \bibinfo {author} {\bibfnamefont {J.~J.}\ \bibnamefont
  {Garc{\'i}a-Ripoll}}, \bibinfo {author} {\bibfnamefont {B.}~\bibnamefont
  {Peropadre}}, \bibinfo {author} {\bibfnamefont {J.-L.}\ \bibnamefont
  {Orgiazzi}}, \bibinfo {author} {\bibfnamefont {M.~A.}\ \bibnamefont
  {Yurtalan}}, \bibinfo {author} {\bibfnamefont {R.}~\bibnamefont {Belyansky}},
  \bibinfo {author} {\bibfnamefont {C.~M.}\ \bibnamefont {Wilson}},\ and\
  \bibinfo {author} {\bibfnamefont {A.}~\bibnamefont {Lupascu}},\ }\bibfield
  {title} {\bibinfo {title} {{Ultrastrong coupling of a single artificial atom
  to an electromagnetic continuum in the nonperturbative regime}},\ }\href
  {https://doi.org/10.1038/nphys3905} {\bibfield  {journal} {\bibinfo
  {journal} {Nat. Phys.}\ }\textbf {\bibinfo {volume} {13}},\ \bibinfo {pages}
  {39} (\bibinfo {year} {2017})}\BibitemShut {NoStop}%
\bibitem [{\citenamefont {Mezzacapo}\ \emph {et~al.}(2014)\citenamefont
  {Mezzacapo}, \citenamefont {Las~Heras}, \citenamefont {Pedernales},
  \citenamefont {DiCarlo}, \citenamefont {Solano},\ and\ \citenamefont
  {Lamata}}]{Mezzacapo2014}%
  \BibitemOpen
  \bibfield  {author} {\bibinfo {author} {\bibfnamefont {A.}~\bibnamefont
  {Mezzacapo}}, \bibinfo {author} {\bibfnamefont {U.}~\bibnamefont
  {Las~Heras}}, \bibinfo {author} {\bibfnamefont {J.~S.}\ \bibnamefont
  {Pedernales}}, \bibinfo {author} {\bibfnamefont {L.}~\bibnamefont {DiCarlo}},
  \bibinfo {author} {\bibfnamefont {E.}~\bibnamefont {Solano}},\ and\ \bibinfo
  {author} {\bibfnamefont {L.}~\bibnamefont {Lamata}},\ }\bibfield  {title}
  {\bibinfo {title} {{Digital Quantum Rabi and Dicke Models in Superconducting
  Circuits}},\ }\href {https://doi.org/10.1038/srep07482} {\bibfield  {journal}
  {\bibinfo  {journal} {Sci. Rep.}\ }\textbf {\bibinfo {volume} {4}},\ \bibinfo
  {pages} {7482} (\bibinfo {year} {2014})}\BibitemShut {NoStop}%
\bibitem [{\citenamefont {Kumar}\ \emph {et~al.}(2025)\citenamefont {Kumar},
  \citenamefont {Hegade}, \citenamefont {Visuri}, \citenamefont {Bhargava},
  \citenamefont {Hernandez}, \citenamefont {Solano}, \citenamefont
  {Albarr{\'a}n-Arriagada},\ and\ \citenamefont {Barrios}}]{Kumar2025}%
  \BibitemOpen
  \bibfield  {author} {\bibinfo {author} {\bibfnamefont {S.}~\bibnamefont
  {Kumar}}, \bibinfo {author} {\bibfnamefont {N.~N.}\ \bibnamefont {Hegade}},
  \bibinfo {author} {\bibfnamefont {A.-M.}\ \bibnamefont {Visuri}}, \bibinfo
  {author} {\bibfnamefont {B.~A.}\ \bibnamefont {Bhargava}}, \bibinfo {author}
  {\bibfnamefont {J.~F.~R.}\ \bibnamefont {Hernandez}}, \bibinfo {author}
  {\bibfnamefont {E.}~\bibnamefont {Solano}}, \bibinfo {author} {\bibfnamefont
  {F.}~\bibnamefont {Albarr{\'a}n-Arriagada}},\ and\ \bibinfo {author}
  {\bibfnamefont {G.~A.}\ \bibnamefont {Barrios}},\ }\bibfield  {title}
  {\bibinfo {title} {Digital-analog quantum computing of fermion-boson models
  in superconducting circuits},\ }\href
  {https://doi.org/10.1038/s41534-025-01001-4} {\bibfield  {journal} {\bibinfo
  {journal} {npj Quantum Inf.}\ }\textbf {\bibinfo {volume} {11}},\ \bibinfo
  {pages} {43} (\bibinfo {year} {2025})}\BibitemShut {NoStop}%
\bibitem [{\citenamefont {Leppäkangas}\ \emph {et~al.}(2025)\citenamefont
  {Leppäkangas}, \citenamefont {Stadler}, \citenamefont {Golubev},
  \citenamefont {Reiner}, \citenamefont {Reiner}, \citenamefont {Zanker},
  \citenamefont {Wurz}, \citenamefont {Renger}, \citenamefont {Verjauw},
  \citenamefont {Gusenkova}, \citenamefont {Pogorzalek}, \citenamefont
  {Vigneau}, \citenamefont {Yang}, \citenamefont {Kindel}, \citenamefont {Ku},
  \citenamefont {Deppe},\ and\ \citenamefont {Marthaler}}]{Leppaekangas2025}%
  \BibitemOpen
  \bibfield  {author} {\bibinfo {author} {\bibfnamefont {J.}~\bibnamefont
  {Leppäkangas}}, \bibinfo {author} {\bibfnamefont {P.}~\bibnamefont
  {Stadler}}, \bibinfo {author} {\bibfnamefont {D.}~\bibnamefont {Golubev}},
  \bibinfo {author} {\bibfnamefont {R.}~\bibnamefont {Reiner}}, \bibinfo
  {author} {\bibfnamefont {J.-M.}\ \bibnamefont {Reiner}}, \bibinfo {author}
  {\bibfnamefont {S.}~\bibnamefont {Zanker}}, \bibinfo {author} {\bibfnamefont
  {N.}~\bibnamefont {Wurz}}, \bibinfo {author} {\bibfnamefont {M.}~\bibnamefont
  {Renger}}, \bibinfo {author} {\bibfnamefont {J.}~\bibnamefont {Verjauw}},
  \bibinfo {author} {\bibfnamefont {D.}~\bibnamefont {Gusenkova}}, \bibinfo
  {author} {\bibfnamefont {S.}~\bibnamefont {Pogorzalek}}, \bibinfo {author}
  {\bibfnamefont {F.}~\bibnamefont {Vigneau}}, \bibinfo {author} {\bibfnamefont
  {P.}~\bibnamefont {Yang}}, \bibinfo {author} {\bibfnamefont {W.}~\bibnamefont
  {Kindel}}, \bibinfo {author} {\bibfnamefont {H.-S.}\ \bibnamefont {Ku}},
  \bibinfo {author} {\bibfnamefont {F.}~\bibnamefont {Deppe}},\ and\ \bibinfo
  {author} {\bibfnamefont {M.}~\bibnamefont {Marthaler}},\ }\href@noop {}
  {\bibinfo {title} {Quantum algorithms for simulating systems coupled to
  bosonic modes using a hybrid resonator-qubit quantum computer}} (\bibinfo
  {year} {2025}),\ \Eprint {https://arxiv.org/abs/2503.11507} {arXiv:2503.11507
  [quant-ph]} \BibitemShut {NoStop}%
\bibitem [{\citenamefont {Langford}\ \emph {et~al.}(2017)\citenamefont
  {Langford}, \citenamefont {Sagastizabal}, \citenamefont {Kounalakis},
  \citenamefont {Dickel}, \citenamefont {Bruno}, \citenamefont {Luthi},
  \citenamefont {Thoen}, \citenamefont {Endo},\ and\ \citenamefont
  {DiCarlo}}]{Langford2017}%
  \BibitemOpen
  \bibfield  {author} {\bibinfo {author} {\bibfnamefont {N.~K.}\ \bibnamefont
  {Langford}}, \bibinfo {author} {\bibfnamefont {R.}~\bibnamefont
  {Sagastizabal}}, \bibinfo {author} {\bibfnamefont {M.}~\bibnamefont
  {Kounalakis}}, \bibinfo {author} {\bibfnamefont {C.}~\bibnamefont {Dickel}},
  \bibinfo {author} {\bibfnamefont {A.}~\bibnamefont {Bruno}}, \bibinfo
  {author} {\bibfnamefont {F.}~\bibnamefont {Luthi}}, \bibinfo {author}
  {\bibfnamefont {D.~J.}\ \bibnamefont {Thoen}}, \bibinfo {author}
  {\bibfnamefont {A.}~\bibnamefont {Endo}},\ and\ \bibinfo {author}
  {\bibfnamefont {L.}~\bibnamefont {DiCarlo}},\ }\bibfield  {title} {\bibinfo
  {title} {{Experimentally simulating the dynamics of quantum light and matter
  at deep-strong coupling}},\ }\href
  {https://doi.org/10.1038/s41467-017-01061-x} {\bibfield  {journal} {\bibinfo
  {journal} {Nat. Commun.}\ }\textbf {\bibinfo {volume} {8}},\ \bibinfo {pages}
  {1715} (\bibinfo {year} {2017})}\BibitemShut {NoStop}%
\bibitem [{\citenamefont {Than}\ \emph {et~al.}(2025)\citenamefont {Than},
  \citenamefont {Kadam}, \citenamefont {Vikramaditya}, \citenamefont {Nguyen},
  \citenamefont {Liu}, \citenamefont {Davoudi}, \citenamefont {Green},\ and\
  \citenamefont {Linke}}]{Than:2025}%
  \BibitemOpen
  \bibfield  {author} {\bibinfo {author} {\bibfnamefont {A.~T.}\ \bibnamefont
  {Than}}, \bibinfo {author} {\bibfnamefont {S.~V.}\ \bibnamefont {Kadam}},
  \bibinfo {author} {\bibfnamefont {V.}~\bibnamefont {Vikramaditya}}, \bibinfo
  {author} {\bibfnamefont {N.~H.}\ \bibnamefont {Nguyen}}, \bibinfo {author}
  {\bibfnamefont {X.}~\bibnamefont {Liu}}, \bibinfo {author} {\bibfnamefont
  {Z.}~\bibnamefont {Davoudi}}, \bibinfo {author} {\bibfnamefont {A.~M.}\
  \bibnamefont {Green}},\ and\ \bibinfo {author} {\bibfnamefont {N.~M.}\
  \bibnamefont {Linke}},\ }\href@noop {} {\bibinfo {title} {Observation of
  quantum-field-theory dynamics on a spin-phonon quantum computer}} (\bibinfo
  {year} {2025}),\ \Eprint {https://arxiv.org/abs/2509.11477} {arXiv:2509.11477
  [quant-ph]} \BibitemShut {NoStop}%
\bibitem [{\citenamefont {Liu}\ \emph {et~al.}(2026)\citenamefont {Liu},
  \citenamefont {Singh}, \citenamefont {Smith}, \citenamefont {Crane},
  \citenamefont {Martyn}, \citenamefont {Eickbusch}, \citenamefont {Schuckert},
  \citenamefont {Li}, \citenamefont {Sinanan-Singh}, \citenamefont {Soley},
  \citenamefont {Tsunoda}, \citenamefont {Chuang}, \citenamefont {Wiebe},\ and\
  \citenamefont {Girvin}}]{Crane2026}%
  \BibitemOpen
  \bibfield  {author} {\bibinfo {author} {\bibfnamefont {Y.}~\bibnamefont
  {Liu}}, \bibinfo {author} {\bibfnamefont {S.}~\bibnamefont {Singh}}, \bibinfo
  {author} {\bibfnamefont {K.~C.}\ \bibnamefont {Smith}}, \bibinfo {author}
  {\bibfnamefont {E.}~\bibnamefont {Crane}}, \bibinfo {author} {\bibfnamefont
  {J.~M.}\ \bibnamefont {Martyn}}, \bibinfo {author} {\bibfnamefont
  {A.}~\bibnamefont {Eickbusch}}, \bibinfo {author} {\bibfnamefont
  {A.}~\bibnamefont {Schuckert}}, \bibinfo {author} {\bibfnamefont {R.~D.}\
  \bibnamefont {Li}}, \bibinfo {author} {\bibfnamefont {J.}~\bibnamefont
  {Sinanan-Singh}}, \bibinfo {author} {\bibfnamefont {M.~B.}\ \bibnamefont
  {Soley}}, \bibinfo {author} {\bibfnamefont {T.}~\bibnamefont {Tsunoda}},
  \bibinfo {author} {\bibfnamefont {I.~L.}\ \bibnamefont {Chuang}}, \bibinfo
  {author} {\bibfnamefont {N.}~\bibnamefont {Wiebe}},\ and\ \bibinfo {author}
  {\bibfnamefont {S.~M.}\ \bibnamefont {Girvin}},\ }\bibfield  {title}
  {\bibinfo {title} {Hybrid oscillator-qubit quantum processors: Instruction
  set architectures, abstract machine models, and applications},\ }\href
  {https://doi.org/10.1103/4rf7-9tfx} {\bibfield  {journal} {\bibinfo
  {journal} {PRX Quantum}\ }\textbf {\bibinfo {volume} {7}},\ \bibinfo {pages}
  {010201} (\bibinfo {year} {2026})}\BibitemShut {NoStop}%
\bibitem [{\citenamefont {Crane}\ \emph {et~al.}(2024)\citenamefont {Crane},
  \citenamefont {Smith}, \citenamefont {Tomesh}, \citenamefont {Eickbusch},
  \citenamefont {Martyn}, \citenamefont {Kühn}, \citenamefont {Funcke},
  \citenamefont {DeMarco}, \citenamefont {Chuang}, \citenamefont {Wiebe},
  \citenamefont {Schuckert},\ and\ \citenamefont {Girvin}}]{Crane:2024}%
  \BibitemOpen
  \bibfield  {author} {\bibinfo {author} {\bibfnamefont {E.}~\bibnamefont
  {Crane}}, \bibinfo {author} {\bibfnamefont {K.~C.}\ \bibnamefont {Smith}},
  \bibinfo {author} {\bibfnamefont {T.}~\bibnamefont {Tomesh}}, \bibinfo
  {author} {\bibfnamefont {A.}~\bibnamefont {Eickbusch}}, \bibinfo {author}
  {\bibfnamefont {J.~M.}\ \bibnamefont {Martyn}}, \bibinfo {author}
  {\bibfnamefont {S.}~\bibnamefont {Kühn}}, \bibinfo {author} {\bibfnamefont
  {L.}~\bibnamefont {Funcke}}, \bibinfo {author} {\bibfnamefont {M.~A.}\
  \bibnamefont {DeMarco}}, \bibinfo {author} {\bibfnamefont {I.~L.}\
  \bibnamefont {Chuang}}, \bibinfo {author} {\bibfnamefont {N.}~\bibnamefont
  {Wiebe}}, \bibinfo {author} {\bibfnamefont {A.}~\bibnamefont {Schuckert}},\
  and\ \bibinfo {author} {\bibfnamefont {S.~M.}\ \bibnamefont {Girvin}},\
  }\href@noop {} {\bibinfo {title} {Hybrid oscillator-qubit quantum processors:
  Simulating fermions, bosons, and gauge fields}} (\bibinfo {year} {2024}),\
  \Eprint {https://arxiv.org/abs/2409.03747} {arXiv:2409.03747 [quant-ph]}
  \BibitemShut {NoStop}%
\bibitem [{\citenamefont {Shapiro}\ \emph {et~al.}(2025)\citenamefont
  {Shapiro}, \citenamefont {Weber}, \citenamefont {Bode}, \citenamefont
  {Wilhelm},\ and\ \citenamefont {Bagrets}}]{Shapiro2025}%
  \BibitemOpen
  \bibfield  {author} {\bibinfo {author} {\bibfnamefont {D.~S.}\ \bibnamefont
  {Shapiro}}, \bibinfo {author} {\bibfnamefont {Y.}~\bibnamefont {Weber}},
  \bibinfo {author} {\bibfnamefont {T.}~\bibnamefont {Bode}}, \bibinfo {author}
  {\bibfnamefont {F.~K.}\ \bibnamefont {Wilhelm}},\ and\ \bibinfo {author}
  {\bibfnamefont {D.}~\bibnamefont {Bagrets}},\ }\bibfield  {title} {\bibinfo
  {title} {{Digital-analog simulations of Schr\"odinger cat states in the
  Dicke-Ising model}},\ }\href {https://doi.org/10.1103/wbp6-y3vd} {\bibfield
  {journal} {\bibinfo  {journal} {Phys. Rev. A}\ }\textbf {\bibinfo {volume}
  {112}},\ \bibinfo {pages} {042412} (\bibinfo {year} {2025})}\BibitemShut
  {NoStop}%
\bibitem [{\citenamefont {Stavenger}\ \emph {et~al.}(2022)\citenamefont
  {Stavenger}, \citenamefont {Crane}, \citenamefont {Smith}, \citenamefont
  {Kang}, \citenamefont {Girvin},\ and\ \citenamefont
  {Wiebe}}]{stavenger2022bosonic}%
  \BibitemOpen
  \bibfield  {author} {\bibinfo {author} {\bibfnamefont {T.~J.}\ \bibnamefont
  {Stavenger}}, \bibinfo {author} {\bibfnamefont {E.}~\bibnamefont {Crane}},
  \bibinfo {author} {\bibfnamefont {K.}~\bibnamefont {Smith}}, \bibinfo
  {author} {\bibfnamefont {C.~T.}\ \bibnamefont {Kang}}, \bibinfo {author}
  {\bibfnamefont {S.~M.}\ \bibnamefont {Girvin}},\ and\ \bibinfo {author}
  {\bibfnamefont {N.}~\bibnamefont {Wiebe}},\ }\href@noop {} {\bibinfo {title}
  {Bosonic qiskit}} (\bibinfo {year} {2022}),\ \Eprint
  {https://arxiv.org/abs/2209.11153} {arXiv:2209.11153 [quant-ph]} \BibitemShut
  {NoStop}%
\bibitem [{\citenamefont {Vool}\ and\ \citenamefont
  {Devoret}(2017)}]{vool2017}%
  \BibitemOpen
  \bibfield  {author} {\bibinfo {author} {\bibfnamefont {U.}~\bibnamefont
  {Vool}}\ and\ \bibinfo {author} {\bibfnamefont {M.}~\bibnamefont {Devoret}},\
  }\bibfield  {title} {\bibinfo {title} {{Introduction to quantum
  electromagnetic circuits}},\ }\href
  {https://doi.org/https://doi.org/10.1002/cta.2359} {\bibfield  {journal}
  {\bibinfo  {journal} {Int. J. Circuit Theory Appl.}\ }\textbf {\bibinfo
  {volume} {45}},\ \bibinfo {pages} {897} (\bibinfo {year} {2017})}\BibitemShut
  {NoStop}%
\bibitem [{\citenamefont {Rasmussen}\ \emph {et~al.}(2021)\citenamefont
  {Rasmussen}, \citenamefont {Christensen}, \citenamefont {Pedersen},
  \citenamefont {Kristensen}, \citenamefont {B\ae{}kkegaard}, \citenamefont
  {Loft},\ and\ \citenamefont {Zinner}}]{rasmussen2021}%
  \BibitemOpen
  \bibfield  {author} {\bibinfo {author} {\bibfnamefont {S.}~\bibnamefont
  {Rasmussen}}, \bibinfo {author} {\bibfnamefont {K.}~\bibnamefont
  {Christensen}}, \bibinfo {author} {\bibfnamefont {S.}~\bibnamefont
  {Pedersen}}, \bibinfo {author} {\bibfnamefont {L.}~\bibnamefont
  {Kristensen}}, \bibinfo {author} {\bibfnamefont {T.}~\bibnamefont
  {B\ae{}kkegaard}}, \bibinfo {author} {\bibfnamefont {N.}~\bibnamefont
  {Loft}},\ and\ \bibinfo {author} {\bibfnamefont {N.}~\bibnamefont {Zinner}},\
  }\bibfield  {title} {\bibinfo {title} {{Superconducting Circuit
  Companion---an Introduction with Worked Examples}},\ }\href
  {https://doi.org/10.1103/PRXQuantum.2.040204} {\bibfield  {journal} {\bibinfo
   {journal} {PRX Quantum}\ }\textbf {\bibinfo {volume} {2}},\ \bibinfo {pages}
  {040204} (\bibinfo {year} {2021})}\BibitemShut {NoStop}%
\bibitem [{\citenamefont {Ciani}\ \emph {et~al.}(2024)\citenamefont {Ciani},
  \citenamefont {DiVincenzo},\ and\ \citenamefont {Terhal}}]{ciani2024}%
  \BibitemOpen
  \bibfield  {author} {\bibinfo {author} {\bibfnamefont {A.}~\bibnamefont
  {Ciani}}, \bibinfo {author} {\bibfnamefont {D.~P.}\ \bibnamefont
  {DiVincenzo}},\ and\ \bibinfo {author} {\bibfnamefont {B.~M.}\ \bibnamefont
  {Terhal}},\ }\href {https://doi.org/https://doi.org/10.59490/tb.85} {\emph
  {\bibinfo {title} {{Lecture {N}otes on {Q}uantum {E}lectrical {C}ircuits}}}}\
  (\bibinfo  {publisher} {TU Delft OPEN Publishing},\ \bibinfo {address}
  {Delft},\ \bibinfo {year} {2024})\BibitemShut {NoStop}%
\bibitem [{\citenamefont {Blais}\ \emph {et~al.}(2004)\citenamefont {Blais},
  \citenamefont {Huang}, \citenamefont {Wallraff}, \citenamefont {Girvin},\
  and\ \citenamefont {Schoelkopf}}]{blais2004}%
  \BibitemOpen
  \bibfield  {author} {\bibinfo {author} {\bibfnamefont {A.}~\bibnamefont
  {Blais}}, \bibinfo {author} {\bibfnamefont {R.-S.}\ \bibnamefont {Huang}},
  \bibinfo {author} {\bibfnamefont {A.}~\bibnamefont {Wallraff}}, \bibinfo
  {author} {\bibfnamefont {S.~M.}\ \bibnamefont {Girvin}},\ and\ \bibinfo
  {author} {\bibfnamefont {R.~J.}\ \bibnamefont {Schoelkopf}},\ }\bibfield
  {title} {\bibinfo {title} {{Cavity quantum electrodynamics for
  superconducting electrical circuits: An architecture for quantum
  computation}},\ }\href {https://doi.org/10.1103/PhysRevA.69.062320}
  {\bibfield  {journal} {\bibinfo  {journal} {Phys. Rev. A}\ }\textbf {\bibinfo
  {volume} {69}},\ \bibinfo {pages} {062320} (\bibinfo {year}
  {2004})}\BibitemShut {NoStop}%
\bibitem [{\citenamefont {Campagne-Ibarcq}\ \emph {et~al.}(2020)\citenamefont
  {Campagne-Ibarcq}, \citenamefont {Eickbusch}, \citenamefont {Touzard},
  \citenamefont {Zalys-Geller}, \citenamefont {Frattini}, \citenamefont
  {Sivak}, \citenamefont {Reinhold}, \citenamefont {Puri}, \citenamefont
  {Shankar}, \citenamefont {Schoelkopf}, \citenamefont {Frunzio}, \citenamefont
  {Mirrahimi},\ and\ \citenamefont {Devoret}}]{Campagne-Ibarcq2020}%
  \BibitemOpen
  \bibfield  {author} {\bibinfo {author} {\bibfnamefont {P.}~\bibnamefont
  {Campagne-Ibarcq}}, \bibinfo {author} {\bibfnamefont {A.}~\bibnamefont
  {Eickbusch}}, \bibinfo {author} {\bibfnamefont {S.}~\bibnamefont {Touzard}},
  \bibinfo {author} {\bibfnamefont {E.}~\bibnamefont {Zalys-Geller}}, \bibinfo
  {author} {\bibfnamefont {N.~E.}\ \bibnamefont {Frattini}}, \bibinfo {author}
  {\bibfnamefont {V.~V.}\ \bibnamefont {Sivak}}, \bibinfo {author}
  {\bibfnamefont {P.}~\bibnamefont {Reinhold}}, \bibinfo {author}
  {\bibfnamefont {S.}~\bibnamefont {Puri}}, \bibinfo {author} {\bibfnamefont
  {S.}~\bibnamefont {Shankar}}, \bibinfo {author} {\bibfnamefont {R.~J.}\
  \bibnamefont {Schoelkopf}}, \bibinfo {author} {\bibfnamefont
  {L.}~\bibnamefont {Frunzio}}, \bibinfo {author} {\bibfnamefont
  {M.}~\bibnamefont {Mirrahimi}},\ and\ \bibinfo {author} {\bibfnamefont
  {M.~H.}\ \bibnamefont {Devoret}},\ }\bibfield  {title} {\bibinfo {title}
  {{Quantum error correction of a qubit encoded in grid states of an
  oscillator}},\ }\href {https://doi.org/10.1038/s41586-020-2603-3} {\bibfield
  {journal} {\bibinfo  {journal} {Nature}\ }\textbf {\bibinfo {volume} {584}},\
  \bibinfo {pages} {368} (\bibinfo {year} {2020})}\BibitemShut {NoStop}%
\bibitem [{\citenamefont {Eickbusch}\ \emph {et~al.}(2022)\citenamefont
  {Eickbusch}, \citenamefont {Sivak}, \citenamefont {Ding}, \citenamefont
  {Elder}, \citenamefont {Jha}, \citenamefont {Venkatraman}, \citenamefont
  {Royer}, \citenamefont {Girvin}, \citenamefont {Schoelkopf},\ and\
  \citenamefont {Devoret}}]{Eickbusch2022}%
  \BibitemOpen
  \bibfield  {author} {\bibinfo {author} {\bibfnamefont {A.}~\bibnamefont
  {Eickbusch}}, \bibinfo {author} {\bibfnamefont {V.}~\bibnamefont {Sivak}},
  \bibinfo {author} {\bibfnamefont {A.~Z.}\ \bibnamefont {Ding}}, \bibinfo
  {author} {\bibfnamefont {S.~S.}\ \bibnamefont {Elder}}, \bibinfo {author}
  {\bibfnamefont {S.~R.}\ \bibnamefont {Jha}}, \bibinfo {author} {\bibfnamefont
  {J.}~\bibnamefont {Venkatraman}}, \bibinfo {author} {\bibfnamefont
  {B.}~\bibnamefont {Royer}}, \bibinfo {author} {\bibfnamefont {S.~M.}\
  \bibnamefont {Girvin}}, \bibinfo {author} {\bibfnamefont {R.~J.}\
  \bibnamefont {Schoelkopf}},\ and\ \bibinfo {author} {\bibfnamefont {M.~H.}\
  \bibnamefont {Devoret}},\ }\bibfield  {title} {\bibinfo {title} {{Fast
  universal control of an oscillator with weak dispersive coupling to a
  qubit}},\ }\href {https://doi.org/10.1038/s41567-022-01776-9} {\bibfield
  {journal} {\bibinfo  {journal} {Nat. Phys.}\ }\textbf {\bibinfo {volume}
  {18}},\ \bibinfo {pages} {1464} (\bibinfo {year} {2022})}\BibitemShut
  {NoStop}%
\bibitem [{\citenamefont {Sivak}\ \emph {et~al.}(2023)\citenamefont {Sivak},
  \citenamefont {Eickbusch}, \citenamefont {Royer}, \citenamefont {Singh},
  \citenamefont {Tsioutsios}, \citenamefont {Ganjam}, \citenamefont {Miano},
  \citenamefont {Brock}, \citenamefont {Ding}, \citenamefont {Frunzio},
  \citenamefont {Girvin}, \citenamefont {Schoelkopf},\ and\ \citenamefont
  {Devoret}}]{Sivak2023}%
  \BibitemOpen
  \bibfield  {author} {\bibinfo {author} {\bibfnamefont {V.~V.}\ \bibnamefont
  {Sivak}}, \bibinfo {author} {\bibfnamefont {A.}~\bibnamefont {Eickbusch}},
  \bibinfo {author} {\bibfnamefont {B.}~\bibnamefont {Royer}}, \bibinfo
  {author} {\bibfnamefont {S.}~\bibnamefont {Singh}}, \bibinfo {author}
  {\bibfnamefont {I.}~\bibnamefont {Tsioutsios}}, \bibinfo {author}
  {\bibfnamefont {S.}~\bibnamefont {Ganjam}}, \bibinfo {author} {\bibfnamefont
  {A.}~\bibnamefont {Miano}}, \bibinfo {author} {\bibfnamefont {B.~L.}\
  \bibnamefont {Brock}}, \bibinfo {author} {\bibfnamefont {A.~Z.}\ \bibnamefont
  {Ding}}, \bibinfo {author} {\bibfnamefont {L.}~\bibnamefont {Frunzio}},
  \bibinfo {author} {\bibfnamefont {S.~M.}\ \bibnamefont {Girvin}}, \bibinfo
  {author} {\bibfnamefont {R.~J.}\ \bibnamefont {Schoelkopf}},\ and\ \bibinfo
  {author} {\bibfnamefont {M.~H.}\ \bibnamefont {Devoret}},\ }\bibfield
  {title} {\bibinfo {title} {{Real-time quantum error correction beyond
  break-even}},\ }\href {https://doi.org/10.1038/s41586-023-05782-6} {\bibfield
   {journal} {\bibinfo  {journal} {Nature}\ }\textbf {\bibinfo {volume}
  {616}},\ \bibinfo {pages} {50} (\bibinfo {year} {2023})}\BibitemShut
  {NoStop}%
\bibitem [{\citenamefont {Valadares}\ \emph {et~al.}(2026)\citenamefont
  {Valadares}, \citenamefont {Dorogov}, \citenamefont {Krisnanda},
  \citenamefont {Loke}, \citenamefont {Huang}, \citenamefont {Song},\ and\
  \citenamefont {Gao}}]{valadares2026}%
  \BibitemOpen
  \bibfield  {author} {\bibinfo {author} {\bibfnamefont {F.}~\bibnamefont
  {Valadares}}, \bibinfo {author} {\bibfnamefont {A.}~\bibnamefont {Dorogov}},
  \bibinfo {author} {\bibfnamefont {T.}~\bibnamefont {Krisnanda}}, \bibinfo
  {author} {\bibfnamefont {M.~C.}\ \bibnamefont {Loke}}, \bibinfo {author}
  {\bibfnamefont {N.-N.}\ \bibnamefont {Huang}}, \bibinfo {author}
  {\bibfnamefont {P.}~\bibnamefont {Song}},\ and\ \bibinfo {author}
  {\bibfnamefont {Y.~Y.}\ \bibnamefont {Gao}},\ }\href@noop {} {\bibinfo
  {title} {Flux-activated resonant control of a bosonic quantum memory}}
  (\bibinfo {year} {2026}),\ \Eprint {https://arxiv.org/abs/2602.18122}
  {arXiv:2602.18122 [quant-ph]} \BibitemShut {NoStop}%
\bibitem [{\citenamefont {S\'en\'echal}\ \emph {et~al.}(2002)\citenamefont
  {S\'en\'echal}, \citenamefont {Perez},\ and\ \citenamefont
  {Plouffe}}]{Senechal2002}%
  \BibitemOpen
  \bibfield  {author} {\bibinfo {author} {\bibfnamefont {D.}~\bibnamefont
  {S\'en\'echal}}, \bibinfo {author} {\bibfnamefont {D.}~\bibnamefont
  {Perez}},\ and\ \bibinfo {author} {\bibfnamefont {D.}~\bibnamefont
  {Plouffe}},\ }\bibfield  {title} {\bibinfo {title} {{Cluster perturbation
  theory for Hubbard models}},\ }\href
  {https://doi.org/10.1103/PhysRevB.66.075129} {\bibfield  {journal} {\bibinfo
  {journal} {Phys. Rev. B}\ }\textbf {\bibinfo {volume} {66}},\ \bibinfo
  {pages} {075129} (\bibinfo {year} {2002})}\BibitemShut {NoStop}%
\bibitem [{\citenamefont {Potthoff}(2003)}]{Potthoff2003}%
  \BibitemOpen
  \bibfield  {author} {\bibinfo {author} {\bibfnamefont {M.}~\bibnamefont
  {Potthoff}},\ }\bibfield  {title} {\bibinfo {title} {Self-energy-functional
  approach to systems of correlated electrons},\ }\href
  {https://doi.org/10.1140/epjb/e2003-00121-8} {\bibfield  {journal} {\bibinfo
  {journal} {The European Physical Journal B - Condensed Matter and Complex
  Systems}\ }\textbf {\bibinfo {volume} {32}},\ \bibinfo {pages} {429}
  (\bibinfo {year} {2003})}\BibitemShut {NoStop}%
\bibitem [{\citenamefont {Potthoff}\ \emph {et~al.}(2003)\citenamefont
  {Potthoff}, \citenamefont {Aichhorn},\ and\ \citenamefont
  {Dahnken}}]{PotthoffAichhornDahnken2003}%
  \BibitemOpen
  \bibfield  {author} {\bibinfo {author} {\bibfnamefont {M.}~\bibnamefont
  {Potthoff}}, \bibinfo {author} {\bibfnamefont {M.}~\bibnamefont {Aichhorn}},\
  and\ \bibinfo {author} {\bibfnamefont {C.}~\bibnamefont {Dahnken}},\
  }\bibfield  {title} {\bibinfo {title} {{Variational Cluster Approach to
  Correlated Electron Systems in Low Dimensions}},\ }\href
  {https://doi.org/10.1103/PhysRevLett.91.206402} {\bibfield  {journal}
  {\bibinfo  {journal} {Phys. Rev. Lett.}\ }\textbf {\bibinfo {volume} {91}},\
  \bibinfo {pages} {206402} (\bibinfo {year} {2003})}\BibitemShut {NoStop}%
\bibitem [{\citenamefont {Payeur}\ and\ \citenamefont
  {S\'en\'echal}(2011)}]{Payeur2011}%
  \BibitemOpen
  \bibfield  {author} {\bibinfo {author} {\bibfnamefont {A.}~\bibnamefont
  {Payeur}}\ and\ \bibinfo {author} {\bibfnamefont {D.}~\bibnamefont
  {S\'en\'echal}},\ }\bibfield  {title} {\bibinfo {title} {{Variational cluster
  approximation study of the one-dimensional Holstein-Hubbard model at half
  filling}},\ }\href {https://doi.org/10.1103/PhysRevB.83.033104} {\bibfield
  {journal} {\bibinfo  {journal} {Phys. Rev. B}\ }\textbf {\bibinfo {volume}
  {83}},\ \bibinfo {pages} {033104} (\bibinfo {year} {2011})}\BibitemShut
  {NoStop}%
\bibitem [{\citenamefont {Wecker}\ \emph
  {et~al.}(2015{\natexlab{a}})\citenamefont {Wecker}, \citenamefont
  {Hastings},\ and\ \citenamefont {Troyer}}]{Wecker2015}%
  \BibitemOpen
  \bibfield  {author} {\bibinfo {author} {\bibfnamefont {D.}~\bibnamefont
  {Wecker}}, \bibinfo {author} {\bibfnamefont {M.~B.}\ \bibnamefont
  {Hastings}},\ and\ \bibinfo {author} {\bibfnamefont {M.}~\bibnamefont
  {Troyer}},\ }\bibfield  {title} {\bibinfo {title} {{Progress towards
  practical quantum variational algorithms}},\ }\href
  {https://doi.org/10.1103/PhysRevA.92.042303} {\bibfield  {journal} {\bibinfo
  {journal} {Phys. Rev. A}\ }\textbf {\bibinfo {volume} {92}},\ \bibinfo
  {pages} {042303} (\bibinfo {year} {2015}{\natexlab{a}})}\BibitemShut
  {NoStop}%
\bibitem [{\citenamefont {Bauer}\ \emph {et~al.}(2016)\citenamefont {Bauer},
  \citenamefont {Wecker}, \citenamefont {Millis}, \citenamefont {Hastings},\
  and\ \citenamefont {Troyer}}]{Bauer2016}%
  \BibitemOpen
  \bibfield  {author} {\bibinfo {author} {\bibfnamefont {B.}~\bibnamefont
  {Bauer}}, \bibinfo {author} {\bibfnamefont {D.}~\bibnamefont {Wecker}},
  \bibinfo {author} {\bibfnamefont {A.~J.}\ \bibnamefont {Millis}}, \bibinfo
  {author} {\bibfnamefont {M.~B.}\ \bibnamefont {Hastings}},\ and\ \bibinfo
  {author} {\bibfnamefont {M.}~\bibnamefont {Troyer}},\ }\bibfield  {title}
  {\bibinfo {title} {{Hybrid Quantum-Classical Approach to Correlated
  Materials}},\ }\href {https://doi.org/10.1103/PhysRevX.6.031045} {\bibfield
  {journal} {\bibinfo  {journal} {Phys. Rev. X}\ }\textbf {\bibinfo {volume}
  {6}},\ \bibinfo {pages} {031045} (\bibinfo {year} {2016})}\BibitemShut
  {NoStop}%
\bibitem [{\citenamefont {Cade}\ \emph {et~al.}(2020)\citenamefont {Cade},
  \citenamefont {Mineh}, \citenamefont {Montanaro},\ and\ \citenamefont
  {Stanisic}}]{Cade2020}%
  \BibitemOpen
  \bibfield  {author} {\bibinfo {author} {\bibfnamefont {C.}~\bibnamefont
  {Cade}}, \bibinfo {author} {\bibfnamefont {L.}~\bibnamefont {Mineh}},
  \bibinfo {author} {\bibfnamefont {A.}~\bibnamefont {Montanaro}},\ and\
  \bibinfo {author} {\bibfnamefont {S.}~\bibnamefont {Stanisic}},\ }\bibfield
  {title} {\bibinfo {title} {{Strategies for solving the Fermi-Hubbard model on
  near-term quantum computers}},\ }\href
  {https://doi.org/10.1103/PhysRevB.102.235122} {\bibfield  {journal} {\bibinfo
   {journal} {Phys. Rev. B}\ }\textbf {\bibinfo {volume} {102}},\ \bibinfo
  {pages} {235122} (\bibinfo {year} {2020})}\BibitemShut {NoStop}%
\bibitem [{\citenamefont {Endo}\ \emph {et~al.}(2020)\citenamefont {Endo},
  \citenamefont {Kurata},\ and\ \citenamefont {Nakagawa}}]{Endo2020}%
  \BibitemOpen
  \bibfield  {author} {\bibinfo {author} {\bibfnamefont {S.}~\bibnamefont
  {Endo}}, \bibinfo {author} {\bibfnamefont {I.}~\bibnamefont {Kurata}},\ and\
  \bibinfo {author} {\bibfnamefont {Y.~O.}\ \bibnamefont {Nakagawa}},\
  }\bibfield  {title} {\bibinfo {title} {{Calculation of the Green's function
  on near-term quantum computers}},\ }\href
  {https://doi.org/10.1103/PhysRevResearch.2.033281} {\bibfield  {journal}
  {\bibinfo  {journal} {Phys. Rev. Res.}\ }\textbf {\bibinfo {volume} {2}},\
  \bibinfo {pages} {033281} (\bibinfo {year} {2020})}\BibitemShut {NoStop}%
\bibitem [{\citenamefont {Bishop}\ \emph {et~al.}(2025)\citenamefont {Bishop},
  \citenamefont {Bagrets},\ and\ \citenamefont {Wilhelm}}]{bishop2023quantum}%
  \BibitemOpen
  \bibfield  {author} {\bibinfo {author} {\bibfnamefont {G.}~\bibnamefont
  {Bishop}}, \bibinfo {author} {\bibfnamefont {D.}~\bibnamefont {Bagrets}},\
  and\ \bibinfo {author} {\bibfnamefont {F.~K.}\ \bibnamefont {Wilhelm}},\
  }\bibfield  {title} {\bibinfo {title} {{Quantum algorithm for
  Green's-function measurements in the Fermi-Hubbard model}},\ }\href
  {https://doi.org/10.1103/PhysRevA.111.062610} {\bibfield  {journal} {\bibinfo
   {journal} {Phys. Rev. A}\ }\textbf {\bibinfo {volume} {111}},\ \bibinfo
  {pages} {062610} (\bibinfo {year} {2025})}\BibitemShut {NoStop}%
\bibitem [{\citenamefont {Wecker}\ \emph
  {et~al.}(2015{\natexlab{b}})\citenamefont {Wecker}, \citenamefont {Hastings},
  \citenamefont {Wiebe}, \citenamefont {Clark}, \citenamefont {Nayak},\ and\
  \citenamefont {Troyer}}]{Troyer_fermions_2015}%
  \BibitemOpen
  \bibfield  {author} {\bibinfo {author} {\bibfnamefont {D.}~\bibnamefont
  {Wecker}}, \bibinfo {author} {\bibfnamefont {M.~B.}\ \bibnamefont
  {Hastings}}, \bibinfo {author} {\bibfnamefont {N.}~\bibnamefont {Wiebe}},
  \bibinfo {author} {\bibfnamefont {B.~K.}\ \bibnamefont {Clark}}, \bibinfo
  {author} {\bibfnamefont {C.}~\bibnamefont {Nayak}},\ and\ \bibinfo {author}
  {\bibfnamefont {M.}~\bibnamefont {Troyer}},\ }\bibfield  {title} {\bibinfo
  {title} {{Solving strongly correlated electron models on a quantum
  computer}},\ }\href {https://doi.org/10.1103/PhysRevA.92.062318} {\bibfield
  {journal} {\bibinfo  {journal} {Phys. Rev. A}\ }\textbf {\bibinfo {volume}
  {92}},\ \bibinfo {pages} {062318} (\bibinfo {year}
  {2015}{\natexlab{b}})}\BibitemShut {NoStop}%
\bibitem [{\citenamefont {Popov}\ and\ \citenamefont
  {Fedotov}(1988)}]{popov1988functional}%
  \BibitemOpen
  \bibfield  {author} {\bibinfo {author} {\bibfnamefont {V.~N.}\ \bibnamefont
  {Popov}}\ and\ \bibinfo {author} {\bibfnamefont {S.}~\bibnamefont
  {Fedotov}},\ }\bibfield  {title} {\bibinfo {title} {The
  functional-integration method and diagram technique for spin systems},\
  }\href {http://www.jetp.ras.ru/cgi-bin/dn/e_067_03_0535} {\bibfield
  {journal} {\bibinfo  {journal} {Zh. Eksp. Teor. Fiz}\ }\textbf {\bibinfo
  {volume} {94}},\ \bibinfo {pages} {183} (\bibinfo {year} {1988})}\BibitemShut
  {NoStop}%
\bibitem [{\citenamefont {Emary}\ and\ \citenamefont
  {Brandes}(2003)}]{emary2003chaos}%
  \BibitemOpen
  \bibfield  {author} {\bibinfo {author} {\bibfnamefont {C.}~\bibnamefont
  {Emary}}\ and\ \bibinfo {author} {\bibfnamefont {T.}~\bibnamefont
  {Brandes}},\ }\bibfield  {title} {\bibinfo {title} {{Chaos and the quantum
  phase transition in the Dicke model}},\ }\href
  {https://doi.org/10.1103/PhysRevE.67.066203} {\bibfield  {journal} {\bibinfo
  {journal} {Phys. Rev. E}\ }\textbf {\bibinfo {volume} {67}},\ \bibinfo
  {pages} {066203} (\bibinfo {year} {2003})}\BibitemShut {NoStop}%
\bibitem [{\citenamefont {Eastham}\ and\ \citenamefont
  {Littlewood}(2001)}]{eastham2001bose}%
  \BibitemOpen
  \bibfield  {author} {\bibinfo {author} {\bibfnamefont {P.}~\bibnamefont
  {Eastham}}\ and\ \bibinfo {author} {\bibfnamefont {P.}~\bibnamefont
  {Littlewood}},\ }\bibfield  {title} {\bibinfo {title} {{Bose condensation of
  cavity polaritons beyond the linear regime: The thermal equilibrium of a
  model microcavity}},\ }\href {https://doi.org/10.1103/PhysRevB.64.235101}
  {\bibfield  {journal} {\bibinfo  {journal} {Phys. Rev. B}\ }\textbf {\bibinfo
  {volume} {64}},\ \bibinfo {pages} {235101} (\bibinfo {year}
  {2001})}\BibitemShut {NoStop}%
\bibitem [{\citenamefont {Dalla~Torre}\ \emph {et~al.}(2016)\citenamefont
  {Dalla~Torre}, \citenamefont {Shchadilova}, \citenamefont {Wilner},
  \citenamefont {Lukin},\ and\ \citenamefont {Demler}}]{PhysRevA.94.061802}%
  \BibitemOpen
  \bibfield  {author} {\bibinfo {author} {\bibfnamefont {E.~G.}\ \bibnamefont
  {Dalla~Torre}}, \bibinfo {author} {\bibfnamefont {Y.}~\bibnamefont
  {Shchadilova}}, \bibinfo {author} {\bibfnamefont {E.~Y.}\ \bibnamefont
  {Wilner}}, \bibinfo {author} {\bibfnamefont {M.~D.}\ \bibnamefont {Lukin}},\
  and\ \bibinfo {author} {\bibfnamefont {E.}~\bibnamefont {Demler}},\
  }\bibfield  {title} {\bibinfo {title} {{Dicke phase transition without total
  spin conservation}},\ }\href {https://doi.org/10.1103/PhysRevA.94.061802}
  {\bibfield  {journal} {\bibinfo  {journal} {Phys. Rev. A}\ }\textbf {\bibinfo
  {volume} {94}},\ \bibinfo {pages} {061802} (\bibinfo {year}
  {2016})}\BibitemShut {NoStop}%
\bibitem [{\citenamefont {Kirton}\ \emph {et~al.}(2019)\citenamefont {Kirton},
  \citenamefont {Roses}, \citenamefont {Keeling},\ and\ \citenamefont {{Dalla
  Torre}}}]{kirton2018introduction}%
  \BibitemOpen
  \bibfield  {author} {\bibinfo {author} {\bibfnamefont {P.}~\bibnamefont
  {Kirton}}, \bibinfo {author} {\bibfnamefont {M.~M.}\ \bibnamefont {Roses}},
  \bibinfo {author} {\bibfnamefont {J.}~\bibnamefont {Keeling}},\ and\ \bibinfo
  {author} {\bibfnamefont {E.~G.}\ \bibnamefont {{Dalla Torre}}},\ }\bibfield
  {title} {\bibinfo {title} {{Introduction to the Dicke Model: From Equilibrium
  to Nonequilibrium, and Vice Versa}},\ }\href
  {https://doi.org/10.1002/qute.201800043} {\bibfield  {journal} {\bibinfo
  {journal} {Adv. Quantum Technol.}\ }\textbf {\bibinfo {volume} {2}},\
  \bibinfo {pages} {1800043} (\bibinfo {year} {2019})}\BibitemShut {NoStop}%
\bibitem [{\citenamefont {Shapiro}\ \emph {et~al.}(2020)\citenamefont
  {Shapiro}, \citenamefont {Pogosov},\ and\ \citenamefont
  {Lozovik}}]{PhysRevA.102.023703}%
  \BibitemOpen
  \bibfield  {author} {\bibinfo {author} {\bibfnamefont {D.~S.}\ \bibnamefont
  {Shapiro}}, \bibinfo {author} {\bibfnamefont {W.~V.}\ \bibnamefont
  {Pogosov}},\ and\ \bibinfo {author} {\bibfnamefont {Y.~E.}\ \bibnamefont
  {Lozovik}},\ }\bibfield  {title} {\bibinfo {title} {{Universal fluctuations
  and squeezing in a generalized Dicke model near the superradiant phase
  transition}},\ }\href {https://doi.org/10.1103/PhysRevA.102.023703}
  {\bibfield  {journal} {\bibinfo  {journal} {Phys. Rev. A}\ }\textbf {\bibinfo
  {volume} {102}},\ \bibinfo {pages} {023703} (\bibinfo {year}
  {2020})}\BibitemShut {NoStop}%
\bibitem [{\citenamefont {Lutterbach}\ and\ \citenamefont
  {Davidovich}(1997)}]{lutterbach1997}%
  \BibitemOpen
  \bibfield  {author} {\bibinfo {author} {\bibfnamefont {L.~G.}\ \bibnamefont
  {Lutterbach}}\ and\ \bibinfo {author} {\bibfnamefont {L.}~\bibnamefont
  {Davidovich}},\ }\bibfield  {title} {\bibinfo {title} {{Method for Direct
  Measurement of the Wigner Function in Cavity QED and Ion Traps}},\ }\href
  {https://doi.org/10.1103/PhysRevLett.78.2547} {\bibfield  {journal} {\bibinfo
   {journal} {Phys. Rev. Lett.}\ }\textbf {\bibinfo {volume} {78}},\ \bibinfo
  {pages} {2547} (\bibinfo {year} {1997})}\BibitemShut {NoStop}%
\bibitem [{\citenamefont {Vlastakis}\ \emph {et~al.}(2013)\citenamefont
  {Vlastakis}, \citenamefont {Kirchmair}, \citenamefont {Leghtas},
  \citenamefont {Nigg}, \citenamefont {Frunzio}, \citenamefont {Girvin},
  \citenamefont {Mirrahimi}, \citenamefont {Devoret},\ and\ \citenamefont
  {Schoelkopf}}]{vlastakis2013}%
  \BibitemOpen
  \bibfield  {author} {\bibinfo {author} {\bibfnamefont {B.}~\bibnamefont
  {Vlastakis}}, \bibinfo {author} {\bibfnamefont {G.}~\bibnamefont
  {Kirchmair}}, \bibinfo {author} {\bibfnamefont {Z.}~\bibnamefont {Leghtas}},
  \bibinfo {author} {\bibfnamefont {S.~E.}\ \bibnamefont {Nigg}}, \bibinfo
  {author} {\bibfnamefont {L.}~\bibnamefont {Frunzio}}, \bibinfo {author}
  {\bibfnamefont {S.~M.}\ \bibnamefont {Girvin}}, \bibinfo {author}
  {\bibfnamefont {M.}~\bibnamefont {Mirrahimi}}, \bibinfo {author}
  {\bibfnamefont {M.~H.}\ \bibnamefont {Devoret}},\ and\ \bibinfo {author}
  {\bibfnamefont {R.~J.}\ \bibnamefont {Schoelkopf}},\ }\bibfield  {title}
  {\bibinfo {title} {{Deterministically Encoding Quantum Information Using
  100-Photon Schr{\"o}dinger Cat States}},\ }\href
  {https://doi.org/10.1126/science.1243289} {\bibfield  {journal} {\bibinfo
  {journal} {Science}\ }\textbf {\bibinfo {volume} {342}},\ \bibinfo {pages}
  {607} (\bibinfo {year} {2013})}\BibitemShut {NoStop}%
\bibitem [{\citenamefont {Fl{\"u}hmann}\ \emph {et~al.}(2019)\citenamefont
  {Fl{\"u}hmann}, \citenamefont {Nguyen}, \citenamefont {Marinelli},
  \citenamefont {Negnevitsky}, \citenamefont {Mehta},\ and\ \citenamefont
  {Home}}]{Flühmann2019}%
  \BibitemOpen
  \bibfield  {author} {\bibinfo {author} {\bibfnamefont {C.}~\bibnamefont
  {Fl{\"u}hmann}}, \bibinfo {author} {\bibfnamefont {T.~L.}\ \bibnamefont
  {Nguyen}}, \bibinfo {author} {\bibfnamefont {M.}~\bibnamefont {Marinelli}},
  \bibinfo {author} {\bibfnamefont {V.}~\bibnamefont {Negnevitsky}}, \bibinfo
  {author} {\bibfnamefont {K.}~\bibnamefont {Mehta}},\ and\ \bibinfo {author}
  {\bibfnamefont {J.~P.}\ \bibnamefont {Home}},\ }\bibfield  {title} {\bibinfo
  {title} {{Encoding a qubit in a trapped-ion mechanical oscillator}},\ }\href
  {https://doi.org/10.1038/s41586-019-0960-6} {\bibfield  {journal} {\bibinfo
  {journal} {Nature}\ }\textbf {\bibinfo {volume} {566}},\ \bibinfo {pages}
  {513} (\bibinfo {year} {2019})}\BibitemShut {NoStop}%
\bibitem [{\citenamefont {Marcus}\ and\ \citenamefont
  {Vandoren}(2019)}]{marcus_2019}%
  \BibitemOpen
  \bibfield  {author} {\bibinfo {author} {\bibfnamefont {E.}~\bibnamefont
  {Marcus}}\ and\ \bibinfo {author} {\bibfnamefont {S.}~\bibnamefont
  {Vandoren}},\ }\bibfield  {title} {\bibinfo {title} {A new class of
  {SYK}-like models with maximal chaos},\ }\href
  {https://doi.org/10.1007/JHEP01(2019)166} {\bibfield  {journal} {\bibinfo
  {journal} {J. High Energy Phys.}\ }\textbf {\bibinfo {volume} {2019}},\
  \bibinfo {pages} {166}}\BibitemShut {NoStop}%
\bibitem [{\citenamefont {Wang}(2020)}]{wang_2020}%
  \BibitemOpen
  \bibfield  {author} {\bibinfo {author} {\bibfnamefont {Y.}~\bibnamefont
  {Wang}},\ }\bibfield  {title} {\bibinfo {title} {{Solvable Strong-Coupling
  Quantum-Dot Model with a Non-Fermi-Liquid Pairing Transition}},\ }\href
  {https://doi.org/10.1103/PhysRevLett.124.017002} {\bibfield  {journal}
  {\bibinfo  {journal} {Phys. Rev. Lett.}\ }\textbf {\bibinfo {volume} {124}},\
  \bibinfo {pages} {017002} (\bibinfo {year} {2020})}\BibitemShut {NoStop}%
\bibitem [{\citenamefont {Maldacena}\ and\ \citenamefont
  {Stanford}(2016)}]{maldacena_2016}%
  \BibitemOpen
  \bibfield  {author} {\bibinfo {author} {\bibfnamefont {J.}~\bibnamefont
  {Maldacena}}\ and\ \bibinfo {author} {\bibfnamefont {D.}~\bibnamefont
  {Stanford}},\ }\bibfield  {title} {\bibinfo {title} {Remarks on the
  {Sachdev}-{Ye}-{Kitaev} model},\ }\href
  {https://doi.org/10.1103/PhysRevD.94.106002} {\bibfield  {journal} {\bibinfo
  {journal} {Phys. Rev. D}\ }\textbf {\bibinfo {volume} {94}},\ \bibinfo
  {pages} {106002} (\bibinfo {year} {2016})}\BibitemShut {NoStop}%
\bibitem [{\citenamefont {Gu}\ \emph {et~al.}(2020)\citenamefont {Gu},
  \citenamefont {Kitaev}, \citenamefont {Sachdev},\ and\ \citenamefont
  {Tarnopolsky}}]{gu_2020}%
  \BibitemOpen
  \bibfield  {author} {\bibinfo {author} {\bibfnamefont {Y.}~\bibnamefont
  {Gu}}, \bibinfo {author} {\bibfnamefont {A.}~\bibnamefont {Kitaev}}, \bibinfo
  {author} {\bibfnamefont {S.}~\bibnamefont {Sachdev}},\ and\ \bibinfo {author}
  {\bibfnamefont {G.}~\bibnamefont {Tarnopolsky}},\ }\bibfield  {title}
  {\bibinfo {title} {Notes on the complex {Sachdev-Ye-Kitaev} model},\ }\href
  {https://doi.org/10.1007/JHEP02(2020)157} {\bibfield  {journal} {\bibinfo
  {journal} {J. High Energy Phys.}\ }\textbf {\bibinfo {volume} {2020}},\
  \bibinfo {pages} {157}}\BibitemShut {NoStop}%
\bibitem [{\citenamefont {Phillips}\ \emph {et~al.}(2022)\citenamefont
  {Phillips}, \citenamefont {Hussey},\ and\ \citenamefont
  {Abbamonte}}]{phillips_2022}%
  \BibitemOpen
  \bibfield  {author} {\bibinfo {author} {\bibfnamefont {P.~W.}\ \bibnamefont
  {Phillips}}, \bibinfo {author} {\bibfnamefont {N.~E.}\ \bibnamefont
  {Hussey}},\ and\ \bibinfo {author} {\bibfnamefont {P.}~\bibnamefont
  {Abbamonte}},\ }\bibfield  {title} {\bibinfo {title} {Stranger than metals},\
  }\href {https://doi.org/10.1126/science.abh4273} {\bibfield  {journal}
  {\bibinfo  {journal} {Science}\ }\textbf {\bibinfo {volume} {377}},\ \bibinfo
  {pages} {eabh4273} (\bibinfo {year} {2022})}\BibitemShut {NoStop}%
\bibitem [{\citenamefont {Marsiglio}(2020)}]{marsiglio_2020}%
  \BibitemOpen
  \bibfield  {author} {\bibinfo {author} {\bibfnamefont {F.}~\bibnamefont
  {Marsiglio}},\ }\bibfield  {title} {\bibinfo {title} {Eliashberg theory: A
  short review},\ }\href {https://doi.org/10.1016/j.aop.2020.168102} {\bibfield
   {journal} {\bibinfo  {journal} {Ann. Phys}\ }\textbf {\bibinfo {volume}
  {417}},\ \bibinfo {pages} {168102} (\bibinfo {year} {2020})}\BibitemShut
  {NoStop}%
\bibitem [{\citenamefont {Hauck}\ \emph {et~al.}(2020)\citenamefont {Hauck},
  \citenamefont {Klug}, \citenamefont {Esterlis},\ and\ \citenamefont
  {Schmalian}}]{hauck_2020}%
  \BibitemOpen
  \bibfield  {author} {\bibinfo {author} {\bibfnamefont {D.}~\bibnamefont
  {Hauck}}, \bibinfo {author} {\bibfnamefont {M.~J.}\ \bibnamefont {Klug}},
  \bibinfo {author} {\bibfnamefont {I.}~\bibnamefont {Esterlis}},\ and\
  \bibinfo {author} {\bibfnamefont {J.}~\bibnamefont {Schmalian}},\ }\bibfield
  {title} {\bibinfo {title} {{Eliashberg equations for an electron–phonon
  version of the Sachdev–Ye–Kitaev model: Pair breaking in non-Fermi liquid
  superconductors}},\ }\href {https://doi.org/10.1016/j.aop.2020.168120}
  {\bibfield  {journal} {\bibinfo  {journal} {Ann. Phys.}\ }\textbf {\bibinfo
  {volume} {417}},\ \bibinfo {pages} {168120} (\bibinfo {year}
  {2020})}\BibitemShut {NoStop}%
\bibitem [{\citenamefont {Esterlis}\ \emph {et~al.}(2021)\citenamefont
  {Esterlis}, \citenamefont {Guo}, \citenamefont {Patel},\ and\ \citenamefont
  {Sachdev}}]{esterlis_2021}%
  \BibitemOpen
  \bibfield  {author} {\bibinfo {author} {\bibfnamefont {I.}~\bibnamefont
  {Esterlis}}, \bibinfo {author} {\bibfnamefont {H.}~\bibnamefont {Guo}},
  \bibinfo {author} {\bibfnamefont {A.~A.}\ \bibnamefont {Patel}},\ and\
  \bibinfo {author} {\bibfnamefont {S.}~\bibnamefont {Sachdev}},\ }\bibfield
  {title} {\bibinfo {title} {Large-{N} theory of critical {Fermi} surfaces},\
  }\href {https://doi.org/10.1103/PhysRevB.103.235129} {\bibfield  {journal}
  {\bibinfo  {journal} {Phys. Rev. B}\ }\textbf {\bibinfo {volume} {103}},\
  \bibinfo {pages} {235129} (\bibinfo {year} {2021})}\BibitemShut {NoStop}%
\bibitem [{\citenamefont {Guo}\ \emph {et~al.}(2022)\citenamefont {Guo},
  \citenamefont {Patel}, \citenamefont {Esterlis},\ and\ \citenamefont
  {Sachdev}}]{guo_2022}%
  \BibitemOpen
  \bibfield  {author} {\bibinfo {author} {\bibfnamefont {H.}~\bibnamefont
  {Guo}}, \bibinfo {author} {\bibfnamefont {A.~A.}\ \bibnamefont {Patel}},
  \bibinfo {author} {\bibfnamefont {I.}~\bibnamefont {Esterlis}},\ and\
  \bibinfo {author} {\bibfnamefont {S.}~\bibnamefont {Sachdev}},\ }\bibfield
  {title} {\bibinfo {title} {{Large-N theory of critical Fermi surfaces. {II}.
  Conductivity}},\ }\href {https://doi.org/10.1103/PhysRevB.106.115151}
  {\bibfield  {journal} {\bibinfo  {journal} {Phys. Rev. B}\ }\textbf {\bibinfo
  {volume} {106}},\ \bibinfo {pages} {115151} (\bibinfo {year}
  {2022})}\BibitemShut {NoStop}%
\bibitem [{\citenamefont {Inkof}\ \emph {et~al.}()\citenamefont {Inkof},
  \citenamefont {Schalm},\ and\ \citenamefont {Schmalian}}]{inkof_2022}%
  \BibitemOpen
  \bibfield  {author} {\bibinfo {author} {\bibfnamefont {G.-A.}\ \bibnamefont
  {Inkof}}, \bibinfo {author} {\bibfnamefont {K.}~\bibnamefont {Schalm}},\ and\
  \bibinfo {author} {\bibfnamefont {J.}~\bibnamefont {Schmalian}},\ }\bibfield
  {title} {\bibinfo {title} {{Quantum critical Eliashberg theory, the
  Sachdev-Ye-Kitaev superconductor and their holographic duals}},\ }\href
  {https://doi.org/10.1038/s41535-022-00460-8} {\bibfield  {journal} {\bibinfo
  {journal} {npj Quantum Mater.}\ }\textbf {\bibinfo {volume} {7}},\ \bibinfo
  {pages} {1}}\BibitemShut {NoStop}%
\bibitem [{\citenamefont {Valentinis}\ \emph {et~al.}(0 04)\citenamefont
  {Valentinis}, \citenamefont {Inkof},\ and\ \citenamefont
  {Schmalian}}]{valentinis_2023}%
  \BibitemOpen
  \bibfield  {author} {\bibinfo {author} {\bibfnamefont {D.}~\bibnamefont
  {Valentinis}}, \bibinfo {author} {\bibfnamefont {G.~A.}\ \bibnamefont
  {Inkof}},\ and\ \bibinfo {author} {\bibfnamefont {J.}~\bibnamefont
  {Schmalian}},\ }\bibfield  {title} {\bibinfo {title} {{{BCS} to incoherent
  superconductivity crossover in the Yukawa-Sachdev-Ye-Kitaev model on a
  lattice}},\ }\href {https://doi.org/10.1103/PhysRevB.108.L140501} {\bibfield
  {journal} {\bibinfo  {journal} {Phys. Rev. B}\ }\textbf {\bibinfo {volume}
  {108}},\ \bibinfo {pages} {L140501} (\bibinfo {year}
  {2023-10-04})}\BibitemShut {NoStop}%
\bibitem [{\citenamefont {Patel}\ \emph {et~al.}(2023)\citenamefont {Patel},
  \citenamefont {Guo}, \citenamefont {Esterlis},\ and\ \citenamefont
  {Sachdev}}]{patel_2023}%
  \BibitemOpen
  \bibfield  {author} {\bibinfo {author} {\bibfnamefont {A.~A.}\ \bibnamefont
  {Patel}}, \bibinfo {author} {\bibfnamefont {H.}~\bibnamefont {Guo}}, \bibinfo
  {author} {\bibfnamefont {I.}~\bibnamefont {Esterlis}},\ and\ \bibinfo
  {author} {\bibfnamefont {S.}~\bibnamefont {Sachdev}},\ }\bibfield  {title}
  {\bibinfo {title} {Universal theory of strange metals from spatially random
  interactions},\ }\href {https://doi.org/10.1126/science.abq6011} {\bibfield
  {journal} {\bibinfo  {journal} {Science}\ }\textbf {\bibinfo {volume}
  {381}},\ \bibinfo {pages} {790} (\bibinfo {year} {2023})}\BibitemShut
  {NoStop}%
\bibitem [{\citenamefont {Li}\ \emph {et~al.}(2024)\citenamefont {Li},
  \citenamefont {Valentinis}, \citenamefont {Patel}, \citenamefont {Guo},
  \citenamefont {Schmalian}, \citenamefont {Sachdev},\ and\ \citenamefont
  {Esterlis}}]{li_2024}%
  \BibitemOpen
  \bibfield  {author} {\bibinfo {author} {\bibfnamefont {C.}~\bibnamefont
  {Li}}, \bibinfo {author} {\bibfnamefont {D.}~\bibnamefont {Valentinis}},
  \bibinfo {author} {\bibfnamefont {A.~A.}\ \bibnamefont {Patel}}, \bibinfo
  {author} {\bibfnamefont {H.}~\bibnamefont {Guo}}, \bibinfo {author}
  {\bibfnamefont {J.}~\bibnamefont {Schmalian}}, \bibinfo {author}
  {\bibfnamefont {S.}~\bibnamefont {Sachdev}},\ and\ \bibinfo {author}
  {\bibfnamefont {I.}~\bibnamefont {Esterlis}},\ }\bibfield  {title} {\bibinfo
  {title} {{Strange Metal and Superconductor in the Two-Dimensional
  Yukawa-Sachdev-Ye-Kitaev Model}},\ }\href
  {https://doi.org/10.1103/PhysRevLett.133.186502} {\bibfield  {journal}
  {\bibinfo  {journal} {Phys. Rev. Lett.}\ }\textbf {\bibinfo {volume} {133}},\
  \bibinfo {pages} {186502} (\bibinfo {year} {2024})}\BibitemShut {NoStop}%
\bibitem [{\citenamefont {Valentinis}\ \emph {et~al.}(3 18)\citenamefont
  {Valentinis}, \citenamefont {Schmalian}, \citenamefont {Sachdev},\ and\
  \citenamefont {Patel}}]{valentinis_2026}%
  \BibitemOpen
  \bibfield  {author} {\bibinfo {author} {\bibfnamefont {D.}~\bibnamefont
  {Valentinis}}, \bibinfo {author} {\bibfnamefont {J.}~\bibnamefont
  {Schmalian}}, \bibinfo {author} {\bibfnamefont {S.}~\bibnamefont {Sachdev}},\
  and\ \bibinfo {author} {\bibfnamefont {A.~A.}\ \bibnamefont {Patel}},\
  }\bibfield  {title} {\bibinfo {title} {{Superlinear Hall angle and carrier
  mobility from non-Boltzmann magnetotransport in the spatially disordered
  Yukawa-Sachdev-Ye-Kitaev model on a square lattice}},\ }\href
  {https://doi.org/10.1103/32ts-qh8d} {\bibfield  {journal} {\bibinfo
  {journal} {Phys. Rev. Res.}\ }\textbf {\bibinfo {volume} {8}},\ \bibinfo
  {pages} {013299} (\bibinfo {year} {2026-03-18})}\BibitemShut {NoStop}%
\bibitem [{\citenamefont {Davis}\ and\ \citenamefont
  {Wang}(2023)}]{Davis:2023}%
  \BibitemOpen
  \bibfield  {author} {\bibinfo {author} {\bibfnamefont {A.}~\bibnamefont
  {Davis}}\ and\ \bibinfo {author} {\bibfnamefont {Y.}~\bibnamefont {Wang}},\
  }\bibfield  {title} {\bibinfo {title} {{Quantum chaos and phase transition in
  the Yukawa--Sachdev-Ye-Kitaev model}},\ }\href
  {https://doi.org/10.1103/PhysRevB.107.205122} {\bibfield  {journal} {\bibinfo
   {journal} {Phys. Rev. B}\ }\textbf {\bibinfo {volume} {107}},\ \bibinfo
  {pages} {205122} (\bibinfo {year} {2023})}\BibitemShut {NoStop}%
\bibitem [{\citenamefont {Solis}\ \emph {et~al.}(2026)\citenamefont {Solis},
  \citenamefont {Windey}, \citenamefont {Bandyopadhyay}, \citenamefont
  {Legramandi},\ and\ \citenamefont {Hauke}}]{Hauke:2026}%
  \BibitemOpen
  \bibfield  {author} {\bibinfo {author} {\bibfnamefont {D.~P.}\ \bibnamefont
  {Solis}}, \bibinfo {author} {\bibfnamefont {A.}~\bibnamefont {Windey}},
  \bibinfo {author} {\bibfnamefont {S.}~\bibnamefont {Bandyopadhyay}}, \bibinfo
  {author} {\bibfnamefont {A.}~\bibnamefont {Legramandi}},\ and\ \bibinfo
  {author} {\bibfnamefont {P.}~\bibnamefont {Hauke}},\ }\bibfield  {title}
  {\bibinfo {title} {{From single-particle to many-body chaos in the
  Yukawa-Sachdev-Ye-Kitaev model: Theory and a cavity-QED proposal}},\ }\href
  {https://doi.org/10.1103/wntd-53rd} {\bibfield  {journal} {\bibinfo
  {journal} {Phys. Rev. B}\ }\textbf {\bibinfo {volume} {113}},\ \bibinfo
  {pages} {184121} (\bibinfo {year} {2026})}\BibitemShut {NoStop}%
\bibitem [{\citenamefont {Berry}\ and\ \citenamefont
  {Tabor}(1977)}]{Berry:1977}%
  \BibitemOpen
  \bibfield  {author} {\bibinfo {author} {\bibfnamefont {M.~V.}\ \bibnamefont
  {Berry}}\ and\ \bibinfo {author} {\bibfnamefont {M.}~\bibnamefont {Tabor}},\
  }\bibfield  {title} {\bibinfo {title} {Level clustering in the regular
  spectrum},\ }\href {https://doi.org/10.1098/rspa.1977.0140} {\bibfield
  {journal} {\bibinfo  {journal} {Proc. R. Soc. A}\ }\textbf {\bibinfo {volume}
  {356}},\ \bibinfo {pages} {375} (\bibinfo {year} {1977})}\BibitemShut
  {NoStop}%
\bibitem [{\citenamefont {Bohigas}\ \emph {et~al.}(1984)\citenamefont
  {Bohigas}, \citenamefont {Giannoni},\ and\ \citenamefont
  {Schmit}}]{Bohigas:1984}%
  \BibitemOpen
  \bibfield  {author} {\bibinfo {author} {\bibfnamefont {O.}~\bibnamefont
  {Bohigas}}, \bibinfo {author} {\bibfnamefont {M.~J.}\ \bibnamefont
  {Giannoni}},\ and\ \bibinfo {author} {\bibfnamefont {C.}~\bibnamefont
  {Schmit}},\ }\bibfield  {title} {\bibinfo {title} {Characterization of
  chaotic quantum spectra and universality of level fluctuation laws},\ }\href
  {https://doi.org/10.1103/PhysRevLett.52.1} {\bibfield  {journal} {\bibinfo
  {journal} {Phys. Rev. Lett.}\ }\textbf {\bibinfo {volume} {52}},\ \bibinfo
  {pages} {1} (\bibinfo {year} {1984})}\BibitemShut {NoStop}%
\bibitem [{\citenamefont {Altland}\ \emph {et~al.}(2021)\citenamefont
  {Altland}, \citenamefont {Bagrets}, \citenamefont {Nayak}, \citenamefont
  {Sonner},\ and\ \citenamefont {Vielma}}]{Altland_2021}%
  \BibitemOpen
  \bibfield  {author} {\bibinfo {author} {\bibfnamefont {A.}~\bibnamefont
  {Altland}}, \bibinfo {author} {\bibfnamefont {D.}~\bibnamefont {Bagrets}},
  \bibinfo {author} {\bibfnamefont {P.}~\bibnamefont {Nayak}}, \bibinfo
  {author} {\bibfnamefont {J.}~\bibnamefont {Sonner}},\ and\ \bibinfo {author}
  {\bibfnamefont {M.}~\bibnamefont {Vielma}},\ }\bibfield  {title} {\bibinfo
  {title} {{From operator statistics to wormholes}},\ }\href
  {https://doi.org/10.1103/PhysRevResearch.3.033259} {\bibfield  {journal}
  {\bibinfo  {journal} {Phys. Rev. Res.}\ }\textbf {\bibinfo {volume} {3}},\
  \bibinfo {pages} {033259} (\bibinfo {year} {2021})}\BibitemShut {NoStop}%
\bibitem [{\citenamefont {Chan}\ \emph {et~al.}(2018)\citenamefont {Chan},
  \citenamefont {De~Luca},\ and\ \citenamefont {Chalker}}]{Chalker:2018}%
  \BibitemOpen
  \bibfield  {author} {\bibinfo {author} {\bibfnamefont {A.}~\bibnamefont
  {Chan}}, \bibinfo {author} {\bibfnamefont {A.}~\bibnamefont {De~Luca}},\ and\
  \bibinfo {author} {\bibfnamefont {J.~T.}\ \bibnamefont {Chalker}},\
  }\bibfield  {title} {\bibinfo {title} {Solution of a minimal model for
  many-body quantum chaos},\ }\href {https://doi.org/10.1103/PhysRevX.8.041019}
  {\bibfield  {journal} {\bibinfo  {journal} {Phys. Rev. X}\ }\textbf {\bibinfo
  {volume} {8}},\ \bibinfo {pages} {041019} (\bibinfo {year}
  {2018})}\BibitemShut {NoStop}%
\bibitem [{\citenamefont {Friedman}\ \emph {et~al.}(2019)\citenamefont
  {Friedman}, \citenamefont {Chan}, \citenamefont {De~Luca},\ and\
  \citenamefont {Chalker}}]{Friedman:2019}%
  \BibitemOpen
  \bibfield  {author} {\bibinfo {author} {\bibfnamefont {A.~J.}\ \bibnamefont
  {Friedman}}, \bibinfo {author} {\bibfnamefont {A.}~\bibnamefont {Chan}},
  \bibinfo {author} {\bibfnamefont {A.}~\bibnamefont {De~Luca}},\ and\ \bibinfo
  {author} {\bibfnamefont {J.~T.}\ \bibnamefont {Chalker}},\ }\bibfield
  {title} {\bibinfo {title} {Spectral statistics and many-body quantum chaos
  with conserved charge},\ }\href
  {https://doi.org/10.1103/PhysRevLett.123.210603} {\bibfield  {journal}
  {\bibinfo  {journal} {Phys. Rev. Lett.}\ }\textbf {\bibinfo {volume} {123}},\
  \bibinfo {pages} {210603} (\bibinfo {year} {2019})}\BibitemShut {NoStop}%
\bibitem [{\citenamefont {Yoshimura}\ \emph {et~al.}(2025)\citenamefont
  {Yoshimura}, \citenamefont {Garratt},\ and\ \citenamefont
  {Chalker}}]{Chalker2024}%
  \BibitemOpen
  \bibfield  {author} {\bibinfo {author} {\bibfnamefont {T.}~\bibnamefont
  {Yoshimura}}, \bibinfo {author} {\bibfnamefont {S.~J.}\ \bibnamefont
  {Garratt}},\ and\ \bibinfo {author} {\bibfnamefont {J.~T.}\ \bibnamefont
  {Chalker}},\ }\bibfield  {title} {\bibinfo {title} {{Operator dynamics in
  Floquet many-body systems}},\ }\href
  {https://doi.org/10.1103/PhysRevB.111.094316} {\bibfield  {journal} {\bibinfo
   {journal} {Phys. Rev. B}\ }\textbf {\bibinfo {volume} {111}},\ \bibinfo
  {pages} {094316} (\bibinfo {year} {2025})}\BibitemShut {NoStop}%
\bibitem [{\citenamefont {Schmitz}\ and\ \citenamefont
  {Bagrets}(2026)}]{SchmitzS:2026}%
  \BibitemOpen
  \bibfield  {author} {\bibinfo {author} {\bibfnamefont {S.}~\bibnamefont
  {Schmitz}}\ and\ \bibinfo {author} {\bibfnamefont {D.}~\bibnamefont
  {Bagrets}},\ }\bibfield  {title} {\bibinfo {title} {{Quantum many body chaos
  in the Yukawa-SYK model at weak coupling}}} (\bibinfo {year} {2026}),\
  \bibinfo {note} {in preparation}\BibitemShut {NoStop}%
\bibitem [{\citenamefont {Kargi}\ \emph {et~al.}(2025)\citenamefont {Kargi},
  \citenamefont {Manatuly}, \citenamefont {Sieberer}, \citenamefont
  {Dehollain}, \citenamefont {Henriques}, \citenamefont {Olsacher},
  \citenamefont {Hauke}, \citenamefont {Heyl}, \citenamefont {Zoller},\ and\
  \citenamefont {Langford}}]{Kargi2025quantumchaos}%
  \BibitemOpen
  \bibfield  {author} {\bibinfo {author} {\bibfnamefont {C.}~\bibnamefont
  {Kargi}}, \bibinfo {author} {\bibfnamefont {A.}~\bibnamefont {Manatuly}},
  \bibinfo {author} {\bibfnamefont {L.~M.}\ \bibnamefont {Sieberer}}, \bibinfo
  {author} {\bibfnamefont {J.~P.}\ \bibnamefont {Dehollain}}, \bibinfo {author}
  {\bibfnamefont {F.}~\bibnamefont {Henriques}}, \bibinfo {author}
  {\bibfnamefont {T.}~\bibnamefont {Olsacher}}, \bibinfo {author}
  {\bibfnamefont {P.}~\bibnamefont {Hauke}}, \bibinfo {author} {\bibfnamefont
  {M.}~\bibnamefont {Heyl}}, \bibinfo {author} {\bibfnamefont {P.}~\bibnamefont
  {Zoller}},\ and\ \bibinfo {author} {\bibfnamefont {N.~K.}\ \bibnamefont
  {Langford}},\ }\bibfield  {title} {\bibinfo {title} {Quantum {C}haos and
  {U}niversal {T}rotterisation {B}ehaviours in {D}igital {Q}uantum
  {S}imulations},\ }\href {https://doi.org/10.22331/q-2025-12-02-1924}
  {\bibfield  {journal} {\bibinfo  {journal} {{Quantum}}\ }\textbf {\bibinfo
  {volume} {9}},\ \bibinfo {pages} {1924} (\bibinfo {year} {2025})}\BibitemShut
  {NoStop}%
\end{thebibliography}
\end{document}